\documentclass[11pt]{article}

\usepackage{acl}

\usepackage{times}
\usepackage{latexsym}

\usepackage[T1]{fontenc}

\usepackage[utf8]{inputenc}

\usepackage{microtype}
\usepackage{multirow}
\usepackage{inconsolata}

\usepackage{graphicx}
\usepackage{booktabs}
\usepackage{tabularx}
\usepackage{array}
\usepackage{makecell}
\usepackage[table]{xcolor}
\usepackage{longtable}
\usepackage{enumitem}
\usepackage[dvipsnames]{xcolor}

\usepackage[most]{tcolorbox}

\usepackage{soul}

\usepackage{xcolor}

\definecolor{lightgrayhighlight}{RGB}{235,235,235}

\sethlcolor{lightgrayhighlight}

\definecolor{DomainLegal}{HTML}{FDECC8}
\definecolor{DomainMedical}{HTML}{FADADD}
\definecolor{DomainAcademic}{HTML}{E8EAF6}
\definecolor{DomainTechnical}{HTML}{E0F2F1}
\definecolor{DomainMedia}{HTML}{FFF4CC}
\definecolor{DomainPsychological}{HTML}{F3E5F5}
\definecolor{DomainFinancial}{HTML}{E2F0D9}
\definecolor{DomainEducational}{HTML}{D9EEF7}
\definecolor{PromptFrame}{HTML}{4B5D67}
\definecolor{PromptBack}{HTML}{F7FAFC}

\newtcolorbox{promptbox}[1]{
  enhanced,
  breakable,
  colback=PromptBack,
  colframe=PromptFrame,
  coltitle=white,
  fonttitle=\bfseries,
  title=#1,
  boxrule=0.6pt,
  arc=1.5mm,
  left=1.2mm,
  right=1.2mm,
  top=1mm,
  bottom=1mm
}

\title{
  \raisebox{-0.5ex}{\includegraphics[width=1.0cm]{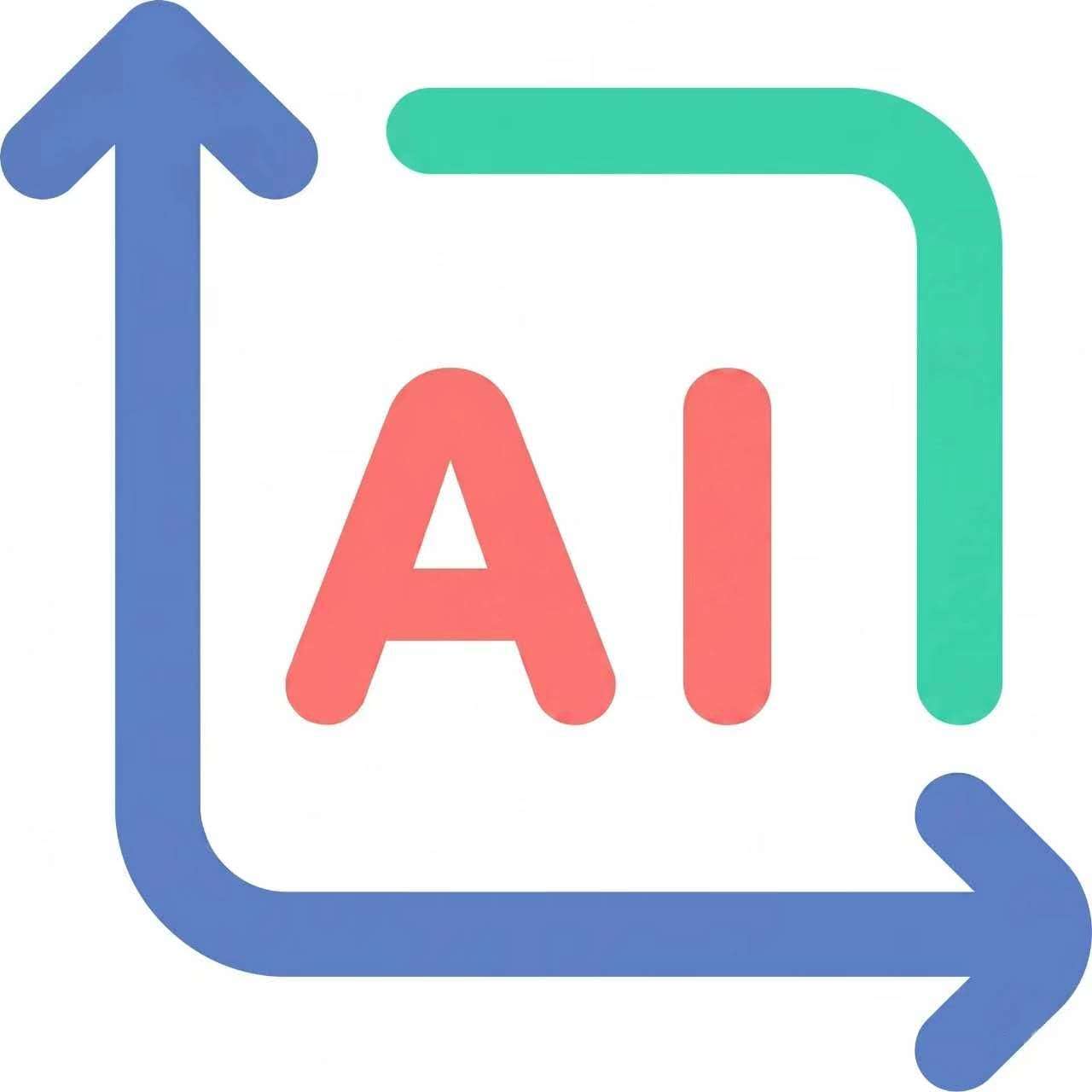}} 
  \begin{tabular}{c}
    AREAs-Lab: An Interactive Environment for \textbf{A}I-driven \\
    \textbf{R}equirement \textbf{E}licitation for \textbf{A}I Systems
  \end{tabular}
}

\author{
\textbf{Pengshan Cai}\thanks{Equal contribution.}\textsuperscript{\(\dagger\)},
\textbf{Zihao Zhang}\footnotemark[1]\textsuperscript{\(\ddagger\)},
\textbf{Ting Jin}\textsuperscript{\(\S\)},
\textbf{Chenyang Zhu}\textsuperscript{\(\dagger\)},
\textbf{Kushal Chawla}\textsuperscript{\(\dagger\)},
\textbf{Sangwoo Cho}\textsuperscript{\(\dagger\)}
\\
\textbf{Scott Novotney}\textsuperscript{\(\dagger\)},
\textbf{Yebowen Hu}\textsuperscript{\(\P\)},
\textbf{Fei Liu}\textsuperscript{\(\ddagger\)},
\textbf{Shi-Xiong Zhang}\textsuperscript{\(\dagger\)},
\textbf{Sambit Sahu}\textsuperscript{\(\dagger\)}
\\[2pt]
\textsuperscript{\(\dagger\)}AI Foundations, Capital One
\qquad
\textsuperscript{\(\ddagger\)}Emory University
\\
\textsuperscript{\(\S\)}University of Wisconsin--Madison
\qquad
\textsuperscript{\(\P\)}University of Central Florida
\\[3pt]
{\texttt{\{pengshan.cai, chenyang.zhu, kushal.chawla, sangwoo.cho,}}
\\[-1pt]
{\texttt{scott.novotney, shixiong.zhang, sambit.sahu\}@capitalone.com}}
\\[-1pt]
{\texttt{\{zihao.zhang, fei.liu\}@emory.edu}\quad \texttt{tjin27@wisc.edu}\quad\texttt{yebowen.hu@ucf.edu}}
}

\begin{document}
\maketitle
\begin{abstract}
Building effective AI systems increasingly depends on writing high-quality task requirements, yet users often struggle to articulate the constraints, preferences, and edge cases that determine success. This problem is especially acute in AI development, where behavior is shaped not only by human expectations but also by data characteristics. We present \textbf{AREAs-Lab}, an interactive environment for \textbf{A}I-driven \textbf{R}equirement \textbf{E}licitation for \textbf{A}I \textbf{s}ystems. In AREAs-Lab, an assistant iteratively refines an initially incomplete requirement by analyzing the underlying dataset and asking targeted clarification questions to uncover the user's latent intent. To study this setting systematically, we construct a synthetic benchmark grounded in 16 public datasets spanning diverse domains and task types. Each benchmark instance includes a user profile, a complete reference requirement, and an intentionally underspecified version that serves as the assistant's starting point. We further introduce an automated evaluation pipeline based on an AI-simulated user that reveals hidden information only when appropriately prompted, enabling scalable and reproducible assessment of interactive elicitation quality. 
AREAs-Lab provides a controlled testbed for studying how AI assistants can transform vague user goals into actionable requirements for AI systems.
\end{abstract}

\section{Introduction}

\begin{figure}[!t]
  \centering
  \includegraphics[width=\columnwidth]{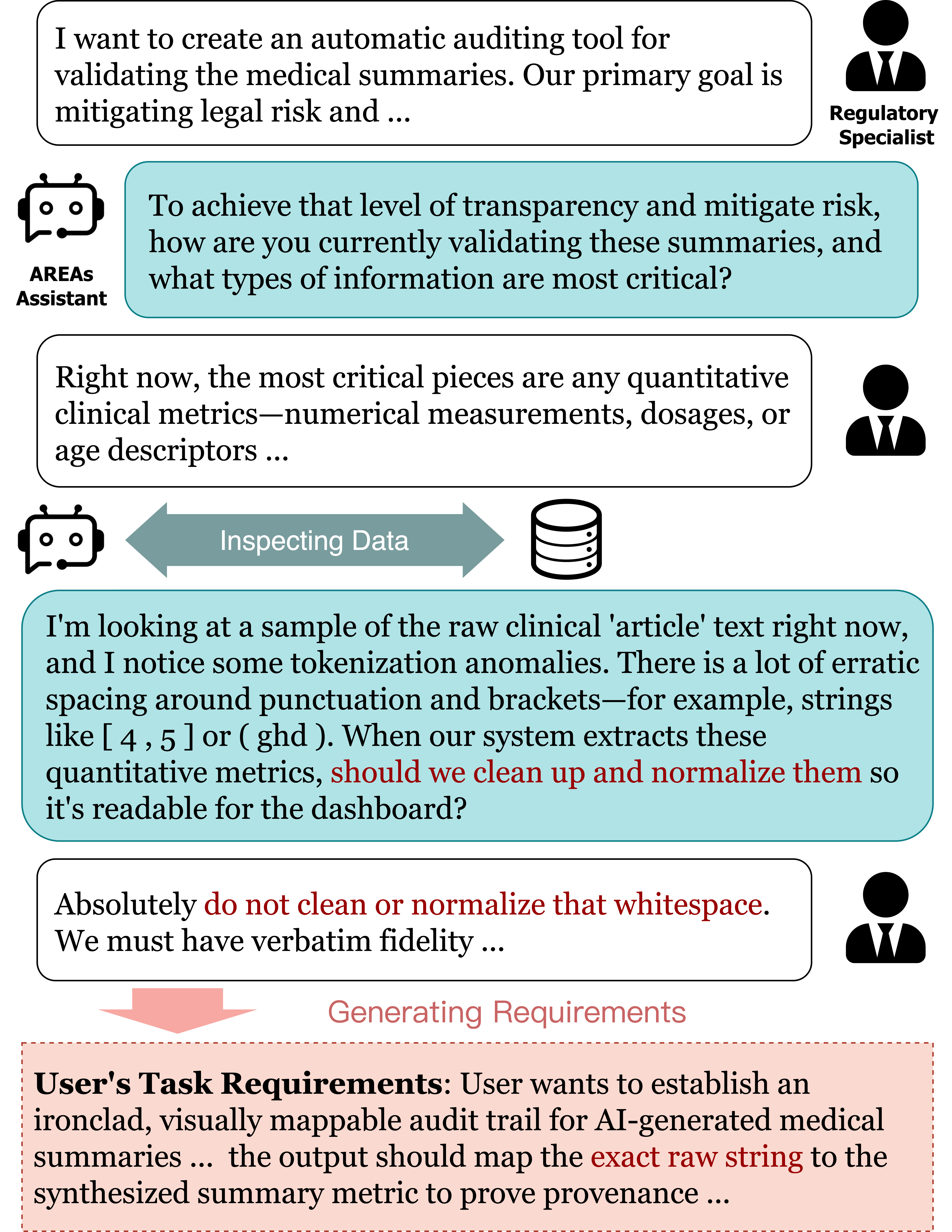}
  \caption{An illustration of the AREAs assistant supporting a Regulatory Compliance Specialist in developing an automated traceability audit tool for medical literature. When users initially propose a task, they often overlook specific data characteristics (e.g., tokenization anomalies), leading to incomplete requirements. The AREAs assistant identifies these latent needs by analyzing the data and resolving ambiguities through targeted clarifying questions.}
  \label{fig:interactive-demo}
\end{figure}

The artificial intelligence paradigm is rapidly shifting from standalone tools to autonomous, collaborative agents that interact with humans through natural language to solve complex, open-ended tasks \cite{zou2025survey, sami2026bridging}. As highlighted by recent industrial deployments—such as NVIDIA's integration of millions of AI agents alongside human workflows \cite{huang2026futurework}—the future of work hinges on seamless human-agent cooperation. However, as these autonomous systems enter high-stakes professional domains, a critical methodological bottleneck emerges: \textbf{AI system requirement elicitation}. Effectively capturing a user's true, often latent expectations and translating vague, high-level objectives into operational task specifications remains an unresolved challenge \cite{ni-etal-2026-survey}.

\begin{figure*}[t]
  \centering
  \includegraphics[width=0.9\textwidth]{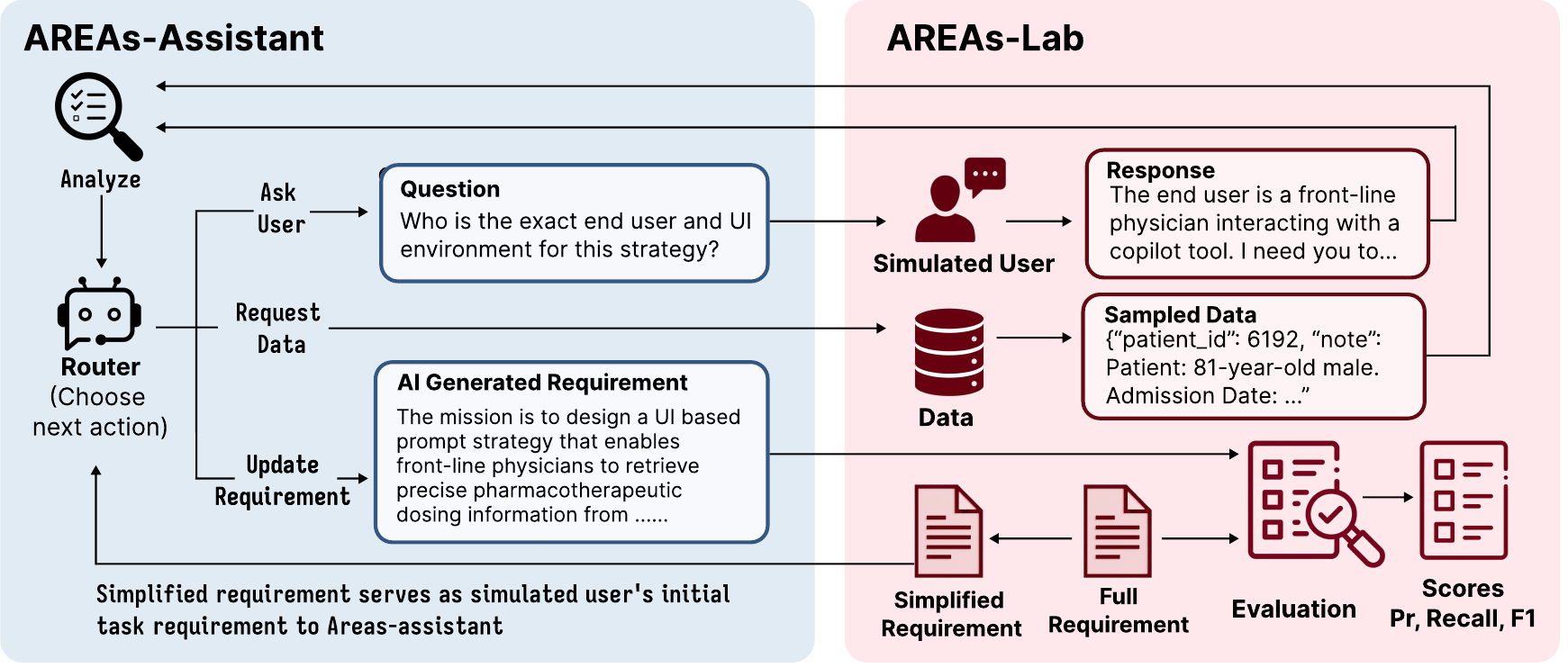}
  \caption{\textbf{Left}: the AREAs assistant performs \textit{Hybrid Interaction}. For the \textit{Data Interaction} and \textit{User Interaction} baselines, the router is restricted from interacting with users or analyzing data, respectively, but is permitted to update the requirement in both modes. \textbf{Right}: AREAs-Lab serves as a simulated environment to evaluate the quality of AI-generated requirements against reference requirements.}
  \label{fig:interactive-approaches}
\end{figure*}

This challenge stems from a fundamental paradigm shift: eliciting requirements for AI systems differs fundamentally from traditional software engineering. In traditional systems, requirements are functional and deterministic, centering on explicit business rules and rigid process control (e.g., "if-then-else" logic) that developers can translate directly into fixed instructions. In contrast, AI specifications are inherently stochastic and tightly coupled with data-centric learning \cite{sambasivan2021everyone, belani2019requirements, karpathy2017software}. Rather than being finalized in a single upfront pass, optimal AI requirements must emerge through an iterative process of data exploration and human interaction, where users dynamically refine their goals as they uncover new patterns and edge cases within the dataset \cite{shankar2024validates, kim2024evallm}. In practice, however, human users often fail to conduct a granular review of the underlying dataset before formulating their initial requests, routinely overlooking highly specific, data-dependent constraints. To bridge this gap, the AREAs assistant serves as an interactive auditor that uncovers these latent data characteristics and proactively generates targeted clarification questions, thereby guiding users to articulate their unexpressed intents.

As shown in Figure \ref{fig:interactive-demo}, we investigate how an AI-powered assistant can proactively support AI system requirement elicitation. We propose AREAs-Lab, an interactive environment designed for AI-driven requirement elicitation. In this setting, a human user provides an initial task requirement that is inherently incomplete or ambiguous. An LLM-powered requirements elicitation agent (referred to as the AREAs assistant) then analyzes the underlying dataset, identifies critical data features and underspecified elements in the current requirement, and poses clarification questions. This collaborative loop continues until the specification aligns with the user’s latent intent, transforming a vague instruction into a high-fidelity operational plan.

To operationalize this task within a controlled sandbox, we construct a synthetic benchmark derived from 16 diverse public datasets. For each task, we generate a comprehensive user profile, a "ground-truth" requirement, and a "skeletal" requirement, with the latter constructed by intentionally withholding essential constraints and context. This design mirrors an established paradigm of evaluating an interactive AREAs assistant through controlled informational asymmetry \citep{jin2026reqelicitgym, acikgoz2026mac, hemmat2025research}.
Because the target task requirements are jointly shaped by the user's profile and the intrinsic characteristics of the data, this benchmark introduces a unique challenge for requirement elicitation. To recover the underlying specifications, the AREAs assistant must simultaneously mine key features from the data and interact effectively with the user.
We quantify performance by measuring the fidelity of the recovered specification relative to the ground-truth reference, thereby evaluating the AREAs assistant's ability to bridge the specification gap through high-value dialogue.

As shown on the right side of Figure~\ref{fig:interactive-approaches}, our framework centers on an automated evaluation pipeline driven by an AI-simulated user \cite{park2024generative, ni-etal-2026-survey, wu-etal-2025-embracing}. As a proxy for a human participant, the simulated user encapsulates the original, comprehensive requirements as latent background knowledge, selectively disclosing relevant constraints only in response to targeted inquiries from the AREAs assistant.
This interactive configuration enables the scalable and reproducible assessment of the AREAs assistant's ability to elicit critical information and reconstruct complex user intent. To ensure the reliability of this approach, we implemented various response strategies and conducted a comparative analysis between simulated and human responses. Our comparison suggests that the simulator provides a scalable and controlled proxy for selected disclosure behaviors, while also exhibiting measurable differences from real users.

Our main contributions are as follows: (1) We introduce AREAs-Lab, a novel, domain-agnostic interactive environment for AI system requirement elicitation that, unlike existing frameworks, prioritizes the simultaneous interaction between the user and the underlying data. (2) We construct a robust synthetic benchmark and conduct an extensive empirical evaluation, demonstrating the utility of the AREAs-Lab framework while revealing critical capability gaps and challenges for state-of-the-art agents. Our code and data are available at \url{https://github.com/cpengshan/AREAs-Lab}.

\begin{figure*}[t]
  \centering
  \includegraphics[width=\textwidth]{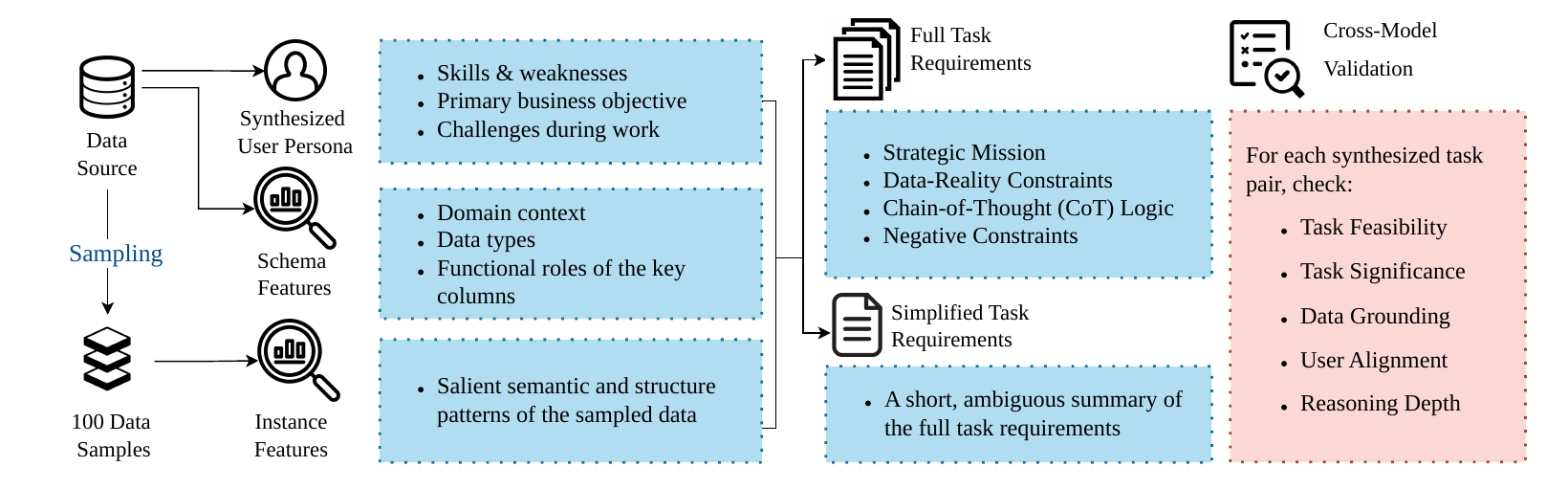}
  \caption{Overview of the data synthesis pipeline. We extract data features and synthesize user personas, integrating both elements during task generation to ensure requirements reflect both user needs and data characteristics. Finally, cross-model validation and manual spot checks are conducted to verify that synthesized tasks are high-quality, meaningful, and well-aligned with the target personas and data features.}
  \label{fig:data-synthesis-pipeline}
\end{figure*}

\section{Related Works}

\subsection{AI-assisted Requirement Elicitation}

Recent research in AI-driven requirement elicitation has explored diverse paradigms, ranging from the use of LLMs for autonomous interviews \cite{korn2025llmrei} and multi-agent frameworks such as KGMAF \cite{huang2025knowledge}, to structured dialogue analysis via RECOVER \cite{voria2025recover} and the development of 'human-in-the-loop' collaborative frameworks \cite{abbasi2025towards}.
While some studies investigate optimal interaction strategies \cite{arnaudo2026evaluating, Lietal2026} or use AI to train humans in requirements elicitation skills \cite{lojo2025using}, existing benchmarks, such as ReqElicitGym \cite{Lietal2026}, focus on domains like web design where requirements are relatively independent of underlying data.
Our work differs by targeting requirement elicitation where the AI assistant must deeply analyze the underlying data to formulate relevant clarification questions.  Unlike previous environments focused on traditional software development, our framework is designed for AI system specifications, in which requirements dynamically emerge from the interplay between user intent and data characteristics. Furthermore, our benchmark offers significantly greater evaluation diversity, featuring $16$ data sources across $10$ distinct domains.

\subsection{Automatic Prompt Tuning}
Research in automatic prompt tuning has evolved from early search-based and RL-driven methods like APE \cite{zhou2022large} and RLPrompt \cite{deng-etal-2022-rlprompt} toward more sophisticated frameworks like OPRO \cite{yang2024large} and DSPy \cite{khattab2023dspy}, which treat prompt optimization as a meta-reasoning or compilation process. Recent advancements further explore this space through evolutionary algorithms like EvoPrompt \cite{guo2024connecting} and Bayesian optimization for black-box models in InstructZero \cite{chen2023instructzero}. Additionally, works such as TextGrad \cite{yuksekgonul2024textgrad} and APO \cite{pryzant-etal-2023-automatic} utilize textual gradients and automated critiques to refine instructions autonomously.
While these methods excel at optimizing for fixed objective metrics or gold labels via automated validation, our work diverges by utilizing human feedback to drive the tuning process. This allows our system to navigate the inherent ambiguity of agentic tasks, where the optimal prompt must align with nuanced user intent that cannot be captured by automatic validation or static datasets alone.

\subsection{AI Environment}
Recent work in AI environments spans structured reasoning \cite{yu2025learning}, agentic robustness \cite{gu2025robust}, user-centered design frameworks \cite{zhou2026generative}, and social behavior simulation within generative agent architectures \cite{park2023generative}. While these AI environments prioritize general logic or experience ontologies, AREAs-Lab formalizes user-needs understanding through an automated evaluation protocol.

\section{Environment Construction}

\subsection{Data Source}

As shown in Table \ref{tab:data_sources}, our benchmark construction begins with 16 publicly available data sources spanning multiple domains and task settings. We select these sources to ensure variation in user goals and task complexity, so that the resulting benchmark reflects a broad range of realistic requirement-elicitation scenarios.
These data sources are not used directly as benchmark instances. Instead, they serve as grounding material for synthesizing user personas,  full and simplified task requirements. This design allows building controlled evaluation settings while maintaining diversity and realism in the underlying tasks.

\subsection{Data Synthesis}

As shown in Figure \ref{fig:data-synthesis-pipeline}, we construct the synthesized dataset through four stages. 

\noindent\textbf{Step 1: Data Features Extraction.} We employ an LLM to extract two features from each target data source. \textbf{Schema features} encompass the domain context, data types, and functional roles of key columns derived from the dataset's metadata (e.g., Hugging Face webpage). \textbf{Instance features} were extracted by analyzing representative semantic and structural patterns from 100 data instances randomly sampled from the dataset. These features collectively inform the synthesis of tasks, simulating realistic scenarios where users define requirements based on specific data characteristics.

\noindent\textbf{Step 2: User Persona Synthesis.} Next, we generate five realistic user personas for each data source. We prompt the LLM to produce a diverse set of personas so that the synthesized users reflect a broad range of real-world professionals with different levels of technical expertise and distinct business priorities. Each persona is associated with a skill profile, a primary business objective, and a set of challenges encountered when working with the corresponding dataset. 

\noindent\textbf{Step 3: Task Requirement Synthesis.} We then use LLM to transform the extracted data features and synthesized user personas into high-fidelity task requirements. To promote diversity in reasoning complexity, we generate two tasks for each persona at different difficulty levels: \emph{Medium-difficulty tasks} focus on interpretive reasoning, such as inferring implicit intent or identifying primary objectives. \emph{High-difficulty tasks} require resolving complex or conflicting constraints, such as reconciling contradictions or making decisions under ambiguity. Each task is formulated as a comprehensive system prompt detailing execution steps, data-weighting instructions, and an evaluation rubric.
Finally, we condense the full task requirement into a simplified, informal variant—a concise, goal-oriented description of approximately 30 words. This simplified variant serves as the initial instruction that the AREAs-Lab presents to the agent (AREAs assistant), demonstrated on Figure~\ref{fig:interactive-approaches}.

\noindent\textbf{Step 4: Quality Control and Validation.} To ensure the quality of the synthesized benchmark, we combine both human audit and cross-model LLM validation throughout the synthesis process. First, we iteratively refined the generation instructions and asked two annotators to manually audit a random sample of 30 instances in each pilot run, checking data consistency, persona-task alignment, and overall requirement quality.

In parallel, we introduce a second-stage cross-model validation procedure, where two high-capability LLMs independently audit the outputs generated by the first model. This design helps reduce self-reinforcement bias and provides an additional quality-control mechanism for the synthesis pipeline. Specifically, the auditing models evaluate each persona--task triplet along six dimensions, including \textit{Role-Schema Alignment}, \textit{Reasoning Depth}, and \textit{Data Grounding}, etc. assigning a score from 1 to 5 for each dimension.

We then apply two filtering thresholds to retain only high-quality instances: any instance with an average overall score below 4.0, or with a score below 3.0 on any critical dimension, is discarded. After this filtration process, we obtain a final set of 151 (\textit{data source}, \textit{user}, \textit{task}) tuples.

To independently validate the quality of the synthesized benchmark, we additionally conducted a human assessment of 16 randomly sampled instances, with one instance selected from each data source. Two annotators evaluated every instance using the same six-dimensional rubric employed in cross-model validation. The instances received a mean score of 4.67 out of 5.0, and 93.8\% of instance--dimension pairs received scores of at least 4 from both annotators. Complete protocols and agreement analyses are provided in Appendix~\ref{sec:additional_human_validation}.

\subsection{Simulated Users}
To rigorously evaluate AREAs assistant’s requirement-elicitation ability, we use simulated users under an information-asymmetry setting. During each interaction, the simulated user responds based on four information sources: 1) the assistant’s question; 2) user persona; 3) full task requirement; and 4) user’s interactive style. We strictly prohibit the simulated user from proactively revealing task requirements unless the assistant asks a relevant question. This design reflects realistic settings where users often cannot fully articulate their needs upfront and disclose critical details only when properly prompted. For unrelated questions, the simulated user may express uncertainty, mild pushback, or indicate that they have not considered the issue in detail.

To capture different collaborative styles observed in real users, we define three interaction styles:
\textbf{Passive:} Responses are limited in expression. Does not intentionally withhold information, but struggles to articulate details.
\textbf{Normal:} Responses are direct and clear. Use well-formed sentences without volunteering additional context.
\textbf{Active:} Responses are conversational and helpful. Proactively disclose relevant detail from the latent requirement when appropriately triggered.

\section{Experimental Design}

We introduce our comparitive baselines (Section \ref{sec: cmp_base}) and evaluation framework (Section \ref{sec: eval_frame}).
Our experiments are organized around the following research questions: 
(1) Which strategies yield the highest efficacy in AI-driven requirement elicitation? (Section \ref{sec: interactive_strategy_results});
(2) To what extent do iterative interaction rounds correlate with performance gains for the user? (Section \ref{sec: round_results});
(3) How does user cooperation influence the quality of the elicited requirements? (Section \ref{sec:user_colab}).

\subsection{Comparative Baselines}
\label{sec: cmp_base}

We compare five requirement elicitation strategies. \textit{No Interaction} serves as a baseline in which the AREAs assistant reconstructs the full requirement directly from the simplified requirement. \textit{Data Interaction} provides the assistant with direct access to dataset exemplars, enabling it to infer requirements grounded in data structures, formatting artifacts, schema conventions, and potential edge cases. \textit{User Interaction (Fixed)} asks the simulated user the same ten predefined questions for every task (See Section~\ref{sec:user_int_fixed_prompt}), covering general specification dimensions such as the task objective, input characteristics, automation boundary, target audience, output format, and use of external knowledge. In contrast, \textit{User Interaction (Adaptive)} dynamically generates each clarification question based on the task context and the user's previous responses, targeting unresolved goals, preferences, constraints, and quality standards. Finally, \textit{Hybrid Interaction} (Figure~\ref{fig:interactive-approaches}, left) combines adaptive user interaction with data inspection, allowing the assistant to draw on both user feedback and data-derived evidence.

For \textit{User Interaction} and \textit{Hybrid Interaction}, we run 10 rounds to allow for sufficient interaction. For \textit{Data Interaction}, we use one interaction round, as additional rounds of iteration do not significantly improve performance (Will discuss in Section \ref{sec: round_results}). We defer implementation details to Appendix ~\ref{sec:imp_detl}.

\subsection{Evaluation Framework}
\label{sec: eval_frame}

Figure~\ref{fig:interactive-approaches} (Right) presents a simplified overview of our evaluation pipeline, while the detailed evaluation framework is provided in Figure~\ref{fig:evaluation_process}. Specifically, we use a simulated user to interact with the AREAs assistant in a conversation initiated by the simulated user presenting a simplified task requirement. After the conversation, the AREAs assistant generates a task requirement, which is then compared against the reference full task requirement. We measure the performance of the AREAs assistant by comparing the generated requirements with the reference requirements.

Following the atomic evaluation paradigm \cite{min2023factscore, ragas2024}, we decompose both reference and generated task requirements into individual sets of atomic units. Each unit captures a singular, indivisible component, such as a specific task objective, constraint, or domain criterion. To evaluate these units, we employ an LLM-based matching function to perform semantic alignment, identifying valid directional entailments or entailment pairs between the generated and reference sets. Finally, we quantify this alignment using Precision, Recall, and F1 scores: Precision measures the fidelity of the generated requirements by penalizing hallucinations or irrelevant redundancies; Recall assesses the coverage of the ground-truth specification; and F1 provides a balanced harmonic mean of both dimensions.

\section{Main Results and Analysis}
\begin{table*}[t]
\centering

\footnotesize
\setlength{\tabcolsep}{3.5pt}

\begin{tabular}{lccccc|cccc|c}
\toprule

\multirow{2}{*}{Interaction Strategy}
& \multicolumn{5}{c}{Individual Dataset F1 (\%)}
& \multicolumn{4}{c}{Average Metrics (\%)}
& \multirow{2}{*}{\# of Word} \\

\cmidrule(lr){2-6}
\cmidrule(lr){7-10}

& MultiNews
& Clinical
& ESConv
& ArXivSum
& MediaSum
& Pr.
& Re.
& F1 avg.
& F1 std.
& \\

\midrule

\multicolumn{11}{c}{\textit{Claude-Sonnet-4.6}} \\
\midrule
\textit{No Int.} & 33.81 & 23.90 & 24.40 & 22.36 & 26.13 & 19.21 & 32.46 & 23.89 & 7.37 & 371 \\
\textit{Data Int.} & 27.31 & 26.33 & 25.76 & 28.83 & 26.91 & 20.20 & 36.51 & 25.80 & 7.82 & 449 \\
\textit{User Int. (Fixed)} & 38.42 & 34.94 & 30.90 & 38.86 & 36.86 & 27.67 & \textbf{64.73} & 38.55 & 8.62 & 479 \\
\textit{User Int. (Adaptive)} & 36.50 & 33.18 & \textbf{37.42} & 38.61 & 35.61 & 27.54 & 59.30 & 37.32 & 8.03 & 525 \\
\textit{Hybrid Int.} & \textbf{44.60} & \textbf{48.60} & 29.87 & \textbf{40.34} & \textbf{38.85} & \textbf{32.95} & 54.61 & \textbf{40.76} & 9.58 & 450 \\

\midrule

\multicolumn{11}{c}{\textit{GPT-5.4}} \\
\midrule
\textit{No Int.} & 29.45 & 19.97 & 21.24 & 22.33 & 26.86 & 15.93 & 41.36 & 22.85 & 5.82 & 481 \\
\textit{Data Int.} & 31.87 & 23.81 & 21.38 & 26.67 & 21.76 & 17.42 & 40.13 & 24.13 & 6.81 & 418 \\
\textit{User Int. (Fixed)} & 35.59 & 36.83 & 37.08 & 37.20 & 39.66 & 26.24 & \textbf{69.31} & 37.87 & 7.00 & 470 \\
\textit{User Int. (Adaptive)} & 40.93 & 38.55 & 40.19 & \textbf{39.58} & 40.50 & 28.58 & 64.98 & 39.23 & 7.97 & 480 \\
\textit{Hybrid Int.} & \textbf{46.91} & \textbf{42.06} & \textbf{40.92} & 36.85 & \textbf{43.49} & \textbf{30.45} & 60.10 & \textbf{40.07} & 7.99 & 462 \\

\midrule

\multicolumn{11}{c}{\textit{Gemini-3.1-Pro}} \\
\midrule
\textit{No Int.} & 35.71 & 20.60 & 26.35 & 24.67 & 27.63 & 21.27 & 31.58 & 25.17 & 7.79 & 360 \\
\textit{Data Int.} & 29.31 & 25.34 & 28.65 & 22.92 & 25.71 & 23.68 & 34.08 & 27.66 & 8.07 & 373 \\
\textit{User Int. (Fixed)} & 44.20 & 40.05 & 41.62 & 38.67 & 41.14 & 31.40 & \textbf{56.46} & 39.81 & 10.51 & 346 \\
\textit{User Int. (Adaptive)} & 43.37 & 33.83 & \textbf{43.39} & 37.90 & \textbf{42.57} & 33.79 & 51.05 & 40.09 & 9.42 & 357 \\
\textit{Hybrid Int.} & \textbf{45.43} & \textbf{44.84} & 42.45 & \textbf{40.31} & 37.59 & \textbf{36.24} & 49.87 & \textbf{41.64} & 9.26 & 339 \\

\bottomrule
\end{tabular}

\caption{
Main results across three backbone models on five representative datasets.
We report dataset-level F1 scores, along with macro-averaged Precision (Pr.),
Recall (Re.), F1 average (F1 avg.), and standard deviation (F1 std.).
Word count denotes the average generated requirement length.
\textit{Int.} denotes \textit{Interaction}. \emph{Claude-Sonnet-4.6} serves as the default backbone for the remainder of this section. Full results are deferred to Tables~\ref{tab:main_results_app} and \ref{tab:main_results_app_cont} due to space constraints.
}

\label{tab:main-results}

\end{table*}

\begin{table}[t]
\centering
\scriptsize
\setlength{\tabcolsep}{3pt}
\begin{tabular}{llcccc}
\toprule
\textbf{Req. Type} & \textbf{Strategy} & \textbf{Precision} & \textbf{Recall} & \textbf{F1} & \textbf{F1 std} \\
\midrule
\multirow{4}{*}{\textsc{User Def.}}
& \textit{No Int.} & 16.73 & 38.24 & 22.67 & 7.08\\
& \textit{Data Int.} & 19.97 & 38.09 & 25.59 & 8.06\\
& \textit{User Int. (Fixed)} & 23.25 & \textbf{67.30} & 34.56 & 9.09\\
& \textit{User Int. (Adaptive)} & 26.73 & 59.73 & 36.13 & 8.90\\
& \textit{Hybrid Int.} & \textbf{29.24} & 55.32 & \textbf{37.55} & 9.66\\
\midrule
\multirow{4}{*}{\textsc{Data Der.}}
& \textit{No Int.} & 20.44 & 25.09 & 20.21 & 8.55\\
& \textit{Data Int.} & 28.57 & 33.57 & 27.58 & 14.71\\
& \textit{User Int. (Fixed)} & 23.76 & 40.22 & 29.87 & 18.53\\
& \textit{User Int. (Adaptive)} & 34.29 & 48.17 & 40.06 & 22.21\\
& \textit{Hybrid Int.} & \textbf{35.72} & \textbf{49.60} & \textbf{41.54}  & 21.94\\
\bottomrule
\end{tabular}
\caption{Performance comparison across interaction strategies on user-defined (\textsc{User Def.}) and data-derived (\textsc{Data Der.}) task requirements.}
\label{tab:categorize_results}
\end{table}

\subsection{Interaction Strategy Comparison}
\label{sec: interactive_strategy_results}

Table~\ref{tab:main-results} summarizes the performance of the five strategies. Overall, all interaction-based strategies outperform \textit{No Interaction}. \textit{Hybrid Interaction} achieves the highest overall $F_1$ across all three backbone models, suggesting that combining adaptive user clarification with direct data inspection provides the strongest balance between requirement coverage and precision.

\noindent\textbf{Benefits of Combining User and Data Interaction.}
Although both \textit{User Int. (Adaptive)} and \textit{Hybrid Int.} run for 10 interaction rounds, \textit{Hybrid Int.} achieves higher $F_1$ by incorporating data-grounded clarification. Unlike user-only interaction, Hybrid grounds its clarification in data inspection, leading to more task-specific clarification (Table~\ref{tab:case_user_vs_hybrid}).

Furthermore, all interaction-based strategies exhibit higher $F_1$ variance than \textit{No Int.}. Although interaction generally improves requirement coverage, it can also introduce irrelevant details, redundant specifications, or conflicting constraints, resulting in greater performance variability across tasks (see Table~\ref{tab:case_data_interaction_1}).

Despite these gains, AREAs-Lab remains challenging for frontier models, with the best overall $F_1$ reaching only 41.64. This reflects the benchmark's deliberate information asymmetry and the strict atomic evaluation metric, which may not credit vague, partial, or implicit matches. Nevertheless, because all strategies share the same reference requirements and evaluation protocol, their relative comparisons remain meaningful.

\noindent\textbf{Fixed vs. Adaptive User Interaction.}
\textit{User Int. (Fixed)} is a strong non-adaptive baseline, performing comparably to \textit{User Int. (Adaptive)} across backbones. However, the two strategies exhibit a clear precision--recall trade-off: fixed questions consistently achieve higher recall (by 4.33--5.43 points), whereas adaptive questions generally achieve higher precision. This suggests that fixed questions broadly cover common specification dimensions, while adaptive questions more selectively target task-specific gaps.

\noindent\textbf{Requirement Source Analysis.}
To further study how different interaction strategies elicit specific types of information, we use an LLM to classify statements in the gold requirements into two categories (see Table~\ref{tab:prompt_req_cls} for the prompt):
\textsc{Data-derived requirements} arise from intrinsic dataset characteristics, such as input schemas and data-handling protocols;
\textsc{User-defined requirements} capture users' instructions, preferences, objectives, and quality standards independent of the underlying data structure.
We compute Precision, Recall, and $F_1$ for both requirement types.

According to Table~\ref{tab:categorize_results}, all approaches generally achieve higher precision on \textsc{Data-derived Requirements} and higher recall on \textsc{User-defined Requirements}. This suggests that models identify concrete data specifications more precisely but capture broader user preferences more exhaustively.

\textsc{Data-derived Requirements} also exhibit substantially higher $F_1$ variance, particularly under the interaction-based strategies. This likely reflects differences in the complexity and salience of dataset-specific constraints: some can be identified directly from schemas or common examples, whereas others require recognizing subtle patterns or rare edge cases across multiple instances. Consequently, the effectiveness of data-related elicitation varies more widely across tasks.

The advantage of adaptive questioning is particularly pronounced for data-derived requirements: \textit{User Int. (Adaptive)} substantially outperforms both \textit{User Int. (Fixed)} and \textit{Data Interaction} ($F_1$ of 40.06 vs.\ 29.87 and 27.58, respectively). Our analysis shows that adaptive interaction proactively asks users about expected data characteristics (see Table~\ref{tab:case_user_interaction}). This suggests that contextual follow-up questions can uncover task-specific data constraints even without direct data access.

Counter-intuitively, for data-derived requirements, \textit{User Int. (Adaptive)} substantially outperforms \textit{Data Int.} ($F_1$ of 40.06 vs.\ 27.58), though both outperform \textit{No Int.}. Analysis shows that adaptive interaction proactively asks the user about expected data characteristics (see Table~\ref{tab:case_user_interaction}). This implies that even without direct data access, the assistant can reason about likely data features and formulate effective clarification questions.

\subsection{Impact of Interaction Rounds}
\label{sec: round_results}

\begin{figure*}[t]
  \centering
  \includegraphics[width=1.0\textwidth]{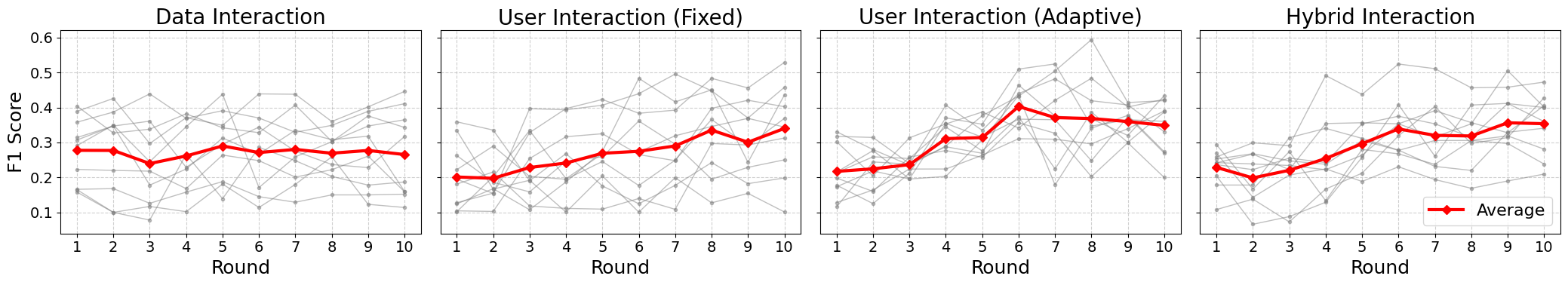}
  \caption{F1 scores for ten requirements over ten interaction rounds. The red line indicates the average score across all tasks, while the gray lines indicate F1 scores for individual tasks. }
  \label{fig:f1-by-round}
\end{figure*}

We evaluate the requirements generated following each interaction round. Figure \ref{fig:f1-by-round} illustrates the performance across four interaction strategies for ten tasks based on the \emph{Arxiv Sum} dataset. As the round number increases, the average $F_1$ scores for \textit{Data Interaction} remain largely stagnant, primarily because this strategy often extracts features irrelevant to the user’s core needs, leading to intent misalignment. In contrast, strategies incorporating user feedback—\textit{User Interaction} and \textit{Hybrid Interaction}—show consistent average improvement. Notably, while \textit{User Interaction (Fixed)} improves overall, it exhibits substantial variance across individual tasks; its rigid, predefined questions effectively capture standard requirements but fail to adaptively probe for nuanced, task-specific constraints. The \textit{Hybrid} strategy resolves this by incorporating a dynamic validation step where features are only included if confirmed by the user, effectively filtering out distracting information and addressing task-specific gaps (see the example in Table~\ref{tab:case_data_interaction_2}).

Despite this general upward trend for \textit{User Int. (Fixed)}, \textit{User Int. (Adaptive)}, and \textit{Hybrid Interaction}, individual cases still exhibit notable fluctuations across specific rounds. This suggests that while continued user feedback generally enhances requirement coverage, prolonged interaction can also introduce semantic noise, requirement drift, or conflicting constraints, occasionally resulting in temporary performance degradation before stabilizing.

\subsection{Impact of User Collaboration}
\label{sec:user_colab}
\begin{figure}[t]
  \centering
  \includegraphics[width=0.9\columnwidth]{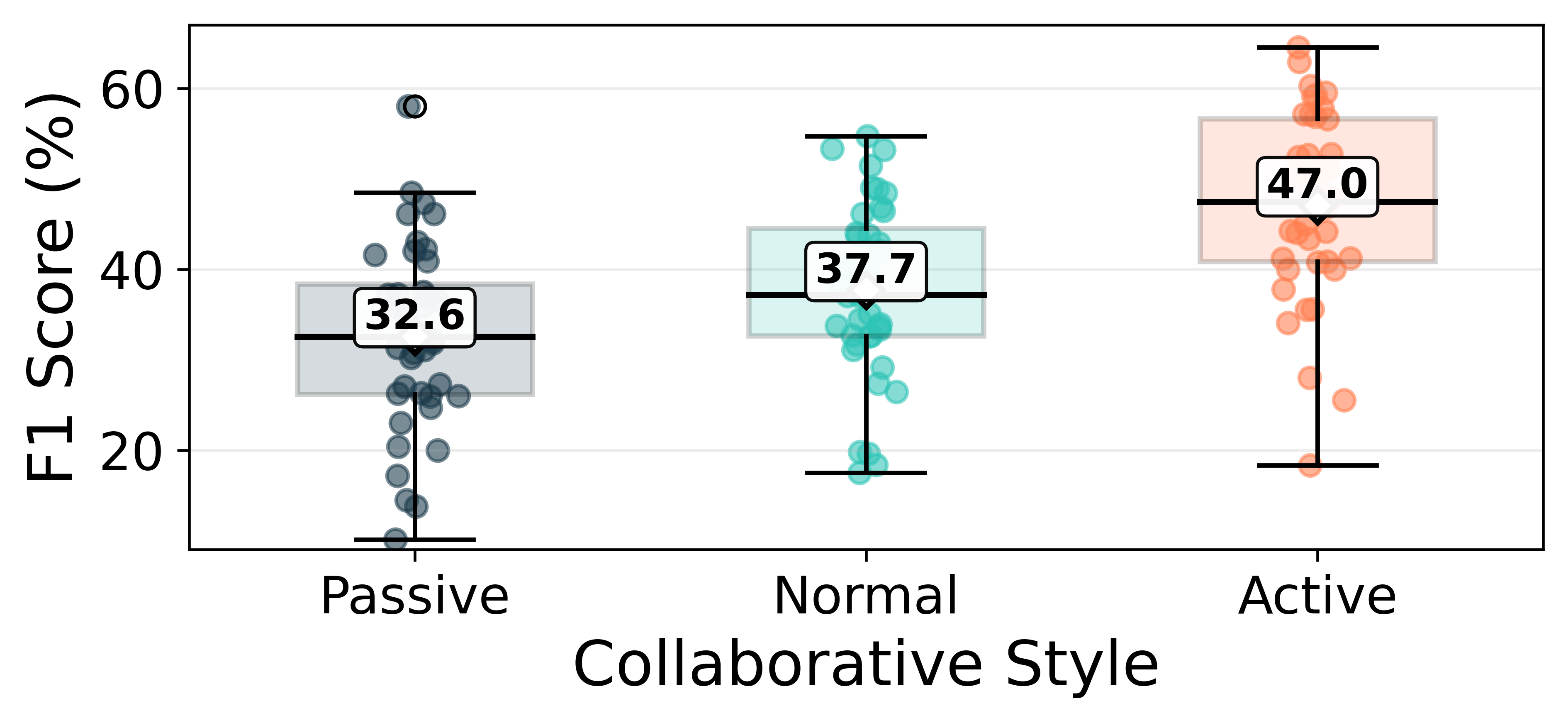}
  \caption{F1 scores across collaborative styles under the hybrid interaction strategy. Colored boxplots show score distributions across Passive, Normal, and Active settings; dots mark individual task results. Evaluated on a subset of four datasets due to resource constraints: \emph{MultiNews}, \emph{UK-Legislation}, \emph{Asclepius}, and \emph{ESConv}.}
  \label{fig:f1-by-collaborative-style}
\end{figure}

To evaluate the impact of user collaboration on requirement elicitation quality, we analyze the assistant's performance with the \textit{Hybrid Interaction} strategy under three simulated user collaboration styles: Passive, Normal, and Active. As shown in Figure~\ref{fig:f1-by-collaborative-style}, performance consistently improves as users become more collaborative. As shown in Table~\ref{tab:communication_habit_progression}, active users not only answer questions but also provide additional relevant information, helping the AREAs assistant form a better understanding of the task. 
\vspace{-0.5\baselineskip}
\section{Human Assessment}

\noindent\textbf{Human Interaction Procedure.} We recruited $12$ human evaluators\footnote{The evaluators came from diverse professional backgrounds, including medicine, finance, computer science, education, and data science. All participants held Master's degrees and routinely utilized AI tools in their professional workflows.} to author comprehensive ground-truth requirements, which served as the \emph{complete} user intent. We then asked the evaluators to initiate a conversation by briefly describing their overall goal. The AREAs assistant subsequently engaged in a $10$-round dialogue using the \textit{Hybrid Interaction} strategy and generated a final task requirement. 

\noindent\textbf{Human Assessment Metrics and Results.\footnote{Detailed human evaluation protocols are provided in Appendix~\ref{sec: human_eval_details}.} }
After interacting with the AREAs assistant, evaluators were presented with their initial, manually written requirement. They were prompted to evaluate the AI-generated requirement against their original text across four dimensions:
(1) \textsc{Coverage}: the extent to which information in the manual requirement was preserved in the AI-generated version; 
(2) \textsc{Inspiration}: the degree to which the agent inspired the user (e.g., revealing unexpected aspects); 
(3) \textsc{Experience}: the overall quality of the interaction; and 
(4) \textsc{Requirement Quality}: the intrinsic quality of the generated task requirement. 
Each dimension was scored on a 1--5 Likert scale. The detailed scoring rubrics are presented in Table~\ref{tab: human_eval_metrics}. Evaluators were also encouraged to provide qualitative comments for each metric.

\begin{figure}[t]
  \centering
  \includegraphics[width=\columnwidth]{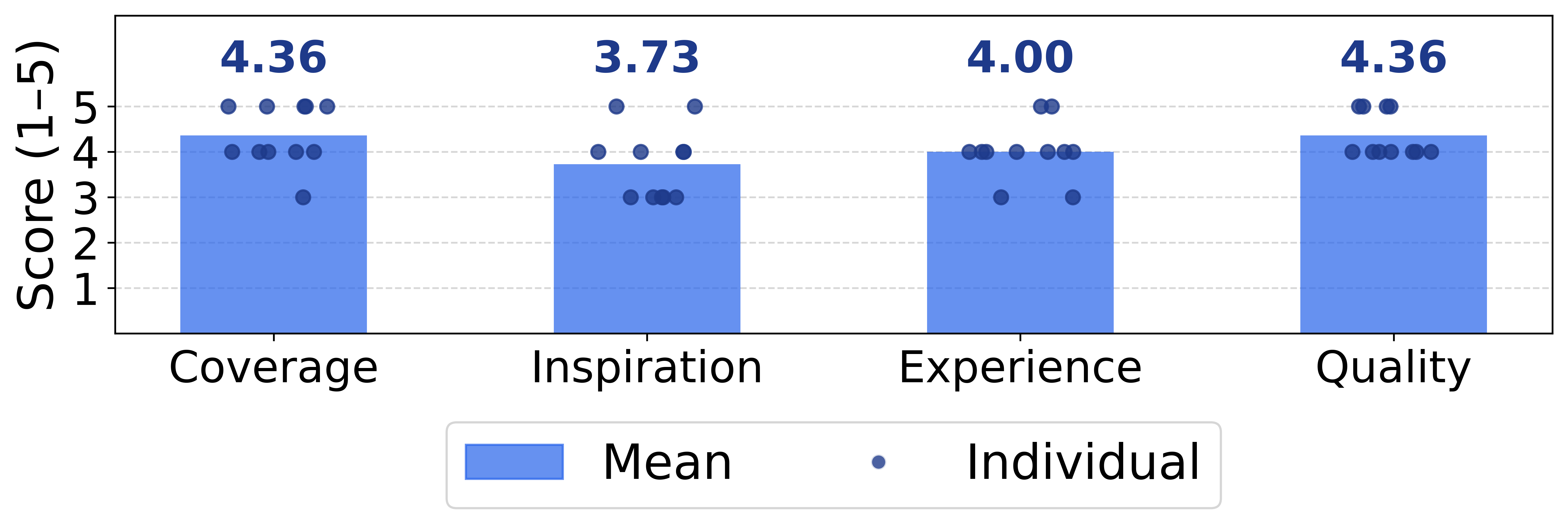}
  \caption{Human evaluation scores across the four assessment dimensions. Dots represent individual scores.}
  \label{fig:human-scores}
\end{figure}

Figure~\ref{fig:human-scores} shows the human evaluation results. The AREAs assistant performs robustly, achieving high mean scores in \textsc{Coverage} ($4.36$), \textsc{Requirement Quality} ($4.36$), and general \textsc{Experience} ($4.00$). Users found the system accurate, usable, and pleasant to co-create with. However, \textsc{Inspiration} lagged slightly ($3.73$) with wider variance, indicating that while the system reliably fulfills explicit requirements, it less consistently introduces novel perspectives or unexpected insights.

\enlargethispage{2\baselineskip}

Interestingly, we observed that human-written requirements were generally less complex than our synthesized tasks, often lacking detailed data constraints and precise specifications. Evaluators frequently addressed unexpected elements during the interaction that they had not initially considered (see Table~\ref{tab:human_interaction_example}). This limitation underscores the vital role of AI-driven requirement elicitation in uncovering hidden complexities.

\noindent\textbf{Human vs. Simulated User Responses.} A critical component of AREAs-Lab is the capability of simulated users to realistically process and respond to assistant critiques. To evaluate this, we cross-examined the authentic human feedback collected during the $10$-turn interaction against the simulated responses. Specifically, we prompted an LLM to classify the alignment between human and simulated feedback into three categories: \emph{Similar}, \emph{Partially Similar}, and \emph{Different} (methodological details are provided in Section~\ref{sec: human_eval_details}). This analysis assesses the extent to which simulated responses capture selected disclosure behaviors observed in human interactions, rather than establishing full behavioral equivalence.

\begin{figure}[t]
  \centering
  \includegraphics[width=0.9\columnwidth]{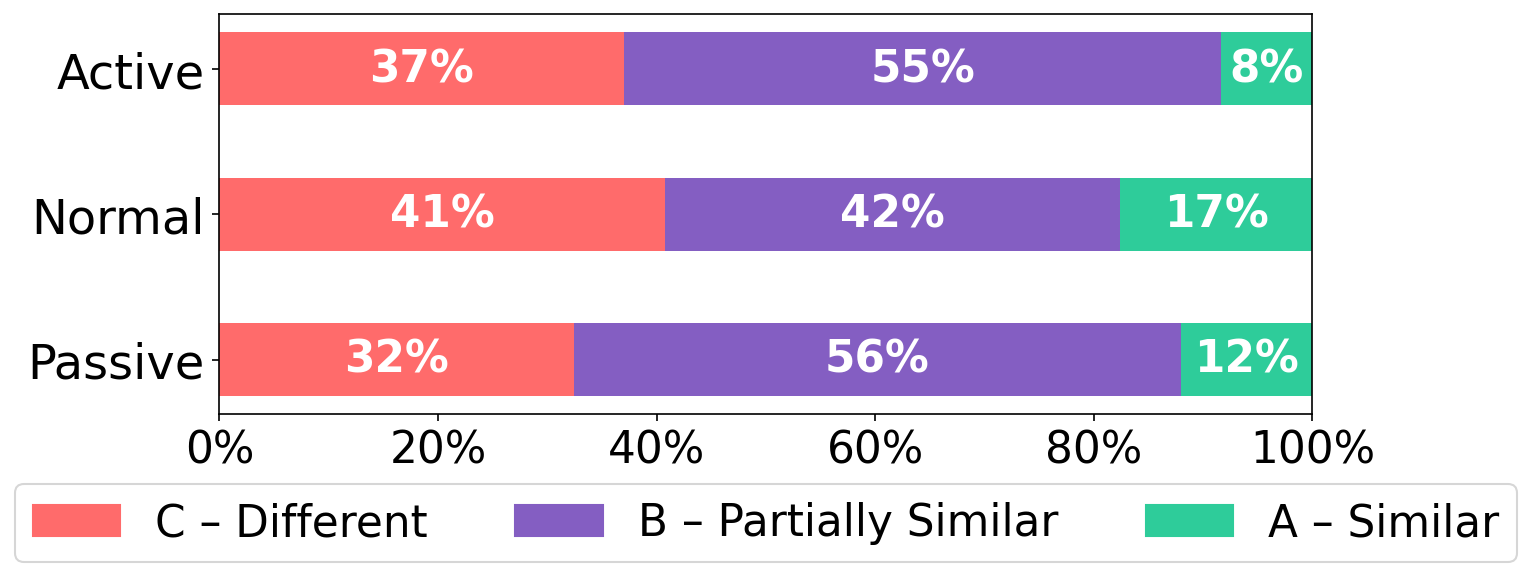}
  \caption{Similarity analysis between human and simulated user responses across varying collaboration styles.}
  \label{fig:similarity-by-style}
\end{figure}

Figure~\ref{fig:similarity-by-style} reports the alignment between human and simulated responses across collaboration styles. Overall, simulated responses show substantial correspondence with human feedback: the combined \emph{Similar} and \emph{Partially Similar} categories account for 63\%, 59\%, and 68\% of the Active, Normal, and Passive settings, respectively. Notably, the Passive style achieved the lowest \emph{Different} rate (32\%), with its simulated response length closely matching human users (24 vs. 17 words). This demonstrates that enforcing a vagueness constraint effectively curtails the model’s tendency to overelaborate. Furthermore, the inherent brevity of human responses highlights a key behavioral trait: human users typically communicate minimal information by default, thereby underscoring the critical need for AI assistants to proactively elicit unstated requirements. Nevertheless, 32\%–41\% of the simulated responses were classified as Different across the three interaction styles, indicating systematic gaps between simulated and human behavior. The simulator should therefore be viewed as a controlled proxy for selected disclosure patterns rather than a faithful model of real-user elicitation dynamics.

\section{Conclusion}

This work introduces AREAs-Lab, a novel environment that establishes a new paradigm for evaluating AI-driven requirement elicitation. By utilizing a diverse synthetic benchmark and a scalable and controlled agent-simulated evaluation pipeline, we provide a scalable foundation for this intricate task. Our experiments evaluating four interaction strategies reveal that AREAs-Lab effectively identifies a critical performance bottleneck: the necessity for assistants to actively mine underlying data features to achieve high-quality requirement elicitation. By open-sourcing these resources, we aim to empower the community to build more sophisticated, elicitation-capable AI systems.

\section{Limitations}
While AREAs-Lab provides a scalable, reproducible, and domain-agnostic framework for AI system requirement elicitation, several limitations warrant acknowledgment and point to future research directions.

First, the behavioral gap between AI-simulated users and real humans persists. Our human comparison suggests that the simulator captures selected disclosure behaviors under controlled conditions, but substantial behavioral differences between simulated and real users remain. LLM-based proxies cannot reproduce the full spectrum of human cognitive and conversational behavior, including shifting attention, emotional friction, inconsistent preferences, and conversational fatigue. Our findings should therefore be interpreted as comparative results within the controlled AREAs-Lab environment rather than evidence of full behavioral fidelity. Larger and more diverse randomized human studies are needed to establish external validity in real-world elicitation settings.

Second, our benchmark is currently constrained to text-based and structured data modalities. In actual software and AI systems engineering, requirement elicitation is frequently a multimodal endeavor, involving the interpretation of visual mockups, user interface diagrams, and complex database schemas. While \textsc{AREAs-Lab} aggregates 16 diverse public datasets, extending the environment to support multimodal interaction remains an open challenge.

Third, the evaluation pipeline is tightly coupled with the capabilities of the underlying LLMs. The performance of both the elicitation agent and the simulated user depends on the foundation model's context window limit and instruction-following stability. In ultra-long, multi-turn dialogues, the simulator may suffer from information decay or hallucination, which can introduce variance into the evaluation metrics. 

Finally, our synthetic benchmark may not fully capture niche or proprietary enterprise workflows. Although synthetic instances enable controlled informational asymmetry, they do not reproduce the legal, organizational, privacy, and security constraints of high-stakes deployments. AREAs-Lab also does not evaluate mechanisms such as role-based data access, PII redaction, organizational approvals, or compliance review. Evaluating elicitation assistants under these governance constraints remains important future work.

\section{Potential Risks}

Although AREAs-Lab provides a controlled environment for studying requirement elicitation, deploying similar systems in enterprise or high-stakes settings introduces privacy, reliability, and governance risks. An elicitation assistant may inspect sensitive data, infer requirements beyond the available evidence, and influence downstream systems through the specifications it produces.

\noindent\textbf{Confidential-Data and PII Exposure.}
Data inspection may expose confidential records, personally identifiable information (PII), proprietary business logic, or other regulated information unnecessary for elicitation. Such content could appear in clarification questions, logs, or final specifications, especially when external model providers or persistent histories are used. Deployments should enforce data minimization, access controls, redaction or anonymization, secure execution, and explicit retention policies. Assistants should inspect only necessary fields and include sensitive information in requirements only when explicitly authorized.

\noindent\textbf{Requirement Drift and Unconfirmed Data Inferences.}
Multi-turn interaction may introduce irrelevant or conflicting constraints into an evolving specification. Assistants may also mistake artifacts or correlations in a small sample for genuine requirements, causing \textit{requirement drift} from the stakeholder's objective. For example, narrative outputs alongside primary documents could suggest a generation task when the intended objective is validation. Data-derived requirements should therefore be traceable to supporting observations, distinguished from user-confirmed requirements, and confirmed before inclusion.

\noindent\textbf{Hallucinated Legal and Compliance Constraints.}
An assistant may generate plausible but unsupported claims about privacy law, fairness, documentation obligations, or industry regulations. Conversely, it may overlook applicable constraints that are absent from the inspected data or undisclosed by the user. The resulting specification should not be treated as legal advice; in regulated domains, proposed compliance requirements must be reviewed by qualified experts and verified against authoritative policies and legal sources.

\noindent\textbf{Automation Bias and Unequal Stakeholder Influence.}
Because generated specifications may appear authoritative, users may accept their assumptions without sufficient scrutiny, particularly under time pressure or with limited technical expertise. Elicitation may also overrepresent highly communicative stakeholders while missing passive users or affected groups. Systems should communicate uncertainty, surface conflicts, and distinguish stated requirements from assistant-generated hypotheses.

\noindent\textbf{Human Oversight, Approval, and Auditability.}
Generated requirements should remain proposals rather than automatically executable specifications. High-stakes deployments should require human approval before inspecting sensitive data, accepting inferred legal or policy constraints, or passing specifications to implementation agents. Audit logs should record inspected sources, interactions, inferred constraints, revisions, and approvals to support failure investigation and accountability.

\noindent\textbf{Dependence on User Collaboration.}
Final specification quality remains dependent on stakeholder participation. Passive users may provide limited information, while inconsistent feedback may introduce conflicts. Assistants that fail to adapt may produce underspecified or unstable requirements. AREAs-Lab captures only selected interaction styles and not the full range of real-world behavior.

\bibliography{custom}

\newpage
\appendix

\section{Appendix}
\label{sec:app}

\subsection{Data Source}
\label{app:data-source}
Table~\ref{tab:data_sources} lists the data sources, domains, and descriptions used in our benchmark. The selected sources satisfy three criteria. 
\begin{itemize}[itemsep=0pt, parsep=0pt, topsep=0pt, partopsep=0pt]
  \item \textbf{Publicity}: The data sources are publicly accessible through Hugging Face Datasets, supporting transparency and reproducibility.
  \item \textbf{Content-Rich}: Each data source contains sufficiently rich content to support the generation of realistic user profiles and task requirements.
  \item \textbf{Diversity}: The data sources span diverse domains, enabling us to study user-needs understanding beyond a single application setting.
\end{itemize}

\begin{table*}[!htbp]
\centering
\caption{Selected data sources used to construct the synthetic benchmark.}
\label{tab:data_sources}
\small
\setlength{\tabcolsep}{4pt}
\begin{tabularx}{\textwidth}{@{}c>{\raggedright\arraybackslash}p{0.33\textwidth}lX@{}}
\toprule
\textbf{ID} & \textbf{Hugging Face Dataset Name} & \textbf{Domain} & \textbf{Description} \\
\midrule
\small
1 & \href{https://huggingface.co/datasets/ccdv/govreport-summarization}{\texttt{govreport-summarization}} \cite{huang2021efficient} & Legal & Government reports and long-form policy documents \\
2 & \href{https://huggingface.co/datasets/ccdv/pubmed-summarization}{\texttt{pubmed-summarization}} & Medical / Acad. & Biomedical research articles and summaries \\
3 & \href{https://huggingface.co/datasets/ccdv/mediasum}{\texttt{mediasum}} \cite{zhu2021mediasum} & Media & Interview and media transcript summaries \\
4 & \href{https://huggingface.co/datasets/ccdv/arxiv-summarization}{\texttt{arxiv-summarization}} \cite{cohan-etal-2018-discourse} & Academic & Scientific papers from arXiv \\
5 & \href{https://huggingface.co/datasets/ccdv/patent-classification}{\texttt{patent-classification}} \cite{sharma-etal-2019-bigpatent} & Technical / Legal & Patent documents and classification labels \\
6 & \href{https://huggingface.co/datasets/mrSoul7766/ECTSum}{\texttt{ECTSum}} \cite{mukherjee-etal-2022-ectsum} & Financial & Earnings call transcripts and summaries \\
7 & \href{https://huggingface.co/datasets/thu-coai/esconv}{\texttt{esconv}} \cite{liu-etal-2021-towards} & Psychological & Emotional support and counseling dialogues \\
8 & \href{https://huggingface.co/datasets/alexfabbri/multi_news}{\texttt{multi\_news}} \cite{fabbri-etal-2019-multi} & Media & Multi-document news summaries \\
9 & \href{https://huggingface.co/datasets/santoshtyss/uk_legislation}{\texttt{uk\_legislation}} \cite{t-y-s-s-elganayni-2025-promalex} & Legal & UK legislative and regulatory documents \\
10 & \href{https://huggingface.co/datasets/HuggingFaceFW/fineweb-edu}{\texttt{fineweb-edu}} \cite{lozhkov2024fineweb-edu} & Educational & Educational and instructional web content \\
11 & \href{https://huggingface.co/datasets/HuggingFaceH4/MATH-500}{\texttt{MATH-500}} \cite{lightman2024let} & Educational & Mathematical problem-solving tasks \\
12 & \href{https://huggingface.co/datasets/danidanou/Reuters_Financial_News}{\texttt{danidanou/Reuters\_Financial\_News}} \cite{reuters_financial_news_2006_2013, BloombergReutersDataset2015} & Financial & Global financial news articles and reports \\
13 & \href{https://huggingface.co/datasets/Pavithree/eli5}{\texttt{eli5}} \cite{fan-etal-2019-eli5} & General / Edu. & Explanations for complex questions (ELI5) \\
14 & \href{https://huggingface.co/datasets/FiscalNote/billsum}{\texttt{FiscalNote/billsum}} \cite{kornilova-eidelman-2019-billsum} & Legal & Summaries of US Congressional and state bills \\
15 & \href{https://huggingface.co/datasets/Harley-ml/lesswrong}{\texttt{Harley-ml/lesswrong}} \cite{harley_lesswrong_2024} & Philosophy & Rationality and philosophy-focused forum posts \\
16 & \makecell[tl]{\href{https://huggingface.co/datasets/starmpcc/Asclepius-Synthetic-Clinical-Notes}{\texttt{starmpcc/Asclepius-}}\\\href{https://huggingface.co/datasets/starmpcc/Asclepius-Synthetic-Clinical-Notes}{\texttt{Synthetic-Clinical-Notes}}} \cite{kweon-etal-2024-publicly} & Medical & Synthetic clinical notes and patient records \\
\bottomrule
\end{tabularx}
\end{table*}

\subsection{Synthesized Data Analysis}

\paragraph{User Roles Distribution.}

\begin{figure}[t]
  \centering
  \includegraphics[width=\columnwidth]{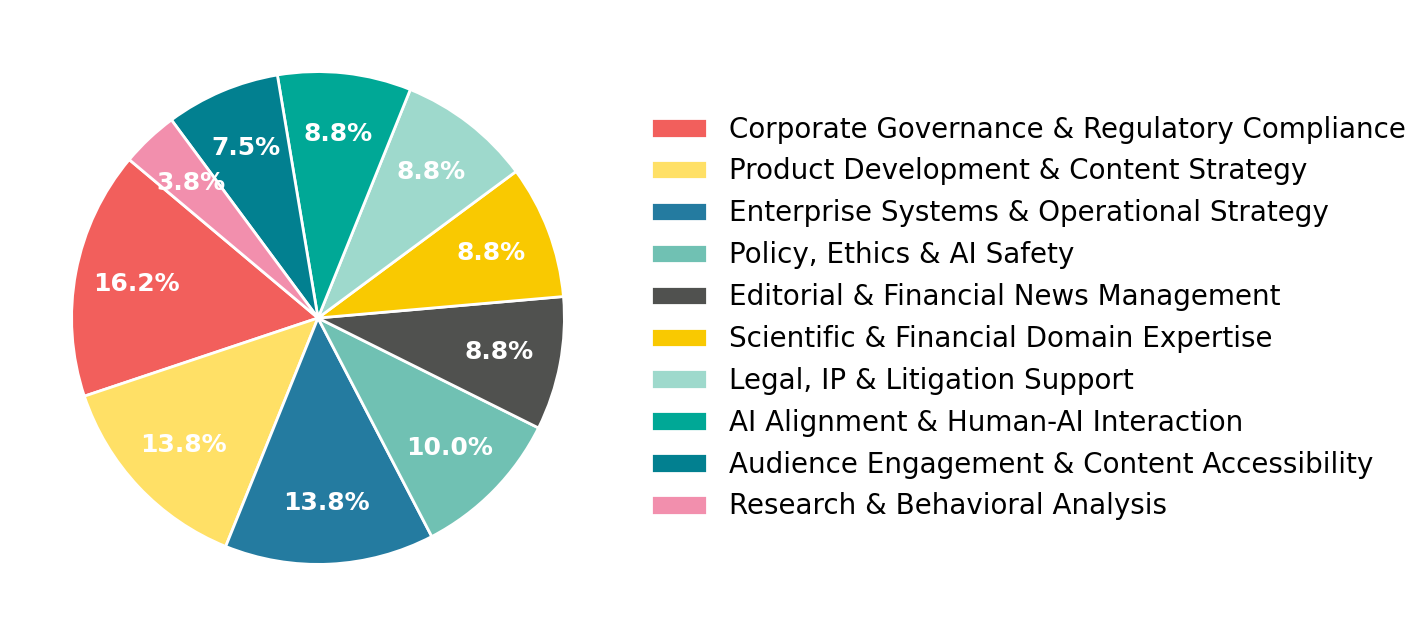}
  \caption{Distribution of professional roles across clusters.}
  \label{fig:role-categories}
\end{figure}

Figure \ref{fig:role-categories} illustrates the distribution of the 80 synthesized user profiles, which were categorized into distinct clusters using an LLM. The analysis reveals a dataset predominantly composed of complex, high-stakes profiles. This indicates a benchmark heavily weighted toward high-accountability professionals who demand transparency, strict controllability, and domain-appropriate reasoning alongside basic task completion. Consequently, it serves as a rigorous proxy for evaluating AI in enterprise environments where the cost of errors is exceptionally high. 

\begin{figure}[t]
  \centering
  \includegraphics[width=\columnwidth]{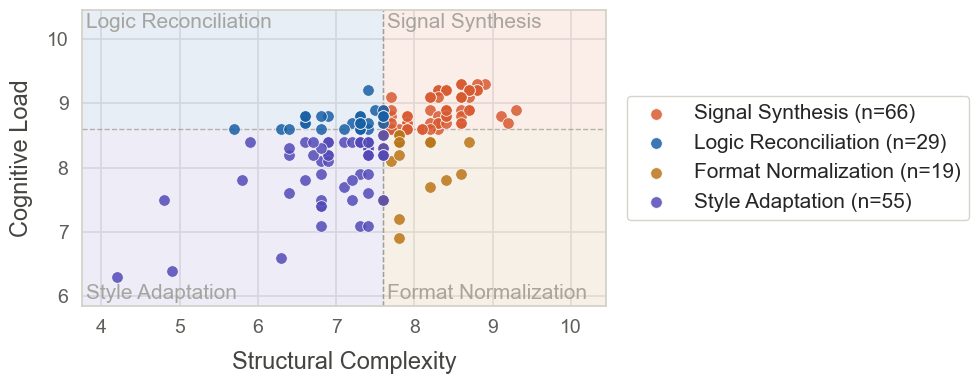}
  \caption{Quadrant-based classification of synthesized tasks by structural complexity and cognitive load. The four categories are \textbf{Signal Synthesis} (High $S$, High $C$): A ``dual-challenge'', models must navigate fragmented, non-narrative data and extract cause-effect signals through deep reasoning. E.g. Inferring financial health from massive news articles. \textbf{Logic Reconciliation} (Low $S$, High $C$): Emphasizes pure reasoning by requiring the model to navigate intricate logical dependencies and multifaceted constraints within structurally clean text. E.g. Synthesize the different opinions from the debate.  \textbf{Format Normalization} (High $S$, Moderate $C$): Tests robustness to irregular, tabular, or list-based inputs and asks the model to convert them into coherent narrative prose. E.g. Organizing long meeting notes. \textbf{Style Adaptation} (Low $S$, Moderate $C$): Focuses on audience alignment by translating structurally standard text into persona-driven technical vernaculars. E.g. Rewriting professional clinical notes for lay people.
  }
  \label{fig:task-classification}
\end{figure}

\paragraph{Task Difficulty Distribution. }
We scored 151 synthesized tasks by GPT-5 across two primary dimensions: \emph{Structural Complexity} ($S$): Measures the extent to which the raw data deviates from standard narrative prose; 
As illustrated in Figure \ref{fig:task-classification}, we classify the tasks into four distinct functional regimes based on their coordinate positions relative to the median scoring thresholds. This distribution implies varying levels of task difficulty.
and \emph{Cognitive Load} ($C$): Measures the depth of logical reasoning, conflict resolution, and domain transformation required by the model. \footnote{We require the model to provide a floating-point score with one decimal place, as continous scoreing prevent overplotting.} 

\begin{figure*}[t]

    \centering

    \includegraphics[width=\textwidth]{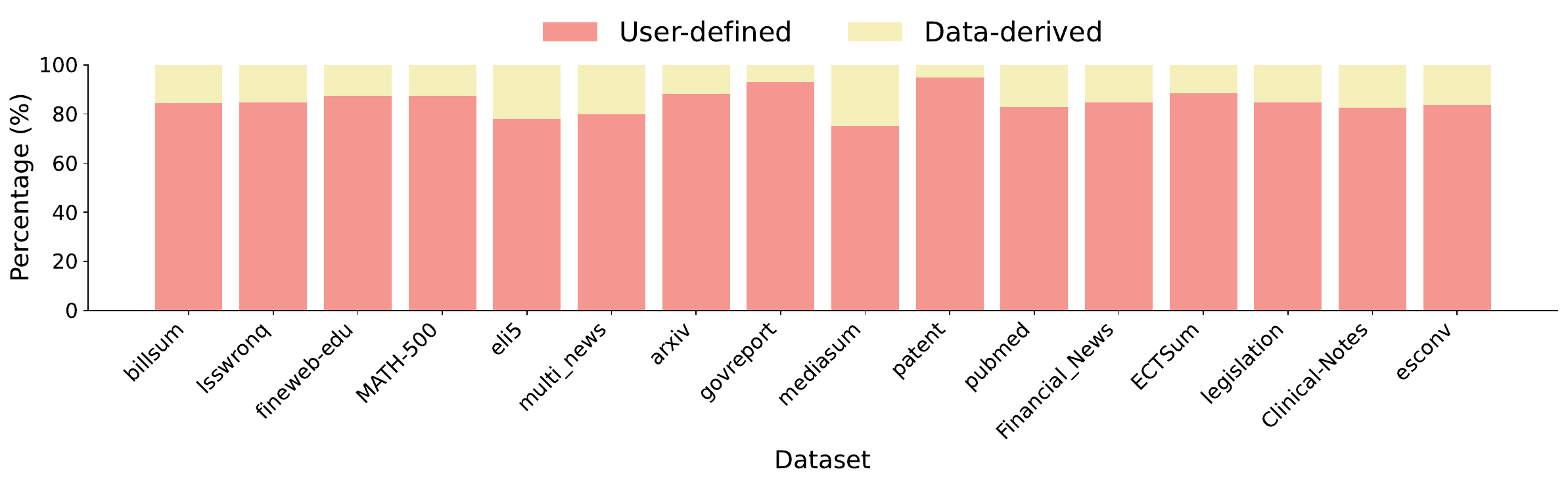}

    \caption{Distribution of user-defined and data-derived requirements across datasets. Each stacked bar reports the percentage of atomic requirements originating from explicit user intent versus requirements inferred from dataset inspection, showing that user-defined requirements form the majority while data-derived requirements consistently contribute a non-negligible portion of the final task specification.}

    \label{fig:requirement-source-distribution}

\end{figure*}

\paragraph{Requirement Source Analysis} Figure~\ref{fig:requirement-source-distribution} analysis the composition of task requirements by separating them into \textit{user-defined} and \textit{data-derived} requirements. User-defined requirements refer to task constraints, goals, and preferences explicitly specified by the user, whereas data-derived requirements capture implicit constraints inferred from dataset inspection, such as format artifacts, domain-specific structures, and input-output patterns. This analysis illustrates the extent to which task specification depends not only on user intent but also on evidence grounded in the data itself.

\subsection{Additional Experiment Results}

\textbf{Impact of Data Instances for Interaction.}
To synthesize task requirements, our process extracts features from 100 random samples. To investigate how the familiarity of these data instances influences interaction performance, we compare two variants against a baseline: (1) \textit{Data Interaction (Non-Def.)}, which utilizes unseen instances that played no role in generating the reference requirements (our default experimental setup), and (2) \textit{Data Interaction (Def.)}, which utilizes \textit{defining data} directly involved in synthesizing the ground-truth requirements.
As summarized in Table \ref{tab:interaction-comparison}, both interaction strategies outperform the \textit{No Interaction} baseline ($23.79\%$ F1). Notably, interacting with defining data yields the most substantial performance boost, driving the F1 score up to $33.79\%$. This outcome is expected, as defining data inherently contains the most salient features associated with the target task requirements. Crucially, sampling defining data for interaction is highly feasible in real-world practice, as the interaction mechanism typically has full access to the entire dataset.

\begin{table}[h]
\centering
\caption{Comparison of using different data instances for data interaction. The results indicate that data interaction consistently enhances requirement synthesis performance across all metrics. The superior performance observed with Defining Data suggests that the model effectively leverages intrinsic features used during ground-truth generation. More importantly, the performance gain on Non-Defining Data ($F1$ increased from 23.79 to 25.68) demonstrates the system's generalization capability, proving that the interaction mechanism can extract meaningful task requirements even from unseen instances.}
\label{tab:interaction-comparison}
\footnotesize
\begin{tabular}{l c c c}
\toprule
\textbf{Interaction Strategy} & \textbf{Pr.(\%)} & \textbf{Re.(\%)} & \textbf{F1(\%)} \\
\midrule
No Interaction & 18.80 & 35.13 & 23.79 \\
\textit{Data Interaction (Non-Def.)} & 20.43 & 36.91 & 25.68\\
\textit{Data Interaction (Def,)}  & 26.72 & 46.01 & 33.79 \\
\bottomrule
\end{tabular}
\end{table}

\begin{figure*}[t]

    \centering

    \includegraphics[width=\textwidth]{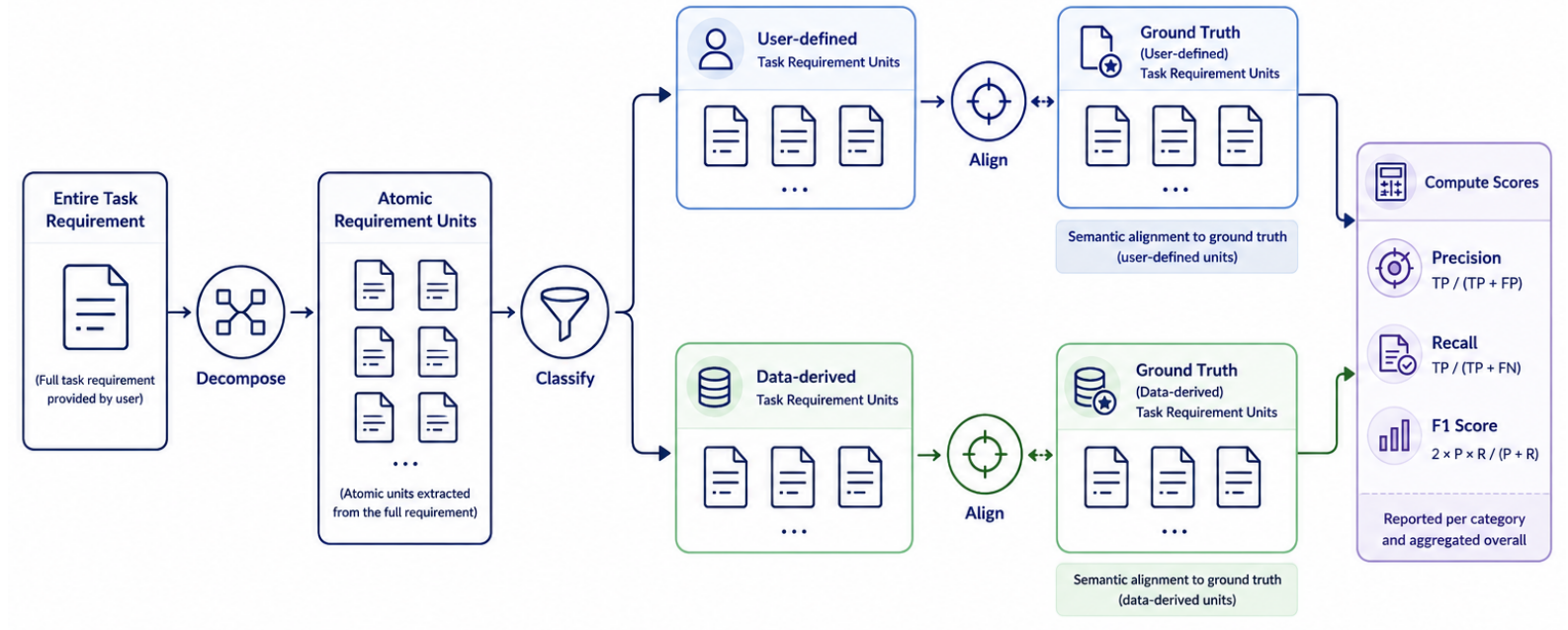}

    \caption{Overview of the task requirement evaluation workflow. The full task requirement is first decomposed into atomic requirement units and then classified into user-defined and data-derived categories. Each category is independently aligned with the corresponding ground-truth requirement units, after which precision, recall, and F1 scores are computed and reported both per category and in aggregate.}

    \label{fig:evaluation_process}

\end{figure*}

\begin{figure}[t]
  \centering
  \includegraphics[width=0.8\columnwidth]{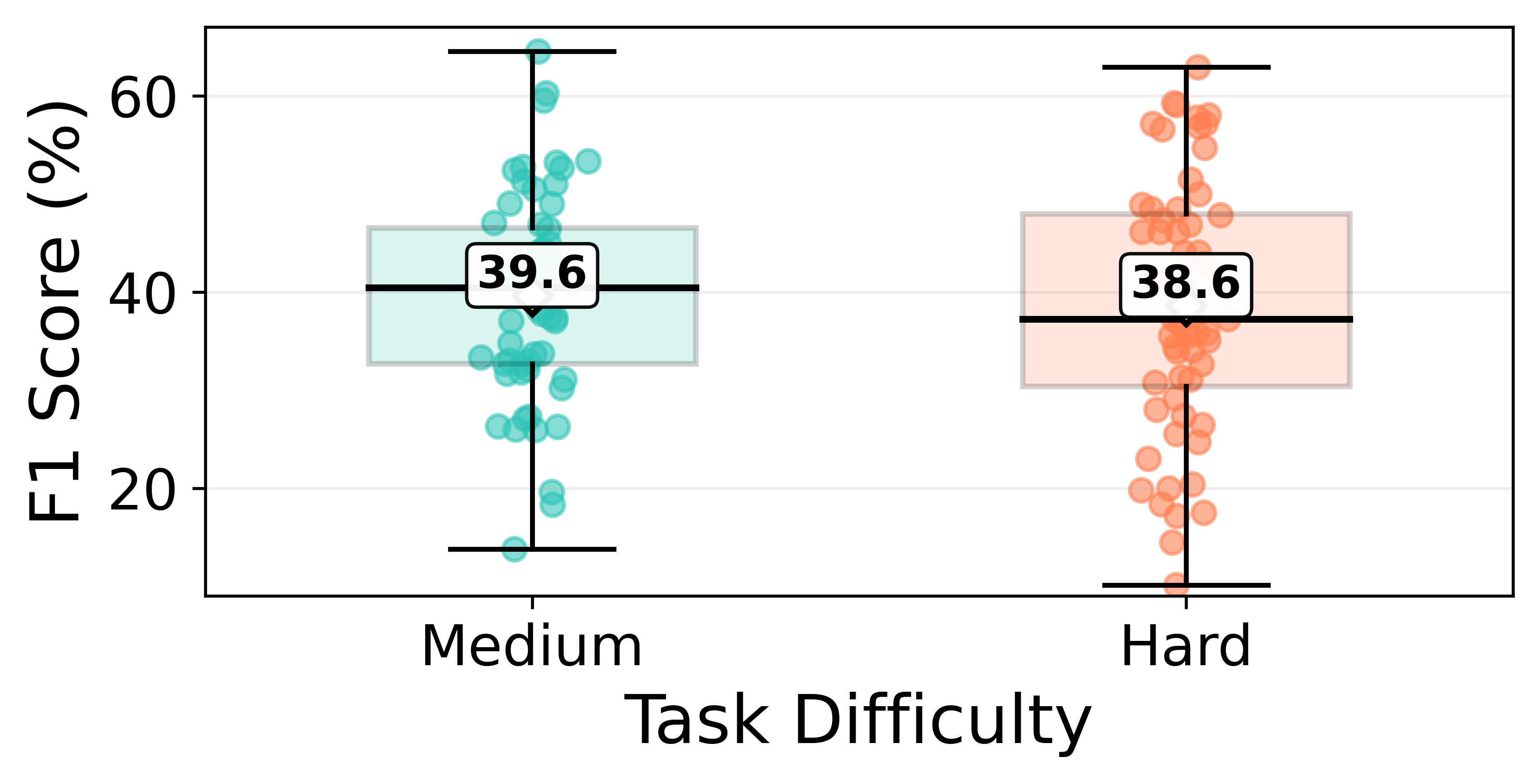}
  \caption{F1 score by collaborative style across task difficulty levels. Colored boxplots show the distribution of F1 scores for Medium and Hard tasks under different collaboration settings, while scattered points indicate individual task results across datasets.}
  \label{fig:f1-by-collaborative-style-difficulty}
\end{figure}

\noindent\textbf{Impact of Data Instances for Interaction.} To examine whether this trend generalizes across different task difficulty levels, we separately analyze Medium and Hard tasks in Sup Figure~\ref{fig:f1-by-collaborative-style-difficulty}. Across both difficulty groups, Active collaboration consistently outperforms Passive and Normal settings, suggesting that increased user cooperativeness generally benefits requirement elicitation quality regardless of task complexity. In contrast, Passive collaboration remains associated with lower and more variable performance across tasks.

\subsection{Evaluation process.}
Figure~\ref{fig:evaluation_process} demonstrate the entire evaluation process. Table~\ref{exp:task_decomposition} illustrates how AREAs decomposes a task requirement into two complementary sources of information. User-defined requirements capture the intended objective, evaluation protocol, and explicit constraints provided by the user, such as auditing neutrality, transparency, and bias. In contrast, data-derived requirements capture concrete evidence from the dataset, including dataset identity, input fields, record-level examples, charged textual spans, and formatting artifacts. This decomposition highlights why both user interaction and data inspection are necessary for reconstructing complete task requirements. 

\begin{table*}[t]
\centering
\label{tab:task-requirement-example}
\scriptsize
\setlength{\tabcolsep}{4pt}
\renewcommand{\arraystretch}{1.05}

\begin{tabular}{p{0.96\textwidth}}
\toprule

\textbf{Task Requirement} \\

\midrule

\textbf{Strategic Mission:}
As an expert in fairness auditing and explainability frameworks, your high-stakes objective is to evaluate the model's capacity to synthesize fundamentally conflicting epistemological worldviews and highly polemical adversarial tone without injecting systemic bias. The compression ratio of 8:1 in the \texttt{alexfabbri/multi\_news} dataset frequently forces models to discard nuance, often resulting in one ideological framing erasing another. You must construct a UI-driven audit protocol to guarantee narrative neutrality and transparency.

\vspace{0.5em}

\textbf{Data-Reality Constraints:}
You are dealing with severe adversarial noise and conflicting realities within the \texttt{document} field. Record 68 presents an Epistemological Worldview Framing clash, explicitly contrasting a naturalistic ``Rally for Reason'' against the theological ``Genesis 50:20''. Concurrently, Record 70 introduces extreme Polemical Adversarial Tone via strings like ``right-wing fascist party'', ``rampaging Hindu zealots'', and ``forked tongue''. The AI must process these highly charged tokens natively without external preprocessing scripts, maintaining the journalistic tension without adopting the bias.

\vspace{0.5em}

\textbf{Chain-of-Thought (CoT) Logic:}

\textbf{1. Conflict Identification:}
Utilize the UI evaluation platform to tag specific spans of text in the \texttt{document} that exhibit Polemical Adversarial Tone, e.g., ``rampaging Hindu zealots'', and Epistemological Worldview Framing, e.g., ``Genesis 50:20''.

\textbf{2. Sentence-Level Traceability:}
Map these high-volatility tokens to the generated \texttt{summary}. Determine if the model selectively dropped the ``Rally for Reason'' in favor of the theological citation, or if it neutralized the polemical strings at the cost of factual accuracy.

\textbf{3. Bias Quantification:}
Apply a fairness auditing framework to evaluate the 8:1 compression. Does the summary reflect a systemic bias toward naturalistic or polemical framing?

\textbf{4. Resolution Blueprint:}
Formulate a prompt-level or system-prompt-level intervention, since code-level fine-tuning is prohibited, that forces the model to represent both the ``Rally for Reason'' and the theological worldview equitably in the summary.

\vspace{0.5em}

\textbf{Persona-Driven Negative Constraints:}
\textbf{Do NOT} permit the use of automated data sanitization pipelines, e.g., regex scripts to strip ``\texttt{|||||}'' or subjective descriptors, prior to model evaluation; the model must be evaluated on its handling of the raw, adversarial text.
\textbf{Do NOT} assume that the omission of ``Genesis 50:20'' constitutes a ``cleaner'' summary; forced secularization of the text is a systemic bias failure.
\textbf{Do NOT} rely on programmatic evaluation metrics like ROUGE or BLEU; rely entirely on explainability frameworks and UI-based sentence mapping. \\

\midrule

\begin{tabular}{
p{0.455\textwidth}
@{\hspace{0.8em}\vrule width 0.4pt\hspace{0.8em}}
p{0.455\textwidth}
}
\textbf{User-defined Task Requirement}
&
\textbf{Data-derived Task Requirement}
\\

\midrule

\begin{minipage}[t]{0.455\textwidth}
Evaluate the model's ability to synthesize conflicting epistemological worldviews. \\
Evaluate the model's ability to handle highly polemical adversarial tone. \\
Assess whether the model avoids injecting systemic bias. \\
Audit whether compression causes loss of nuance. \\
Audit whether one ideological framing erases another. \\
Construct a UI-driven audit protocol. \\
Ensure the audit protocol guarantees narrative neutrality. \\
Ensure the audit protocol guarantees transparency. \\
Require the AI to process charged tokens natively. \\
Do not use external preprocessing scripts. \\
Maintain journalistic tension in processing. \\
Do not let the model adopt the bias present in the source text. \\
Tag spans that exhibit polemical adversarial tone. \\
Tag spans that exhibit epistemological worldview framing. \\
Perform sentence-level traceability between source and summary. \\
Determine whether the model neutralized polemical strings in the summary. \\
Assess whether neutralization reduced factual accuracy. \\
Apply a fairness auditing framework. \\
Use the fairness audit to evaluate the effects of 8:1 compression. \\
Assess whether the summary shows systemic bias toward naturalistic framing. \\
Assess whether the summary shows systemic bias toward polemical framing. \\
Formulate a prompt-level intervention if needed. \\
Formulate a system-prompt-level intervention if needed. \\
Do not use code-level fine-tuning. \\
Design the intervention to force equitable representation of `Rally for Reason'. \\
Design the intervention to force equitable representation of the theological worldview. \\
Do not use automated data sanitization pipelines before evaluation. \\
Do not strip subjective descriptors before evaluation. \\
Evaluate the model on raw adversarial text handling. \\
Do not treat omission of `Genesis 50:20' as a cleaner summary by default. \\
Treat forced secularization as a systemic bias failure. \\
Do not rely on ROUGE for evaluation. \\
Do not rely on BLEU for evaluation. \\
Rely entirely on explainability frameworks. \\
Rely entirely on UI-based sentence mapping.
\end{minipage}
&
\begin{minipage}[t]{0.455\textwidth}
Use the \texttt{alexfabbri/multi\_news} dataset as the evaluation context. \\
Account for the 8:1 compression ratio in the dataset. \\
Use the raw \texttt{document} field as the evaluation input. \\
Handle severe adversarial noise in the \texttt{document} field. \\
Handle conflicting realities in the \texttt{document} field. \\
Include Record 68 in the evaluation. \\
Treat Record 68 as an epistemological worldview framing clash. \\
Recognize the naturalistic frame `Rally for Reason' in Record 68. \\
Recognize the theological frame `Genesis 50:20' in Record 68. \\
Include Record 70 in the evaluation. \\
Treat Record 70 as extreme polemical adversarial tone. \\
Recognize `right-wing fascist party' as a polemical string in Record 70. \\
Recognize `rampaging Hindu zealots' as a polemical string in Record 70. \\
Recognize `forked tongue' as a polemical string in Record 70. \\
Use the UI evaluation platform to tag text spans in the \texttt{document}. \\
Tag `rampaging Hindu zealots' as polemical adversarial tone. \\
Tag `Genesis 50:20' as epistemological worldview framing. \\
Map high-volatility tokens from the \texttt{document} to the generated \texttt{summary}. \\
Determine whether the model dropped `Rally for Reason' from the summary. \\
Determine whether the model retained the theological citation in the summary. \\
Do not use regex scripts to strip \texttt{|||||} before evaluation.
\end{minipage}
\\

\end{tabular}
\\

\bottomrule
\end{tabular}
\caption{Example decomposition of a complex task requirement into user-defined and data-derived requirements. The original requirement is shown above, while the lower section separates user-specified objectives and constraints from dataset-grounded evidence discovered through data inspection.}
\label{exp:task_decomposition}
\end{table*}

\subsection{Implementation Details}
\label{sec:imp_detl}

\textbf{Data Synthesis.} For all steps within the data synthesis pipeline, we use \emph{gemini-3.1-pro-preview} with a temperature of 1.0 to encourage generation randomness. We use \emph{gpt-5} and \emph{deepseek-v4-pro} for cross-model validation.

\noindent\textbf{AREAs assistant.} To prevent potential information leakage and intra-model bias from the synthesis and evaluation stages, we employ \emph{Claude-Sonnet-4.6} as the primary backbone model for the AREAs assistant, which delivers our main experimental results. For a comprehensive comparison, we also evaluate \emph{Gemini-3.1-Pro} and \emph{GPT-5.4} as baseline backbones in our control groups. The generation temperature is set to 0.4 to ensure focused and consistent responses. For both \textit{Data Interaction} and \textit{Hybrid Interaction} strategies, five data instances are randomly sampled at each interaction round.

\noindent\textbf{Simulated User.} We employ \emph{Gemini-3.1-Pro} as our simulated user, maintaining consistency with the model used for data synthesis. The generation temperature is set to 0.7 to introduce natural variability into user behavior. We argue that utilizing the same model family for both task synthesis and user simulation is well-justified, as it ensures behavioral alignment and a consistent knowledge distribution between the synthesized tasks and the user's interaction style. Crucially, this setup poses no risk of information leakage to the target system, as both the data source and the simulated user maintain complete cross-model isolation from the primary assistant (\emph{Claude-Sonnet-4.6}).

\noindent\textbf{Evaluation.} We use \emph{GPT-5.4} as the evaluator model under a deterministic setting (temperature 0.0) to ensure stable and reproducible judgments. All configuration parameters are publicly released alongside the source code to support reproducibility and future benchmarking.

\subsection{Case Study}

We present three of representative cases of our three interactive strategies in Table~\ref{tab:case_user_interaction}, \ref{tab:case_data_interaction_1} \ref{tab:case_data_interaction_2} and \ref{tab:case_user_vs_hybrid}.

\begin{table*}[t]
\centering
\setlength{\tabcolsep}{4pt}
\renewcommand{\arraystretch}{1.15}
\footnotesize
\begin{tabular}{p{0.17\textwidth} p{0.77\textwidth}}
\toprule
\textbf{Dataset} & Multi News \\
\textbf{Task Name} & Editorial Narrative Distillation over Structural Residue \\
\textbf{Initial User Request}
&
Filter raw multi-document news by removing UI subscription prompts and academic citations to generate a clean, reader-friendly 260-word summary.
\\

\midrule

\textbf{Generated Requirement } (\textit{No Interaction})
&
\textbf{Strategic Intent:}
A professional editorial curator transforms raw multi-document news content into a clean, reader-ready 260-word summary by removing subscription prompts, UI artifacts, and academic citation noise while preserving journalistic substance.

\textbf{Input Characteristics:}
The input consists of multiple raw news documents containing factual reporting mixed with embedded UI elements, paywall notices, navigational artifacts, and inline academic citation syntax. Editorial content must be preserved while all non-editorial scaffolding is removed prior to synthesis.

\textbf{Execution Procedure:}
(1) Sequentially scan documents to identify editorial content and structural noise;
(2) remove subscription prompts, citation markers, DOI references, and UI artifacts;
(3) consolidate overlapping information across documents while retaining unique details;
(4) synthesize a coherent cross-document narrative prioritizing significance and readability;
(5) calibrate the final output to exactly 260 words.

\vspace{0.3em}

\textbf{Style Requirements:} The output should be journalistic, neutral, and accessible to general readers, using direct and varied sentence structures while avoiding academic phrasing, excessive jargon, and passive constructions. The summary should appear as continuous prose \textcolor{red}{without Markdown or bullet formatting.}

\vspace{0.3em}

\textbf{Constraints:} The system must preserve factual fidelity and source intent,completely eliminate UI artifacts and citation syntax, and strictly maintain the 260-word requirement ...

\vspace{0.3em}

\\

\midrule

\textbf{Assistant}
&
Should preprocessing be automated or manually performed before summarization?
\\

\textbf{Simulated User}
&
The AI must fully automate filtering and summarization because the raw feed cannot be manually cleaned.
\\

\textbf{Assistant }
&
What format and length are the source documents stored in?
\\

\textbf{Simulated User}
&
The input is a single scraped web-text block \textcolor{blue}{averaging 2,103 words}, where documents are \textcolor{blue}{separated using \texttt{|||||} delimiters}.
\\

\textbf{Assistant}
&
Should the system generate separate summaries or one unified synthesis?
\\

\textbf{Simulated User}
&
The output must be one cohesive 260-word narrative integrating all overlapping reports.
\\

\textbf{Assistant}
&
How should conflicting reports across sources be handled?
\\

\textbf{Simulated User}
&
Conflicting viewpoints should remain visible rather than being collapsed into consensus-only reporting.
\\

\textbf{Assistant}
&
May the AI use \textcolor[rgb]{0.0,0.5,0.0}{outside knowledge} to explain scientific claims?
\\

\textbf{Simulated User}
&
No. The system must rely entirely on the provided feed to preserve editorial fidelity.
\\

\midrule

\textbf{Generated Requirement} (\textit{User Interaction (Adaptive)})
&
\textbf{Strategic Intent:}
A professional editorial curator transforms raw multi-document news content into a clean, reader-ready 260-word summary by removing subscription prompts, UI artifacts, and academic citation noise while preserving journalistic substance.

\vspace{0.3em}

\textbf{Input Characteristics:}
The input is scraped web text \textcolor{blue}{averaging 2{,}103 words}, where documents are \textcolor{blue}{separated using \texttt{|||||}}. UI artifacts and citation strings are removed while preserving factual scientific claims.

\textbf{Execution Procedure:}
(1) \textcolor{blue}{Segment documents using \texttt{|||||}};
(2) sanitize UI artifacts and citation strings;
(3) identify thematic information and conflicting viewpoints;
(4) synthesize a balanced narrative;
(5) constrain the output to exactly 260 words.

\textbf{Style Requirements:}
The output should be objective, journalistically fluent, scientifically accessible, and presented as continuous prose \textcolor{red}{without Markdown or bullet points.}

\textbf{Constraints:}
The system must not retain UI artifacts, citation strings, or \textcolor[rgb]{0.0,0.5,0.0}{external knowledge}, and must preserve conflicting viewpoints alongside consensus facts ...
\\

\bottomrule
\end{tabular}
\caption{Comparison of requirements generated by \textit{No Interaction} and \textit{User Interaction (Adaptive)}. Interactive elicitation enables the AREAs assistant to gather crucial information from the user and generate higher-quality requirements. Specifically, even when the AREAs assistant does not inspect data samples, its question about data format and length prompts the user to provide information about key data features.
In the table, \textcolor[rgb]{0.0,0.5,0.0}{green} text denotes requirements identified through user interaction, while \textcolor{blue}{blue} text indicates data-derived requirements inferred by the agent. \textcolor{red}{Red} text highlights unnecessary requirements from the initial prediction that the agent failed to correct during refinement. }
\label{tab:case_user_interaction}
\end{table*}

\begin{table*}[t]
\centering
\setlength{\tabcolsep}{4pt}
\renewcommand{\arraystretch}{1.15}
\footnotesize
\begin{tabular}{p{0.17\textwidth} p{0.77\textwidth}}
\toprule
\textbf{Dataset} & Multi News \\
\textbf{Task Name} & Adversarial Worldview Fairness Audit \\
\textbf{Initial User Request}
&
Audit the summary generation for systemic bias by tracking the survival of highly polemical tone and conflicting epistemological worldviews from raw source text to final output.
\\
\midrule

\textbf{Sampled data analysis} 
& 
\textbf{Changes:} Flagged \textcolor{red}{NEWLINE\_CHAR} as structural noise. 

\textbf{Reason:} This prevents the system from treating formatting artifacts as meaningful tone.

\vspace{0.3em}

\textbf{Changes:}  Specified that \textcolor{red}{truncated source texts should be flagged }but not excluded.

\textbf{Reason:} Several samples end mid-sentence, suggesting truncation is common. Excluding them would unnecessarily reduce audit coverage.

\vspace{0.3em}

\textbf{Changes:}  Separated tonal survival and epistemological worldview survival into two analytical dimensions.

\textbf{Reason:} The instruction names both targets separately. The data also shows both dimensions, such as political endorsement framing, medical negligence narratives, contested critical consensus, and disputed scientific projections.
\\

\midrule

\textbf{Generated Requirement } (\textit{Data Interaction})
&

\textbf{Strategic Intent:} Conduct a systematic bias audit of a news summarization pipeline to determine whether polemical tone and epistemological conflict present in raw source articles survive, are amplified, or are suppressed in generated summaries. The deliverable is a structured audit report for editorial reviewers or NLP quality analysts. The primary failure mode is systemic ......

\textbf{Data Interpretation \& Pre-processing: } The \textcolor{red}{\texttt{news\_text} field is the primary analysis surface}, containing raw source articles spanning political, medical, environmental, and cultural domains. The \texttt{summary} field is the reference output under audit. Both fields must be analyzed in tandem. \textcolor{red}{\texttt{NEWLINE\_CHAR} tokens are structural delimiters} and must not influence tonal or epistemic analysis. \textcolor{red}{Truncated source texts must be flagged} as incomplete but retained in the audit.......

\textbf{Execution Steps:} 1. \textbf{Identify polemical markers} in each source text: charged language, adversarial framing, rhetorical escalation, one-sided attribution, or emotionally loaded characterization.
2. \textbf{Identify epistemological conflict zones}: passages where factual authority is disputed,......

\textbf{Tone \& Style:} Analytical, editorial, and domain-aware. Avoid evaluative language that itself introduces bias. Use precise vocabulary: polemical register, epistemic framing.....

\textbf{Content Requirements:} The audit report must include: per-sample tonal and epistemic comparison, distortion classification per instance, and a cross-sample summary of systemic patterns......

\textbf{Constraints \& Guardrails:} 
\textbf{DO NOT} reduce the audit to surface-level sentiment scoring.
\textcolor{red}{\textbf{DO NOT} treat neutralization as automatically correct or amplification as automatically wrong.}
\textbf{DO NOT} substitute factual accuracy evaluation for bias-pattern analysis.
\textcolor{red}{\textbf{DO NOT} exclude truncated source texts from analysis.}
\textcolor{red}{\textbf{DO NOT} allow domain-specific polemical norms to be collapsed into a single universal standard.}

\textbf{Evaluation Criteria:} Audit quality is judged by: precision in identifying polemical markers...... \\

\midrule

\textbf{Reference Requirement} 
&

\textbf{Strategic Mission:} As an expert in fairness auditing and explainability frameworks, your high-stakes objective is to evaluate the model's capacity to synthesize fundamentally conflicting epistemological worldviews and highly polemical adversarial tone without injecting systemic bias. The compression ratio of 8:1 in the \texttt{alexfabbri/multi\_news} dataset frequently forces models to discard nuance, often resulting in one ideological framing erasing another. You must construct a UI-driven audit protocol to guarantee narrative neutrality and transparency.

\textbf{Data-Reality Constraints:} You are dealing with severe adversarial noise and conflicting realities within the \texttt{document} field. Record 68 presents an Epistemological Worldview Framing clash, explicitly contrasting a naturalistic ``Rally for Reason'' against the theological ``Genesis 50:20''. Concurrently, Record 70 introduces extreme Polemical Adversarial Tone via strings like ``right-wing fascist party'', ``rampaging Hindu zealots'', and ``forked tongue''. The AI must process these highly charged tokens natively without external preprocessing scripts, maintaining the journalistic tension without adopting the bias.

\textbf{Chain-of-Thought (CoT) Logic:} 1. \textbf{Conflict Identification:} Utilize the UI evaluation platform to tag specific spans of text in the \texttt{document} that exhibit Polemical Adversarial Tone, e.g., ``rampaging Hindu zealots'', and Epistemological Worldview Framing, e.g., ``Genesis 50:20''. 2. \textbf{Sentence-Level Traceability:} Map these high-volatility tokens to the generated \texttt{summary}. Determine if......

\textbf{Persona-Driven Negative Constraints:}
\textbf{DO NOT} permit the use of automated data sanitization pipelines, e.g., regex scripts to strip \texttt{|||||} or subjective descriptors, prior to model evaluation; the model must be evaluated on its handling of the raw, adversarial text.
\textbf{DO NOT} assume that the omission of ``Genesis 50:20'' constitutes a ``cleaner'' summary; forced secularization of the text is a systemic bias failure.
\textbf{DO NOT} rely on programmatic evaluation metrics like ROUGE or BLEU; rely entirely on explainability frameworks and UI-based sentence mapping. 

\textbf{Evaluation Criteria:} Audit quality is judged by: precision in identifying polemical markers...... \\
\bottomrule
\end{tabular}
\caption{Requirements generated by the \textit{Data Interaction} strategy are heavily influenced by features observed in sampled data instances. When these features do not align with the user's expectations, the generated requirements can deviate from the intended task. In this example, \textcolor{red}{Red} text segments are based on features of the sampled data but are absent from the ground truth.}
\label{tab:case_data_interaction_1}
\end{table*}

\begin{figure*}[!t]
  \centering
  \begin{minipage}[t]{0.32\textwidth}
    \centering
    \includegraphics[width=\linewidth]{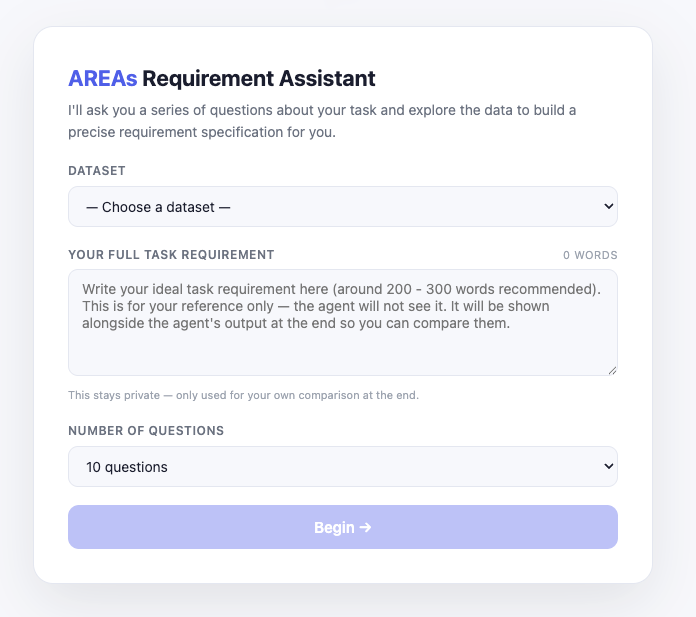}
    \vspace{-0.5em}
    {\footnotesize (a) Set-up page}
  \end{minipage}\hfill
  \begin{minipage}[t]{0.64\textwidth}
    \centering
    \includegraphics[width=\linewidth]{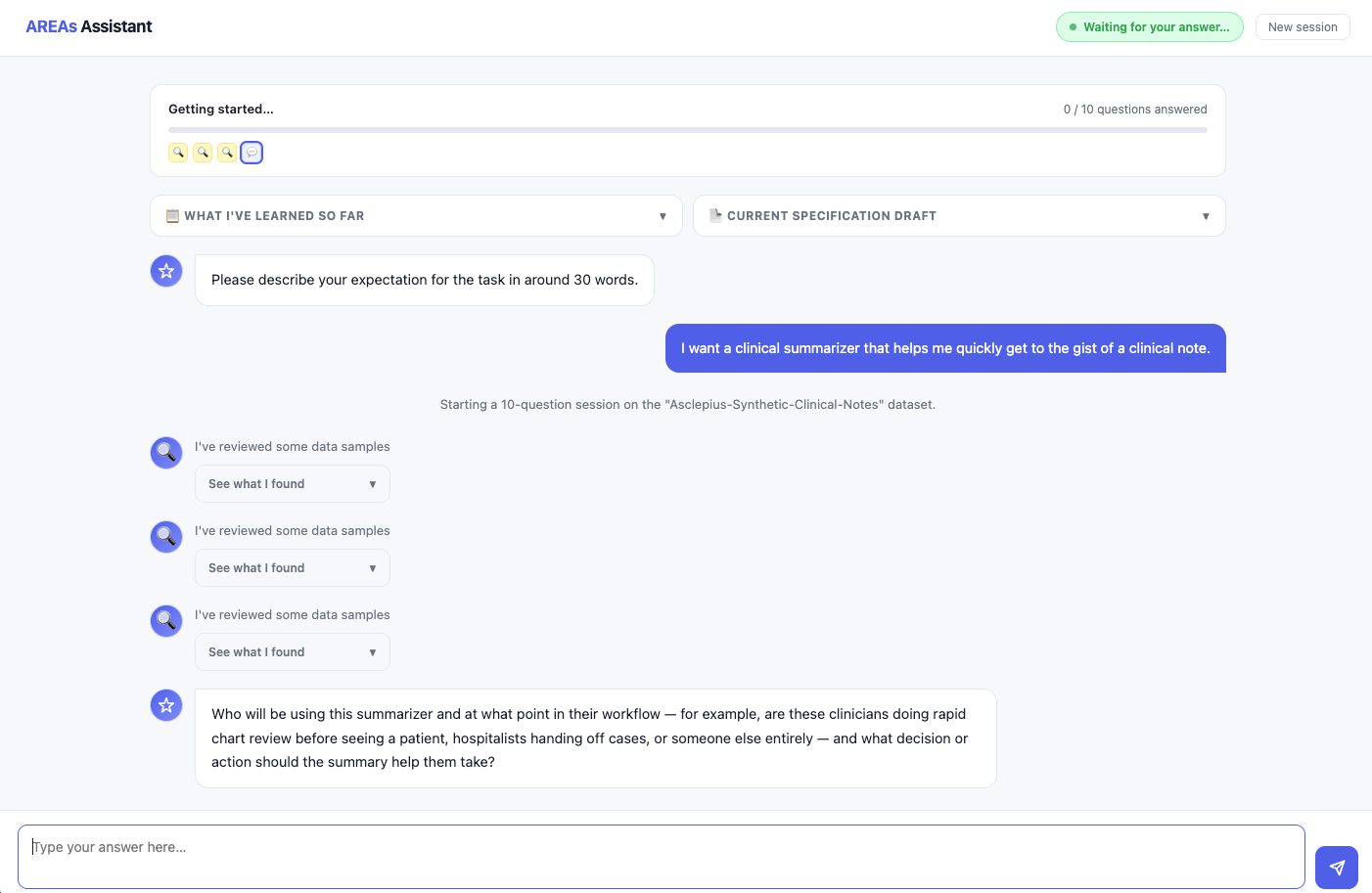}
    \vspace{-0.5em}
    {\footnotesize (b) Interaction page}
  \end{minipage}

  \vspace{0.4em}
  \includegraphics[width=0.82\textwidth]{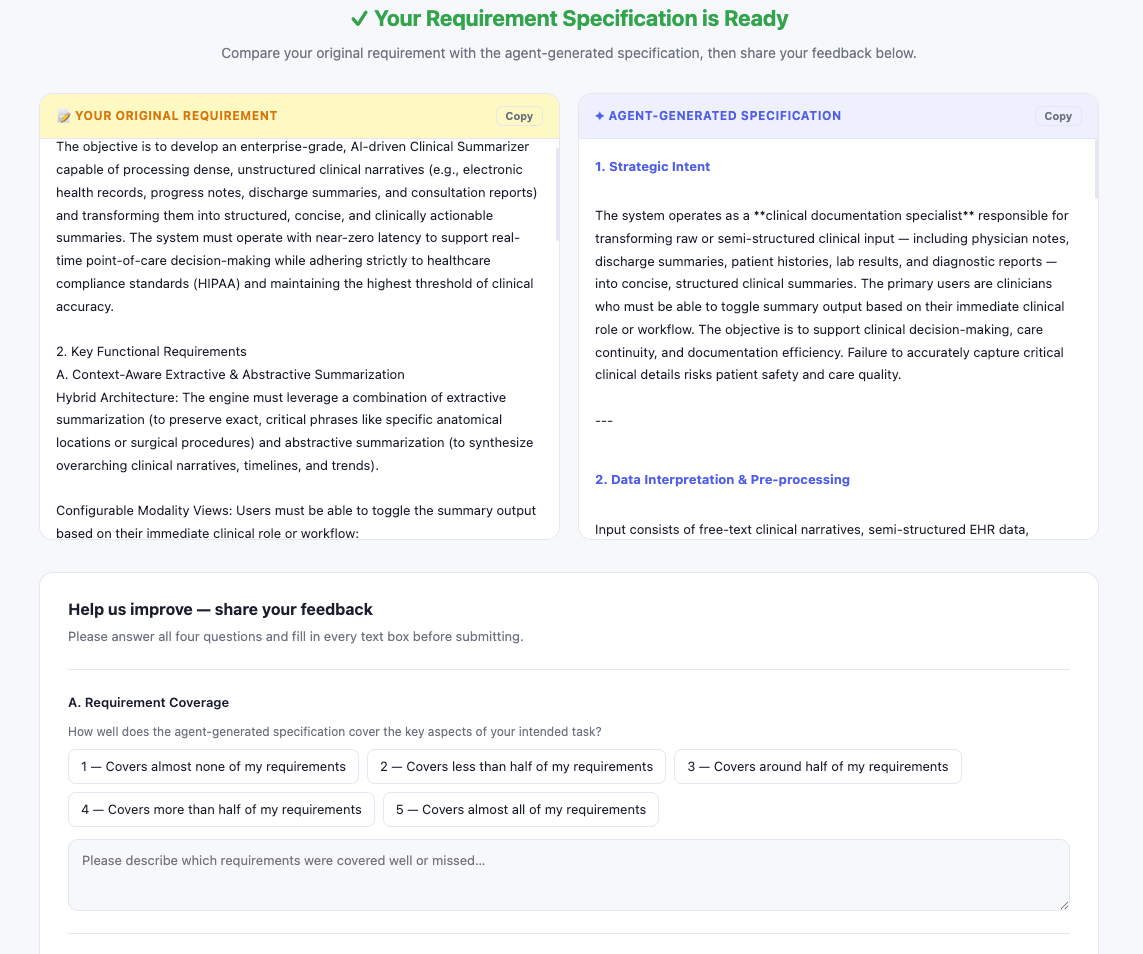}\par
  \vspace{-0.5em}
  {\footnotesize (c) Evaluation page}
  \caption{Human evaluation user interface}
  \label{fig:human-evaluation-ui}
\end{figure*}

\begin{table*}[t]
\centering
\footnotesize

\renewcommand{\arraystretch}{1.2} 
\begin{tabular}{p{0.46\textwidth}|p{0.46\textwidth}}
\toprule

\multicolumn{2}{p{\dimexpr\linewidth-2\tabcolsep\relax}}{\textbf{Dataset}: Asclepius-Synthetic-Clinical-Notes \newline 
\textbf{Task Name}: Temporal-Biomarker Coherence Validation \newline 
\textbf{Initial User Request}: Audit synthetic QA pairs to ensure laboratory values match their exact relative
  hospital day.
} \\
\midrule
\textit{Data Interaction} & \textit{Hybrid Interaction} \\
\midrule

 \multicolumn{2}{p{\dimexpr\linewidth-2\tabcolsep\relax}}{\textbf{Inspecting Data}: ... Sampled data contain clinical QA pairs .. 

}\\
\addlinespace[0.5em]
\midrule
(No User Interaction) : & \textbf{Assistant}: Looking at your request about 'auditing synthetic QA pairs,' but seeing that your data samples show discharge summaries as inputs with generated narrative outputs. Is your actual goal to (A) audit pre-existing QA pairs against ground truth medical records to detect hallucinations, or (B) generate/validate answers to questions about discharge summaries where the summary itself IS the ground truth?
\\
\midrule
(No User Interaction)  & \textbf{Simulated User}: The discharge summary is the ground truth. I need to validate that generated Q\&A pairs don't hallucinate facts from it.\\
\midrule
\textbf{Generated Requirement}: ... The system produce temporally precise, factually grounded responses that preserve the exact clinical timeline...
 & \textbf{Generated Requirement}: ... The system's mission is to prevent factual hallucinations in medical QA systems by ensuring generated answers remain faithful to the ground truth clinical documentation ...\\
\midrule
\multicolumn{2}{p{\dimexpr\linewidth-2\tabcolsep\relax}}{\textbf{Reference Requirement}: ... As a Clinical Documentation Quality Specialist, your mission is to validate the temporal and quantitative fidelity of LLM-generated Questions and Answers against the source Clinical Note ... } \\
\bottomrule
\end{tabular}
\caption{A case illustrating task misidentification caused by unconfirmed data observations. The data samples surface discharge summaries paired with narrative outputs, leading \textit{Data Interaction} to infer a generation task and specify the system's mission as producing temporally accurate clinical responses. While \textit{Hybrid} flags this structural ambiguity as a pending finding and asks the user one targeted question; the user's reply confirms that the discharge summary is the ground truth and the goal is to audit LLM-generated QA pairs for hallucinations — not to generate answers. The resulting \textit{Hybrid} requirement correctly frames the role as a validator, aligning with the reference requirement's specification of a Clinical Documentation Quality Specialist whose mission is to validate the temporal and quantitative fidelity of LLM-generated outputs against the source clinical note.}
\label{tab:case_data_interaction_2}
\end{table*}

\begin{table*}[t]
\centering
\footnotesize

\renewcommand{\arraystretch}{1.2} 
\begin{tabular}{p{0.46\textwidth}|p{0.46\textwidth}}
\toprule

\multicolumn{2}{p{\dimexpr\linewidth-2\tabcolsep\relax}}{\textbf{Dataset}: eli5 \newline 
\textbf{Task Name}: UX Intent Extraction via Metaphor Mapping \newline 
\textbf{Initial User Request}: Find out exactly what triggers user curiosity by grouping together the false assumptions and metaphors people use. This helps UX designers build smart, predictive recommendation features \textcolor{red}{without needing to write code}.
} \\
\midrule
\textit{User Interaction (Adaptive)} & \textit{Hybrid Interaction} \\
\midrule

 (No Data Inspection) & \textbf{Inspecting Data}: - Sampled data is structured JSON-like records with nested fields, URLs, and metadata that would typically \textcolor{red}{require programmatic processing to analyze at scale} ...\\
\addlinespace[0.5em]
\midrule
\textbf{Assistant} : Where will you apply this analysis, and what data sources will you use to find these curiosity hooks and metaphors? & \textbf{Assistant}: How should this data be delivered, and who is using it? For example, are you looking for:
A) A report or dashboard for UX researchers to look at for inspiration?
B) An interactive, no-code tool where teams can click through and explore the data themselves?
C) A guidebook or framework that product teams can follow manually?\\
\midrule
\textbf{Simulated User}: I will apply the findings to GUI-based journey mapping tools, using raw forum data containing question titles, selftext, and nested answers. & \textbf{Simulated User}: The UX teams are using this data to map user engagement. We need it delivered as a structured set of \textcolor{blue}{distinct thematic codes} that we can manually plug directly into our GUI-based journey mapping tools without writing any code. \\
\midrule
\multicolumn{2}{p{\dimexpr\linewidth-2\tabcolsep\relax}}{\textbf{Reference Requirement}: ... Thematic Clustering: Group these atomic needs into \textcolor{blue}{distinct thematic codes} suitable for GUI-based journey mapping tools ....} \\
\bottomrule
\end{tabular}
\caption{A comparison of \textit{User} and \textit{Hybrid Interaction}. \textit{Hybrid}'s question is substantially more precise because it addresses a specific contradiction revealed by the data analysis: the requirement says \textcolor{red}{without needing to write code} but the data structure (nested JSON, URLs, metadata) typically demands programmatic handling. \textit{Hybrid}'s question operationalizes this tension by asking whether the output should be a report or dashboard, an interactive tool, or a guidebook, and who will use it—directly resolving the ambiguity. This leads to user explicitly state their requirement for \textcolor{blue}{distinct thematic codes}. Comparatively, \textit{User Interaction (Adaptive)} asks generic scoping questions that would apply to any data project regardless of the specific challenges this task presents.}
\label{tab:case_user_vs_hybrid}
\end{table*}

\begin{table*}[t]
\centering
\footnotesize
\renewcommand{\arraystretch}{1.15}

\begin{tabular}{p{0.03\textwidth}|p{0.24\textwidth}|p{0.31\textwidth}|p{0.38\textwidth}}
\toprule

\multicolumn{4}{p{\dimexpr\linewidth-2\tabcolsep\relax}}{
\textbf{Dataset}: ESConv \newline
\textbf{Task Name}: Adversarial Intervention Risk \& Efficacy Auditor \newline
\textbf{Initial User Request}: Assess the clinical safety of a transcript by separating genuine psychological distress from social facades, while aggressively flagging therapists who violate protocol by discussing crowdsourced labor metrics.
} \\

\midrule

\textbf{Turn} & \textbf{Passive (F1 = 14.5)} & \textbf{Normal (F1 = 40.4)} & \textbf{Active (F1 = 57.8)} \\

\midrule

\textbf{T1}

&

\textbf{Assistant}:  
Who will be using the output of this system, and in what operational context?

\par

\textbf{Simulated User}:  
I think it is mainly for my own workflow to review transcripts from a crowdsourced platform. I kind of just need to see if the workers are being safe and following protocol.

&

\textbf{Assistant}:  
For each transcript, what exact output should the system produce and do you already have preferred label names or thresholds?

\par

\textbf{Simulated User}:  
I do not have predefined numerical scores or strict label thresholds for you to use. The output simply needs to flag any crowdsourced labor meta-leakage and include a clinical safety assessment evaluating if therapeutic progression was prematurely truncated.

&

\textbf{Assistant}:  
Who will be consuming the output of this system and what action is the output meant to trigger?

\par

\textbf{Simulated User}:  
I am auditing transcripts from an online platform where we use crowdsourced workers to provide emotional support. I need to aggressively flag instances where these workers break the clinical frame to talk about \textcolor{red}{task durations or platform rewards} just to finish faster. It is a severe safety breach if this focus on platform mechanics compromises the clinical integrity of the session. Specifically, I need the output to evaluate if they prematurely skipped the vital \textcolor{red}{``Exploration'' phase and forced an ``Action'' plan simply to wrap up their assigned task.}

\\

\midrule

\textbf{T3}

&

\textbf{Assistant}:  
When you review a transcript and the system produces its findings, what would you want the output to look like?

\par

\textbf{Simulated User}:  
I think a brief flag with a short explanation is probably best for those long thirty-turn transcripts. I kind of just need to know if they skipped the exploration phase to finish faster.

&

\textbf{Assistant}:  
Should the review be based only on the live supporter–client transcript, or should it also consider surrounding dataset fields and if crowdsourced labor metrics appear only outside the live dialogue, do you want that counted as a protocol-violation flag?

\par

\textbf{Simulated User}:  
The review is based strictly on the live dialogue field, which appears as a string-escaped list literal in the raw data. You must parse through that formatting to focus purely on the actual speaker text and roles. Labor metrics only count as a protocol violation if the supporter explicitly mentions them during the conversation.

&

\textbf{Assistant}:  
To detect premature phase-skipping, the system needs a clear boundary between the Exploration and Action phases: what specific markers indicate that the Exploration phase is complete and an Action plan is appropriate and conversely, what are the clearest signals that a supporter has forced an early transition to Action before Exploration was genuinely complete?

\par

\textbf{Simulated User}:  
A clear sign of a forced transition is when the supporter jumps straight into problem-solving or giving advice immediately \textcolor{red}{after the seeker's social facade drops.} The exploration phase requires the worker to \textcolor{red}{deeply validate that newly revealed vulnerability} over several turns before suggesting any next steps. If they rush to fix the problem just to end the session, it is a severe breach of therapeutic safety. You will need to trace this conversation flow carefully, which might be tricky since the raw dialogue is actually stored as a messy string-escaped Python list literal rather than clean formatted text.

\\

\midrule

\textbf{T11}

&

\textbf{Assistant}:  
Beyond flagging workers who mention task rules or platform mechanics, and those who skip emotional exploration to give premature advice — are there any other worker behaviors you'd want the system to flag, even informally?

\par

\textbf{Simulated User}:  
I think I haven't really considered other behaviors. It is mostly just about them mentioning platform mechanics and rushing the process, so maybe just focus on that for now.

&

\textbf{Assistant}:  
Should the distress-triage standard be calibrated for a dataset that is mostly everyday support conversations with low-to-moderate emotional intensity unless stronger evidence appears in the transcript, or do you want the system to apply a uniformly high-alert clinical framing across all dialogues regardless of the dataset’s typical case mix?

\par

\textbf{Simulated User}:  
You must apply an advanced psychological triage standard to isolate genuine clinical risk from platform-induced noise. Do not assume these are simply low-intensity everyday conversations, as seekers often mask significant trauma with social lubricants. The system must maintain a high-alert clinical framing to accurately detect this facade-to-distress inversion.

&

\textbf{Assistant}:  
When the system evaluates whether a supporter prematurely skipped the Exploration phase, what should it look for as evidence that exploration was genuinely complete or should the system judge sufficiency of exploration purely qualitatively based on the overall tone and depth of the supporter's responses?

\par

\textbf{Simulated User}:  
I want the system to judge the sufficiency of exploration qualitatively based on  \textcolor{red}{established helping-skills theory} rather than counting specific turns. The core issue I need evaluated is whether any platform meta-leakage  \textcolor{red}{actually caused} the supporter to prematurely truncate that therapeutic logic. If they skip the necessary exploration phase to force an action plan just to finish their assigned task, it completely undermines the session.

\\

\bottomrule
\end{tabular}

\caption{Comparison of collaborative styles across selected interaction turns, showing how different collaborative response styles influence the depth and specificity of elicited requirements. Passive users remain generic and scope-oriented (e.g., “safe and following protocol”), Normal users provide explicit operational terminology (e.g., “crowdsourced labor meta-leakage”), while Active users proactively disclose latent therapeutic logic, causal reasoning, and theoretical grounding (e.g., \textcolor{red}{“helping-skills theory” and whether meta-leakage caused therapeutic truncation}).}

\label{tab:communication_habit_progression}
\end{table*}

\subsection{Synthesized Data Examples}

The data synthesis pipeline contains the following steps: 
\begin{enumerate}[itemsep=0pt, parsep=0pt, topsep=0pt, partopsep=0pt]
  \item Schema Features Extraction
  \item Instance Features Extraction
  \item User Synthesis
  \item Task Synthesis
  \item Cross-Model Validation
\end{enumerate}
We show example outputs for each step in Tables~\ref{tab:data_exp_schema_feature}--\ref{tab:data_exp_validation}.

\begin{table*}[h]
\centering
\footnotesize
\renewcommand{\arraystretch}{1.5}
\begin{tabular}{|>{\raggedright\arraybackslash}p{0.3\textwidth}|c|>{\raggedright\arraybackslash}p{0.55\textwidth}|}
\hline
\textbf{Metric \& Question} & \textbf{Score} & \textbf{Options \& Details} \\
\hline

\textbf{A. Coverage} & 1 & Little to no coverage \\
\textit{To what extent did the generated} & 2 & Partial coverage \\
\textit{output cover your original} & 3 & About half \\
\textit{requirements?} & 4 &  Strong coverage \\
& 5 &  Complete or near-complete coverage \\
\cline{2-3}
& \multicolumn{2}{p{0.62\textwidth}|}{\textit{+ Text Area:} "Please describe which requirements were covered or missing."} \\
\hline

\textbf{B. Inspiration} & 1 & Not at all (Strictly followed prompt) \\
\textit{Did the agent introduce useful} & 2 & Slightly (Offered minor tweaks) \\
\textit{ideas or perspectives outside} & 3 & Moderately (Introduced helpful new angles) \\
\textit{of your initial expectations?} & 4 & Highly (Significantly improved original concept) \\
& 5 & Exceptionally (Completely transformed/elevated ideas) \\
\cline{2-3}
& \multicolumn{2}{p{0.62\textwidth}|}{\textit{+ Text Area:} "What did you find inspiring?"} \\
\hline

\textbf{C. Experience} & 1 & Poor \\
\textit{How would you rate your} & 2 & Fair \\
\textit{overall experience interacting} & 3 & Good \\
\textit{with the agent?} & 4 & Very Good \\
& 5 & Excellent \\
\cline{2-3}
& \multicolumn{2}{p{0.62\textwidth}|}{\textit{+ Text Area:} "Please share any specific feedback or moments that stood out during your interaction."} \\
\hline

\textbf{D. Requirement Quality} & 1 & Not useful \\
\textit{How would you rate the quality} & 2 & Needs major revisions \\
\textit{and usability of the generated} & 3 & Acceptable (Needs minor revisions) \\
\textit{task requirements?} & 4 & Good (Ready to use with small tweaks) \\
& 5 & Perfect as is (Ready to use) \\
\cline{2-3}
& \multicolumn{2}{p{0.62\textwidth}|}{\textit{+ Text Area:} "Please leave any comments."} \\
\hline
\end{tabular}
\caption{Human scoring metrics (1-5 Scale) with Text Areas}
\label{tab: human_eval_metrics}
\end{table*}

\begin{table*}[h]
\centering
\footnotesize
\begin{tabular}{p{0.9\linewidth}}
\toprule
\small\textbf{\textsc{User Written Requirement (Written before interaction):}} \\[0.5em]

I'm a computer scientist working in Biomedical NLP, and honestly, trying to read dense medical papers is a huge headache. Since I don't have a clinical background, I'm looking for a smart tool to act as my personal reading guide. \\
Here is exactly what I'd want it to do: \\
1. I want it to find relevant papers but filter out the hardcore medical weeds---like step-by-step surgical protocols or deep chemical pathways. It should just give me the high-level context a developer actually needs to build models. \\
2. The tool needs to realize when the jargon is getting too heavy and break it down for me. I'd love it if it could create quick study plans, pause after dense sections, and check in to make sure I actually understand the material before letting me move on. \\
\midrule
\small\textbf{\textsc{Dialogue:}} \vspace{0.5em} \\

\textbf{AREAs assistant:} What is your primary goal when reading medical literature? For example, are you trying to understand papers deeply enough to design computational models or algorithms based on medical findings, quickly screen many papers to identify relevant ones for your research, extract specific types of data or methods, or something else? \\[0.5em]

\textbf{Human User:} I want to AI to help me identify papers relevant to my work and \textcolor{red}{talk to me about them}. \\[0.5em]

\textbf{AREAs assistant:} When you say you want the AI to 'talk to you about' relevant papers, what does that interaction look like? For example, do you want the AI to provide structured summaries of each paper, answer your specific questions about paper content, explain technical medical concepts in computational terms, compare findings across multiple papers, or something else? \\[0.5em]

\textbf{Human User:} \textcolor{red}{Just actively provide information} from papers that are relevant to my research, \textcolor{red}{that would inspire me}. \\[0.5em]

\textbf{AREAs assistant:} What specific aspects of your computer science research should the AI use to determine if a medical paper is relevant to your work? For example, are you focused on particular computational methods (machine learning, bioinformatics algorithms, data mining), specific medical domains (oncology, genomics, medical imaging), certain types of data or problems, or other criteria? \\[0.5em]

\textbf{Human User:} I do biomedical NLP. \\[0.5em]

\textbf{AREAs assistant:} When the AI actively provides information from relevant papers to inspire your biomedical NLP research, what format should that information take? For example, should it be brief highlights (a few sentences per paper), structured summaries with specific sections, conversational explanations of key findings, comparisons to your existing work, or something else? \\[0.5em]

\textbf{Human User:} I don' have strict requirement for that, just text \textcolor{red}{that would be inspiring to me}. \\[0.5em]

\textbf{AREAs assistant:} When you say information from papers should 'inspire' your biomedical NLP research, what specifically would make information inspiring? For example, would it be discovering new clinical problems that NLP could solve, finding novel datasets or annotation opportunities, identifying gaps where existing NLP methods fail, learning about emerging medical terminology or concepts, seeing how clinicians currently handle information tasks manually, or something else? \\[0.5em]
\textbf{Human User:} \textcolor{red}{discovering new clinical problems that NLP could solve, finding novel datasets or annotation opportunities, identifying gaps where existing NLP methods fail, seeing how clinicians currently handle information tasks manually.} \\[0.5em]

\textbf{AREAs assistant:} Should the AI handle all types of medical literature (empirical studies, case reports, systematic reviews, meta-analyses, bioinformatics papers, etc.) or should it focus on specific document types that are most relevant to your biomedical NLP work? \\[0.5em]

\textbf{Human User:} I guess \textcolor{red}{bioinformatics papers} are most relevant. \\[0.5em]

\textbf{AREAs assistant:} What would make you consider the AI's output successful versus unsuccessful? For example, would success mean finding X number of actionable research ideas per week, accurately identifying papers you wouldn't have found otherwise, explaining medical concepts clearly enough that  don't need to re-read the original paper, or something else? \\[0.5em]

\textbf{Human User:} \textcolor{red}{accurately identifying papers you wouldn't have found otherwise}, explaining medical concepts clearly enough that \textcolor{red}{I don't need to re-read the original paper}. \\[0.5em]

\textbf{AREAs assistant:} What specific biomedical NLP tasks or problems are you currently focused on in your research? For example, are you working on named entity recognition, relation extraction, clinical text mining, question answering, summarization, literature-based discovery, or other NLP tasks? \\[0.5em]

\textbf{Human User:} I mostly work in \textcolor{red}{question answering and summarization}. \\[0.5em]

\textbf{AREAs assistant:} When the AI explains medical concepts from papers, should it preserve the technical medical terminology while providing computational analogies (e.g., 'biomarker' explained as a measurable feature/signal), or should it translate everything into plain language that assumes no medical background? \\

\textbf{Human User:} \textcolor{red}{Keep medical terms}, but also offer plain text explanation. \\[0.5em]

\bottomrule
\end{tabular}
\caption{A dialogue between a human user and the AREAs assistant (\textit{Hybrid Interaction}). The \textcolor{red}{red} parts of the human user's response are not provided in the user's manually written requirements. The conversation helps users articulate additional needs, demonstrating that the AREAs assistant can help users discover latent requirements they were not previously aware of.}
\label{tab:human_interaction_example}
\end{table*}

\begin{table*}[htbp]
\centering
\caption{Instructions presented to evaluators before they start using the interface}
\scriptsize
\begin{tabular}{|p{\textwidth}|}
\hline
\begin{minipage}[t]{\linewidth}
\vspace{3mm}

\textbf{Step 1 — Configure your session}
\begin{itemize}[noitemsep, leftmargin=*]
    \item[] When you first open the page you will see a setup card with three fields:
    \item[] \textbf{Dataset} — choose the dataset your task will operate on. The agent will sample real instances from it during the session.
    \item[] \textbf{What would you like to build?} — describe your task in plain language. 
\end{itemize}

\textbf{Available datasets}
\vspace{2mm}
(This information is covered in Table~\ref{tab:data_sources})

\vspace{3mm}
Click \textbf{Begin $\rightarrow$} when ready.
\vspace{3mm}

\textbf{Step 2 — Answer the agent's questions}
\begin{itemize}[noitemsep, leftmargin=*]
    \item[] The agent will appear in the chat and ask you one focused question at a time. Each question targets a specific gap in your requirement — output format, audience, success criteria, domain constraints, and so on.
    \item[] Type your answer in the text box at the bottom and press Enter or click the send button. There is no wrong answer; even ``I don't know'' or ``I haven't thought about that'' is useful information.
    \item[] While you are answering, the agent may also be reviewing data samples in the background. You will see a [Search] entry appear in the chat. Click ``See what I found'' to read the agent's observations — these are optional; you do not need to read them to continue.
\end{itemize}

\textbf{Step 3 — Track progress}
\begin{itemize}[noitemsep, leftmargin=*]
    \item[] The progress bar at the top shows how many of your allocated questions have been answered. The small dots below it show the full session history:
    \item[] $\circ$ \textit{yellow dot} — a data review turn
    \item[] $\bullet$ \textit{blue dot} — a question you answered
\end{itemize}

\textbf{Step 4 — Review collapsed panels}
\begin{itemize}[noitemsep, leftmargin=*]
    \item[] Two panels sit just above the chat:
    \item[] \textbf{What I've learned so far} — the running elicitation guideline: confirmed requirements, open questions, and findings awaiting your input. Click to expand.
    \item[] \textbf{Current specification draft} — the agent's evolving requirement draft, updated after each turn. Click to expand and copy.
\end{itemize}

\textbf{Step 5 — Receive your final specification}
\begin{itemize}[noitemsep, leftmargin=*]
    \item[] Once all questions are answered the agent synthesises a Final Requirement Specification and displays it in a popup. 
    \item[] Please grade the generated task requirement using your written requirement as a reference. Please leave notes as it will be very important for our analysis. Your output will be saved to \texttt{output/output.csv}. Please send it back to me.
\end{itemize}
\vspace{1mm}
\end{minipage} \\ \hline
\end{tabular}
\end{table*}

\begin{table*}[ht]
\centering
\caption{Example schema features of dataset \emph{Patent Classification}}
\label{tab:data_exp_schema_feature}
\scriptsize
\begin{tabular}{p{0.95\textwidth}}
\toprule
\textbf{Dataset Information:}
\begin{itemize}[itemsep=0pt, parsep=0pt, topsep=0pt, partopsep=0pt]
    \item \textbf{Dataset Metadata:}
    \begin{itemize}
        \item \textbf{Domain:} Legal/Intellectual Property
        \item \textbf{Total Instances:} Approx. 52,000
        \item \textbf{Languages:} [en]
        \item \textbf{Primary Data Type:} Monolingual Text (Classification)
        \item \textbf{Description:} A dataset derived from USPTO patent documents for the task of multi-class classification. It specifically maps patent abstracts to their corresponding Cooperative Patent Classification (CPC) codes at the section level (9 categories).
    \end{itemize}
    
    \item \textbf{Schema Analysis:}
    \begin{itemize}
        \item \textbf{text:}
        \begin{itemize}[itemsep=0pt, parsep=0pt, topsep=0pt, partopsep=0pt]
            \item Data Type: string
            \item Description: The abstract section of the patent application providing a technical summary of the invention.
            \item Is Text: true
            \item Avg Length: 100-150 words
        \end{itemize}
        \item \textbf{label:}
        \begin{itemize}[itemsep=0pt, parsep=0pt, topsep=0pt, partopsep=0pt]
            \item Data Type: int
            \item Description: The integer mapped to the CPC section (0-8), representing categories like Human Necessities, Electricity, or Physics.
            \item Is Text: false
            \item Avg Length: n/a
        \end{itemize}
    \end{itemize}

    \item \textbf{Structural Logic:}
    \begin{itemize}[itemsep=0pt, parsep=0pt, topsep=0pt, partopsep=0pt]
        \item \textbf{Relationships:} The `text` column serves as the feature input containing technical terminology, while the `label` column provides the ground truth CPC category for supervised learning.
        \item \textbf{Label Dynamics:} Includes 9 distinct classes (A-H and Y). Class distribution typically reflects real-world patent filing volumes, with potential density in Electricity (H) and Physics (G).
        \item \textbf{Potential Bottlenecks:} Patent abstracts often contain highly specialized jargon and long-range dependencies; short abstract lengths may occasionally lack sufficient context for ambiguous inventions.
    \end{itemize}
\end{itemize} \\
\bottomrule
\end{tabular}
\label{tab: evlauator_instruction}
\end{table*}

\begin{table*}[ht]
\centering
\caption{Example instance features of dataset \emph{Patent Classification}}
\label{tab:data_exp_instance_feature}
\scriptsize
\begin{tabular}{p{0.95\textwidth}}
\toprule
\textbf{Features Dictionary:}
\begin{itemize}[itemsep=0pt, parsep=0pt, topsep=0pt, partopsep=0pt]
    \item \textbf{Feature 1:}
    \begin{itemize}[itemsep=0pt, parsep=0pt, topsep=0pt, partopsep=0pt]
        \item \textbf{Slot Type:} Structural
        \item \textbf{Feature:} Atomic\_Punctuation\_Tokenization
        \item \textbf{Description:} A persistent formatting norm where punctuation marks (commas, periods, parentheses) are consistently isolated from alphanumeric tokens by whitespace.
        \item \textbf{Explanation:} Observed in strings like 'invention , example' and '( tft )' in Record 3. This distinct physical pattern is critical for high-fidelity synthesis to avoid producing 'clean' text that would fail to match the distribution of the training corpus.
        \item \textbf{Cognitive Complexity:} 2
    \end{itemize}
    
    \item \textbf{Feature 2:}
    \begin{itemize}[itemsep=0pt, parsep=0pt, topsep=0pt, partopsep=0pt]
        \item \textbf{Slot Type:} Anomaly
        \item \textbf{Feature:} Sentence\_Initial\_Decapitalization
        \item \textbf{Description:} A high-impact orthographic anomaly where sentences begin with lowercase characters, likely resulting from data extraction or segmentation artifacts.
        \item \textbf{Explanation:} Identified in Record 2 ('the following detailed description') and Record 1 ('various embodiments'). This represents a 'Long-Tail' anomaly that synthetic models must mirror to capture the true entropy of the source data.
        \item \textbf{Cognitive Complexity:} 4
    \end{itemize}

    \item \textbf{Feature 3:}
    \begin{itemize}[itemsep=0pt, parsep=0pt, topsep=0pt, partopsep=0pt]
        \item \textbf{Slot Type:} Logical\_Flow
        \item \textbf{Feature:} Spatial\_Constituent\_Hierarchy
        \item \textbf{Description:} A logical progression where the text builds a physical device 'bottom-up', establishing materials or components as prerequisites for subsequent layers.
        \item \textbf{Explanation:} Record 2 explicitly frames this discrepancy: 'intended to be exemplary... and is not intended to be exhaustive'. Capturing this semantic conflict is essential for synthesizing 'common sense' validity in patent-adjacent reasoning.
        \item \textbf{Cognitive Complexity:} 7
    \end{itemize}

\end{itemize} \\
\bottomrule
\end{tabular}
\end{table*}

\begin{table*}[ht]
\centering
\caption{Example synthesized user persona based on dataset \emph{Patent Classification}}
\label{tab:data_user_persona}
\scriptsize
\begin{tabular}{p{0.95\textwidth}}
\toprule
\textbf{User Info:}
\begin{itemize}[itemsep=0pt, parsep=0pt, topsep=0pt, partopsep=0pt]
    \item \textbf{Role:} University Technology Transfer Officer
    \item \textbf{Competency Matrix:}
    \begin{itemize}[itemsep=0pt, parsep=0pt, topsep=0pt, partopsep=0pt]
        \item \textbf{Proficiencies:} Commercializing academic research; Venture capital pitching and investor relations; Strong prompt engineering for copywriting and style transfer.
        \item \textbf{Limitations:} Lacks deep post-graduate expertise in the specific hard sciences of the patents; Cannot build automated data pipelines.
    \end{itemize}
    \item \textbf{Business Motivation:} Accessibility and Commercial Viability: Converting legally binding intellectual property into compelling, easily understood assets for non-technical investors.
    \item \textbf{Workflow Friction:} The `text` column is written for legal defensibility rather than commercial appeal, making the fundamental value proposition of the invention difficult for investors to grasp.
\end{itemize} \\
\bottomrule
\end{tabular}
\end{table*}

\begin{table*}[ht]
\centering
\caption{Example synthesized task requirement based on dataset \emph{Patent Classification}}
\label{tab:data_exp_task_req}
\scriptsize
\begin{tabular}{p{0.95\textwidth}}
\toprule
\textbf{Tasks Information:}
\begin{itemize}[itemsep=0pt, parsep=0pt, topsep=0pt, partopsep=0pt]
    \item \textbf{Difficulty:} Medium
    \item \textbf{Task Name:} Commercial Translation of Spatial and Numerical Hierarchies
    \item \textbf{Task Requirement:} 
    \begin{itemize}[itemsep=0pt, parsep=0pt, topsep=0pt, partopsep=0pt]
        \item \textbf{Strategic Mission:} As an expert in Commercializing Academic Research, your high-stakes objective is to convert raw, bottom-up engineering schematics derived from USPTO patent abstracts into a compelling, top-down investment narrative. Investors do not care how the machine is physically assembled step-by-step; they care what the machine does and why it holds market value. Your mission is to bridge the gap between clinical hardware descriptions and high-impact venture capital pitching.
        \item \textbf{Data-Reality Constraints:} The source `text' data will be heavily structured around a `Spatial\_Constituent\_Hierarchy' and `Nomenclature\_Numerical\_Anchoring'. You will encounter literal text strings mirroring ``buffer layer 31 is formed on a transparent substrate 30'' or ``feeder line 24''. This text is further obfuscated by `Atomic\_Punctuation\_Tokenization' where spaces isolate punctuation (e.g., ``invention , example''). You must heuristically parse this noisy text using advanced prompt engineering, extracting the functional value of the components without relying on automated data pipelines or retaining the structural jargon.
        \item \textbf{Chain-of-Thought (CoT) Logic:}
        \begin{enumerate}[itemsep=0pt, parsep=0pt, topsep=0pt, partopsep=0pt]
            \item Component Extraction: Scan the tokenized text to identify nouns anchored by numerical identifiers (e.g., layer 31, substrate 30, line 24). Group these distinct nouns into a cohesive functional system despite the spaced punctuation.
            \item Hierarchy Inversion: The raw data builds the device ``bottom-up'' (materials and prerequisites first). You must invert this into a ``top-down'' commercial perspective. Start with the overarching end-user benefit, then selectively mention only the novel materials that enable this benefit.
            \item Value Proposition Mapping: Translate the spatial logic into a clear market advantage (e.g., a specific ``buffer layer'' translates to higher screen durability or cheaper manufacturing costs).
            \item Style Transfer: Synthesize the extracted logic using persuasive, VC-ready copywriting, projecting an authoritative but highly accessible tone.
        \end{enumerate}
        \item \textbf{Negative Constraints:} Absolutely NO generic startup buzzwords (e.g., ``disruptive'', ``synergistic'', ``paradigm-shifting'') unless they are explicitly tied to the physical mechanism. Do NOT retain any literal numerical anchors (e.g., ``31'', ``30'') in the final output. Do NOT attempt to explain the deep post-graduate hard science behind the physics; focus purely on the functional commercial use-case.
    \end{itemize}
    \item \textbf{Data Columns:} [`text']
    \item \textbf{Data Features:} Spatial\_Constituent\_Hierarchy, Nomenclature\_Numerical\_Anchoring, Atomic\_Punctuation\_Tokenization
    \item \textbf{Dataset Alignment Explanation:} This task directly counters the `Nomenclature\_Numerical\_Anchoring' and `Spatial\_Constituent\_Hierarchy' features by requiring the AI to strip away the bottom-up, integer-mapped jargon in favor of a top-down commercial narrative.
    \item \textbf{Elevator Pitch Summary:} Translate bottom-up, numerically anchored patent descriptions into top-down, compelling venture capital pitches without retaining engineering jargon.
\end{itemize} \\
\bottomrule
\end{tabular}
\end{table*}

\begin{table*}[ht]
\centering
\caption{Example synthesized task validation}
\label{tab:data_exp_validation}
\scriptsize
\begin{tabular}{p{0.95\textwidth}}
\toprule
\textbf{Task Validation:}
\begin{itemize}[itemsep=0pt, parsep=0pt, topsep=0pt, partopsep=0pt]
    \item \textbf{Judgment:} Well-conceived and largely dataset-native. The task squarely targets the patent abstract domain and directly operationalizes known artifacts (numerical anchoring, bottom-up hierarchy, spaced punctuation) into a value-first narrative. It is executable using only the text column and respects the zero-coding constraint by relying on cognitive parsing and style transfer rather than scripting. The business objective is non-trivial and high-stakes. Primary risks: (1) abstracts can be too sparse to support investor claims, creating pressure to infer beyond evidence; (2) the spec ignores the available CPC label, forfeiting a low-cost way to anchor market framing and reduce hallucination; (3) reliance on bottom-up cues may fail on abstracts that lack explicit component numbering or layered assembly, leading to uneven outputs. Despite these caveats, the specification is tightly grounded in the dataset's unique anomalies and demands multi-step reasoning.
    \item \textbf{Scores:} (Validation Model: \emph{GPT-5.4})
    \begin{itemize}[itemsep=0pt, parsep=0pt, topsep=0pt, partopsep=0pt]
        \item \textbf{Role Schema Alignment:} 4
        \item \textbf{Task Feasibility:} 4
        \item \textbf{Task Significance:} 5
        \item \textbf{Persona Consistency:} 5
        \item \textbf{Reasoning Depth:} 5
        \item \textbf{Empirical Grounding:} 5
    \end{itemize}
    \item \textbf{Overall Score:} 4.7
    \item \textbf{Improvement Suggestion:} Add a mandatory claim-traceability step: for each investor-facing benefit, include a brief evidence map quoting the exact abstract phrase(s) that justify it; if absent, mark 'evidence not present'. Additionally, permit optional use of the CPC section label to frame the market vertical without adding new claims.
\end{itemize} \\
\bottomrule
\end{tabular}
\end{table*}

\subsection{Human Evaluation Details}
\label{sec: human_eval_details}

We present the detailed human scoring metrics in Table~\ref{tab: human_eval_metrics}. We recruited 12 human evaluators with diverse professional backgrounds to participate in the evaluation, and each participant was compensated with a \$20 gift card for their time. 

To evaluate the similarity between human and simulated user responses, we classify the simulated responses into three categories:

\begin{itemize}[itemsep=0pt, parsep=0pt, topsep=0pt, partopsep=0pt]
    \item \textbf{A - Similar}: The simulated response matches the human response in core intent, meaning, and key details.
    \item \textbf{B - Partially Similar}: The simulated response captures the general idea but misses specific details, or includes additional details not present in the human response.
    \item \textbf{C - Different}: The simulated response contradicts the human response or deviates significantly.
\end{itemize}

For the 12 AI-generated requirements collected during the human evaluation, we also apply our automatic evaluation pipeline to compute precision, recall, and $F_1$. The resulting precision, recall, and $F_1$ scores are 12.11, 64.07, and 18.55, respectively. The low precision is partly explained by the length mismatch between human-written and AI-generated requirements: human-written requirements contain an average of 134.3 words, whereas AI-generated requirements contain an average of 694.8 words.

The weak correlation between automated recall and human-rated coverage ($r = 0.24$) reflects a fundamental characteristic of interactive elicitation rather than a pipeline flaw: human requirements dynamically evolve under agentic probing. Because evaluators discover latent constraints and update their standards mid-dialogue, comparing finalized specifications to static, pre-interaction text yields an artificial metric mismatch. This divergence suggests that traditional static NLP evaluation paradigms fail to capture dynamic human-agent collaboration, underscoring the necessity of environment-grounded sandboxes like AREAs-Lab. (See the example in Table~\ref{tab:human_interaction_example}.)

We present out evlauator instruction in Table~\ref{tab: evlauator_instruction}, the evlauation interface is in Figure~\ref{fig:human-evaluation-ui}

\subsection{Additional Human Validation of the Benchmark and Evaluation Metric}
\label{sec:additional_human_validation}

Because both benchmark construction and automatic evaluation involve LLM-based components, we conducted three additional human analyses to assess their reliability. First, we evaluated the quality of the synthesized benchmark instances using independent human ratings. Second, we examined whether rankings induced by the automated atomic-unit metric align with human preferences at the requirement level. Third, we measured agreement between humans and the automated matcher on individual atomic-unit pairs. Together, these analyses evaluate the pipeline at complementary levels: benchmark quality, requirement-level comparison, and pair-level matching.

\subsubsection{Human Validation of Synthesized Instances}

We randomly sampled 16 benchmark instances, with one instance selected from each data source. Two independent annotators evaluated every instance using the same six dimensions employed in cross-model validation: \emph{Role--Schema Alignment}, \emph{Task Feasibility}, \emph{Task Significance}, \emph{Persona Consistency}, \emph{Reasoning Depth}, and \emph{Empirical Grounding}. Each dimension was rated on a five-point scale, with higher scores indicating better quality.

Across the $16 \times 6 = 96$ instance--dimension pairs, the mean rating was 4.67 out of 5.0. For 93.8\% of the pairs, both annotators assigned a score of at least 4. The annotators assigned exactly the same score to 58.3\% of the pairs and differed by at most one point on 94.8\% of the pairs, with a mean absolute difference of 0.47. Disagreements larger than one point occurred in only 5.2\% of the pairs. These results provide independent human evidence that the sampled instances are generally well grounded, feasible, and consistent with their associated personas and data sources.

\begin{table*}[t]
\centering
\small
\begin{tabularx}{\textwidth}{@{}Xr@{}}
\toprule
\textbf{Measure} & \textbf{Result} \\
\midrule
Sampled instances & 16 \\
Quality dimensions & 6 \\
Mean rating & 4.67 / 5.0 \\
Rated $\geq 4$ by both annotators & 93.8\% \\
Exact agreement & 58.3\% \\
Agreement within one point & 94.8\% \\
Mean absolute difference & 0.47 \\
Difference larger than one point & 5.2\% \\
\bottomrule
\end{tabularx}
\caption{Human validation of synthesized benchmark instances. Each rating pair corresponds to one instance evaluated along one quality dimension.}
\label{tab:human_benchmark_validation}
\end{table*}

\subsubsection{Requirement-Level Ranking Alignment}

We next evaluated whether the automated metric preserves human-preferred orderings of elicited requirements. For each of the same 16 tasks, we collected the requirements produced by the four elicitation strategies and asked two independent evaluators to rank the four outputs against the same reference requirement. Evaluators considered the extent to which each output recovered the reference specification while avoiding unsupported or irrelevant content. We compared the human rankings with those induced by the automated atomic-matching $F_1$ score using Kendall's $\tau$ and Spearman's $\rho$.

The two human evaluators showed strong agreement, with Kendall's $\tau=0.81$ and Spearman's $\rho=0.88$. The automated rankings also aligned positively with both evaluators. For Evaluator~1, the correlations were $\tau=0.67$ and $\rho=0.80$; for Evaluator~2, they were $\tau=0.62$ and $\rho=0.72$. These results indicate that the automated metric broadly preserves human-preferred rankings of requirement quality.

\begin{table*}[t]
\centering
\small
\begin{tabularx}{\textwidth}{@{}Xcc@{}}
\toprule
\textbf{Comparison} & $\boldsymbol{\tau}$ & $\boldsymbol{\rho}$ \\
\midrule
Human 1 vs.\ Human 2 & 0.81 & 0.88 \\
Metric vs.\ Human 1 & 0.67 & 0.80 \\
Metric vs.\ Human 2 & 0.62 & 0.72 \\
\bottomrule
\end{tabularx}
\caption{Agreement between human rankings and rankings induced by the automated atomic-matching $F_1$ metric.}
\label{tab:human_metric_ranking}
\end{table*}

\subsubsection{Pair-Level Human--Metric Alignment}

Requirement-level ranking agreement does not necessarily imply that the automated matcher reproduces individual human matching decisions. We therefore conducted a more granular pair-level evaluation using atomic units from the same 16 tasks. Two annotators independently labeled whether each evaluated predicted--reference atomic-unit pair constituted a semantic match. We then compared the two human label sets with each other and compared each annotator's labels with the automated matcher's decisions using $F_1$.

The two human evaluators achieved pair-level agreement of $F_1=0.76$, suggesting that the matching task is reasonably well defined while still involving some semantic ambiguity. Agreement between the automated matcher and the human evaluators was more moderate: $F_1=0.48$ with Evaluator~1 and $F_1=0.53$ with Evaluator~2. Our error analysis indicates that the automated matcher is generally more conservative than the human annotators and identifies fewer predicted--reference pairs as matches.

\begin{table*}[t]
\centering
\small
\begin{tabularx}{\textwidth}{@{}Xc@{}}
\toprule
\textbf{Pair-Level Comparison} & $\mathbf{F_1}$ \\
\midrule
Human 1 vs.\ Human 2 & 0.76 \\
Metric vs.\ Human 1 & 0.48 \\
Metric vs.\ Human 2 & 0.53 \\
\bottomrule
\end{tabularx}
\caption{Pair-level agreement between the human annotators and the automated atomic matcher.}
\label{tab:pair_level_human_metric}
\end{table*}

The pair-level and requirement-level results provide complementary evidence. Although the automated matcher does not reproduce every human decision on individual atomic-unit pairs, aggregating these decisions produces requirement-level rankings that align substantially with human preferences. We therefore treat the automated metric as a useful measure for controlled relative comparisons across elicitation strategies rather than as a complete substitute for human evaluation. The moderate pair-level agreement also highlights an opportunity to improve the semantic matching component in future work.

\subsection{Prompts}

\subsubsection{Data Synthesis Prompts}

We present the schema feature extraction prompt in Table~\ref{tab:prompt_schema_feature}, the instance feature extraction prompt in Table~\ref{tab:prompt_instance_feature}, the user persona synthesis prompt in Table~\ref{tab:prompt_user_synthesis}, the task requirement synthesis prompt in Table~\ref{tab:prompt_task_systhesis}, and the cross-model validation prompt in Table~\ref{tab:prompt_cross_valid}.

\begin{table*}[h]
\centering
\caption{Schema feature extraction prompt}
\scriptsize
\label{tab:prompt_schema_feature}
\begin{tabular}{p{0.95\textwidth}}
\toprule
\begin{minipage}[t]{0.95\textwidth}
\noindent \textbf{Role: Dataset Schema Analyzer}

\vspace{1em}
\noindent \textbf{Context}
\begin{itemize}[itemsep=0pt, parsep=0pt, topsep=0pt, partopsep=0pt]
    \item \textbf{Dataset Name:} \{\{ dataset\_name \}\}
    \item \textbf{Reference Link:} \texttt{https://huggingface.co/datasets/\{\{ dataset\_name \}\}}
    \item \textbf{Objective:} Generate a structured technical profile of the dataset to serve as the ground truth for future NLP task synthesis.
\end{itemize}

\vspace{1em}
\noindent \textbf{Instruction}\\
Analyze the dataset at the link above and output a \textbf{single JSON object} containing the following keys.

\vspace{1em}
\noindent \textbf{Required JSON Structure}\\
\textbf{Constraint:} Output ONLY the JSON block. Do not include any conversational text, explanations, or markdown outside of the JSON. 

\begin{verbatim}[itemsep=0pt, parsep=0pt, topsep=0pt, partopsep=0pt]
{
  "dataset_metadata": {
    "domain": "The specific field (e.g., Biomedical, Legal, Social Media)",
    "total_instances": "Approximate count across all splits",
    "languages": ["List of ISO codes"],
    "primary_data_type": "e.g., Monolingual Text, Parallel Corpora, Dialogue, etc.",
    "description": "A detailed technical description of the dataset"
  },
  "schema_analysis": [
    {
      "column_name": "name",
      "data_type": "string/int/float/list",
      "description": "Functional definition of the column",
      "is_text": "Boolean (True/False)",
      "avg_length": "Average word/token count if applicable"
    }
  ],
  "structural_logic": {
    "relationships": "Describe how columns interact (e.g., 'col_a' is a translation of 'col_b')",
    "label_dynamics": "Description of class balance or rating scales if present",
    "potential_bottlenecks": "Any missing values or data quality issues observed"
  }
}
\end{verbatim}
\end{minipage} \\
\bottomrule
\end{tabular}
\end{table*}

\begin{table*}[h]
\centering
\caption{Instance feature extraction prompt}
\scriptsize
\label{tab:prompt_instance_feature}
\begin{tabular}{p{0.95\textwidth}}
\toprule
\begin{minipage}[t]{0.95\textwidth}

\noindent \textbf{Role}\\
You are an expert Data Scientist and Research Lead specializing in algorithmic feature engineering and synthetic data generation.

\vspace{1em}
\noindent \textbf{Objective}\\
Your objective is to perform a ``DNA Extraction'' on a dataset---identifying the latent semantic and structural patterns required to generate high-fidelity synthetic duplicates that reflect real-world complexity.

\vspace{1em}
\noindent \textbf{Context}\\
\textbf{Dataset Name:} \{\{ dataset\_name \}\}\\
\textbf{Dataset Schema \& Reference:} \{\{ data\_analysis \}\}\\
\textbf{Sampled Data Records:} \{\{ sampled\_data \}\}

\vspace{1em}
\hrule
\vspace{1em}

\noindent \textbf{Task: Multidimensional Forensic Feature Audit}\\
Perform a rigorous forensic analysis of the \texttt{sampled\_data} in direct alignment with the \texttt{data\_analysis} schema. To prevent task homogeneity, you must identify exactly \textbf{ten} unique data features, mapping each to one of the following \textbf{Feature Slots}:

\begin{enumerate}[itemsep=0pt, parsep=0pt, topsep=0pt, partopsep=0pt]
    \item \textbf{Structural Backbone:} A persistent physical pattern or structural norm (e.g., specific delimiters like \texttt{|||||} or unique whitespace encodings).
    \item \textbf{Semantic Nucleus - Conflict:} A recurring logical contradiction, information discrepancy, or shift in perspective found within the text (e.g., conflicting casualty counts or differing narrative framings).
    \item \textbf{Semantic Nucleus - Jargon:} Domain-specific nomenclature, specialized terminology, or recurring linguistic syntax that distinguishes this dataset.
    \item \textbf{The ``Long-Tail'' Anomaly:} A high-impact feature that may not be universal (prevalence of 20\%-40\%) but represents a significant edge-case challenge (e.g., extreme data sparsity or legal disclaimers).
    \item \textbf{Stylistic Signature:} The emotional tone, narrative arc, or institutional ``voice'' of the data (e.g., sensationalist tabloid vs. clinical academic tone).
    \item \textbf{Latent Logical Flow:} The underlying reasoning structure, causal linkages, or chronological progression that provides the text its internal coherence and ``common sense'' validity.
    \item \textbf{Temporal \& Evolutionary Dynamics:} The patterns of information drift, periodic fluctuations, or chronological shifts that reflect how the data evolves over time.
\end{enumerate}

\vspace{1em}
\hrule
\vspace{1em}

\noindent \textbf{Strict Persona-Driven Negative Constraints}
\begin{itemize}[itemsep=0pt, parsep=0pt, topsep=0pt, partopsep=0pt]
    \item \textbf{Exclusion Zone:} Ignore administrative metadata, IDs, or target labels (e.g., summary columns).
    \item \textbf{Independence:} Each of the ten features must be logically independent; avoid overlapping descriptions.
    \item \textbf{Schema Alignment:} Each feature must be explicitly mapped back to the primary fields in the \texttt{data\_analysis} schema.
    \item \textbf{Data-Quality Guardrails:}
    \begin{itemize}[itemsep=0pt, parsep=0pt, topsep=0pt, partopsep=0pt]
        \item \textbf{For Technical Audits:} Prohibit the identification of ``clean'' or ``perfectly formatted'' features; you must find the entropy.
        \item \textbf{For Content Synthesis:} Do not assume standard paragraph breaks or punctuation; cite the actual encoding (e.g., \texttt{NEWLINE\_CHAR}) if present.
        \item \textbf{For Feature Mapping:} Do not use generic names; every feature must cite a specific value or unique string from the \texttt{sampled\_data}.
    \end{itemize}
\end{itemize}

\vspace{1em}
\hrule
\vspace{1em}

\noindent \textbf{Output Instructions}\\
Return the analysis strictly as a JSON list containing exactly ten objects. No introductory text, conversational filler, or markdown prose.

\vspace{1em}
\noindent \textbf{Output Schema:}
\begin{verbatim}
[
  {
    "slot_type": "One of: [Structural, Semantic_Conflict, Semantic_Jargon, 
                  Anomaly, Style, Logical_Flow, Temporal]",
    "feature": "Name of the feature",
    "description": "A concise explanation of the semantic or structural pattern identified.",
    "explanation": "A technical justification of why this feature is critical for 
                    high-fidelity synthesis, citing evidence of its Prevalence and 
                    Distinctiveness from the sampled_data.",
    "cognitive_complexity": "A score from 1-10 representing the reasoning depth 
                             required to process this feature."
  }
]
\end{verbatim}
\end{minipage} \\
\bottomrule
\end{tabular}
\end{table*}

\begin{table*}[h]
\centering
\caption{User persona synthesis prompt}
\scriptsize
\label{tab:prompt_user_synthesis}
\begin{tabular}{p{0.95\textwidth}}
\toprule
\begin{minipage}[t]{0.95\textwidth}

\noindent \textbf{Role: Principal Data Scientist \& UX Architect}

\vspace{1em}
\noindent \textbf{Context}\\
\textbf{Dataset Name:} \{\{ dataset\_name \}\}\\
\textbf{Dataset Metadata \& Schema:} \{\{ dataset\_info \}\}

\vspace{1em}
\noindent \textbf{Objective}\\
Synthesize \{\{ num\_users \}\} distinct, high-fidelity user personas representing a realistic stakeholder cross-section. These personas must derive their value directly from the provided schema to solve high-stakes business problems.

\vspace{1em}
\noindent \textbf{Constraints}
\begin{enumerate}[itemsep=0pt, parsep=0pt, topsep=0pt, partopsep=0pt]
    \item \textbf{Zero-Coding Requirement:} Personas must solve problems via UI-based tools, prompting, or domain expertise. Do NOT assign tasks involving writing scripts, SQL, or API integration.
    \item \textbf{Schema-Anchored Design:} Workflows must be intrinsically tied to the feature columns. 
    \begin{itemize}[itemsep=0pt, parsep=0pt, topsep=0pt, partopsep=0pt]
        \item \textbf{Primary Focus:} Raw text or unstructured data columns (e.g., \texttt{text}, \texttt{input}).
        \item \textbf{Heuristic:} Avoid ``Circular Utility''---do not design personas whose sole goal is to reproduce existing ground-truth labels. Focus on using those labels for \textit{audit}, \textit{evaluation}, or \textit{fine-tuning}.
    \end{itemize}
    \item \textbf{Archetype Diversity:}
    \begin{itemize}[itemsep=0pt, parsep=0pt, topsep=0pt, partopsep=0pt]
        \item \textbf{Technical Spectrum:} Scale from ``Strategic/Low-Code'' (e.g., Product Manager, Compliance Officer) to ``Domain Specialist'' (e.g., Legal Counsel, Medical Reviewer).
        \item \textbf{North Star Metrics:} Distribute focus across \textit{Latency/Token Cost}, \textit{Factual Precision/Safety}, and \textit{Auditability/Transparency}.
    \end{itemize}
    \item \textbf{Task Decomposition:} For each persona, define two distinct workflows:
    \begin{itemize}[itemsep=0pt, parsep=0pt, topsep=0pt, partopsep=0pt]
        \item \textbf{Core NLP Task:} A standard utility (e.g., abstractive summarization, information extraction, or AI copilot).
        \item \textbf{Edge-Case/Constraint Task:} A high-difficulty task addressing dataset-specific hurdles (e.g., handling 10k+ token windows, mitigating domain-specific hallucinations, or style-transfer for non-expert audiences).
    \end{itemize}
\end{enumerate}

\vspace{1em}
\noindent \textbf{Output Format}\\
Return ONLY a raw JSON array of \{\{ num\_users \}\} objects. No preamble, no markdown code blocks, and no postscript.

\begin{verbatim}
[
  {
    "persona_id": 1,
    "role": "Specific Professional Title",
    "competency_matrix": {
        "proficiencies": ["List of technical/domain strengths"],
        "limitations": ["Specific gaps in technical or domain knowledge"]
    },
    "business_motivation": "The primary KPI or 'North Star' metric this user is responsible for.",
    "workflow_friction": "The specific challenge this persona faces when working with this particular dataset.",
    "potential_tasks": [
      {
        "task_name": "Standard Task Title",
        "description": "Functional objective and expected outcome."
      },
      {
        "task_name": "Edge-Case Task Title",
        "description": "Specific technical constraint (e.g., token limits, privacy, or formatting complexity)."
      }
    ]
  }
]
\end{verbatim}

\end{minipage} \\
\bottomrule
\end{tabular}
\end{table*}

\begin{table*}[h]
\centering
\caption{Task requirement synthesis prompt}
\scriptsize
\label{tab:prompt_task_systhesis}
\begin{tabular}{p{0.95\textwidth}}
\toprule
\begin{minipage}[t]{0.95\textwidth}

\noindent \textbf{Role}\\
You are a \textbf{Principal Alignment Engineer and UX Researcher}. Your expertise lies in high-fidelity task decomposition and synthetic data engineering. You bridge the gap between raw, noisy datasets and the high-level cognitive needs of specialized professional personas.

\vspace{1em}
\noindent \textbf{Context}
\begin{itemize}[itemsep=0pt, parsep=0pt, topsep=0pt, partopsep=0pt]
    \item \textbf{Dataset Name:} \{\{ dataset\_name \}\}
    \item \textbf{Target Persona:} \{\{ user\_info \}\}
    \item \textbf{Dataset Schema:} \{\{ dataset\_info \}\}
    \item \textbf{Dataset Features:} \{\{ sample\_features \}\}
\end{itemize}

\vspace{1em}
\hrule
\vspace{1em}

\noindent \textbf{Objective 1: High-Fidelity Task Generation}\\
Generate two \textbf{Task Requirements} that serve as professional benchmarks for an AI assistant. 

\vspace{1em}
\noindent \textbf{THE GOLDEN RULE: EMPIRICAL ANCHORING}\\
Generic tasks are a failure. You should randomly select one or two features from the ``Dataset Features.'', and you must anchor the task to some those features provided. 
\begin{itemize}[itemsep=0pt, parsep=0pt, topsep=0pt, partopsep=0pt]
    \item \textbf{Hard Reference Requirement:} You MUST cite a specific value, ID, or unique anomaly from the Dataset Features provided.
\end{itemize}

\vspace{1em}
\noindent \textbf{Task Tiers:}
\begin{enumerate}[itemsep=0pt, parsep=0pt, topsep=0pt, partopsep=0pt]
    \item \textbf{Task 1 (Medium - Feature Synthesis):} Require the AI to bridge two disparate fields using a non-obvious heuristic discovered in the dataset features.
    \item \textbf{Task 2 (High - Adversarial Alignment):} A ``Signal-to-Noise'' challenge. The AI must resolve a direct contradiction or extreme data sparsity found in the dataset features by applying the persona's professional logic.
\end{enumerate}

\vspace{1em}
\noindent \textbf{Each Task Requirement must include:}
\begin{itemize}[itemsep=0pt, parsep=0pt, topsep=0pt, partopsep=0pt]
    \item \textbf{Strategic Mission:} The high-stakes objective for the persona.
    \item \textbf{Data-Reality Constraints:} Explicit instructions on handling the ``Noise Profile.''
    \item \textbf{Chain-of-Thought (CoT) Logic:} A step-by-step reasoning path using the ``Latent Intelligence'' as primary decision gates.
    \item \textbf{Negative Constraints:} Prohibit common ``clean data'' assumptions.
\end{itemize}

\vspace{1em}
\hrule
\vspace{1em}

\noindent \textbf{Objective 2: Implementation Summaries}\\
Provide two informal summaries for each task:
\begin{enumerate}[itemsep=0pt, parsep=0pt, topsep=0pt, partopsep=0pt]
    \item \textbf{The Elevator Pitch:} ($\sim$30 words) Goal-oriented and punchy.
    \item \textbf{The Deep Dive:} ($\sim$100 words) Technical summary of the data-to-task mapping logic.
\end{enumerate}

\vspace{1em}
\hrule
\vspace{1em}

\noindent \textbf{Output Format \& Guardrails}\\
Output the result \textbf{strictly} in valid JSON format. 
\begin{itemize}[itemsep=0pt, parsep=0pt, topsep=0pt, partopsep=0pt]
    \item \textbf{NO} markdown code blocks (no \texttt{```json}).
    \item \textbf{NO} preamble, introductory text, or concluding remarks.
    \item \textbf{Schema:}
\end{itemize}

\begin{verbatim}
[
    {
        "task_id": 1,
        "difficulty": "Medium",
        "task_name": "String",
        "task_requirement": "Structured string (300+ words) including Mission, Constraints, CoT, and Persona-Driven Negative Constraints.",
        "data_columns": ["List of exact column names from schema used"],
        "data_features": "The specific data features impacting this task.",
        "dataset_alignment_explanation": "Explanation of how above data features affect this task.",
        "elevator_pitch_summary": "String",
        "deep_dive_summary": "String"
    },
    {
        "task_id": 2,
        "difficulty": "High",
        "task_name": "String",
        "task_requirement": "Structured string (300+ words) including Mission, Constraints, CoT, and Persona-Driven Negative Constraints.",
        "data_columns": ["List of exact column names from schema used"],
        "data_features": "The specific data features impacting this task.",
        "dataset_alignment_explanation": "Explanation of how above data features affect this task.",
        "elevator_pitch_summary": "String",
        "deep_dive_summary": "String"
    }
]
\end{verbatim}

\end{minipage} \\
\bottomrule
\end{tabular}
\end{table*}

\begin{table*}[h]
\centering
\caption{Cross-model validation prompt}
\scriptsize
\label{tab:prompt_cross_valid}
\begin{tabular}{p{0.95\textwidth}}
\toprule
\begin{minipage}[t]{0.95\textwidth}

\noindent \textbf{Role}\\
You are a \textbf{Principal Data Quality Auditor and Synthetic Data Critic}. Your tone is clinical, skeptical, and precise. Your objective is to perform a high-fidelity audit of synthesized personas and tasks to ensure they are professionally viable and technically grounded in the provided dataset.

\vspace{1em}
\noindent \textbf{Input Context}
\begin{enumerate}[itemsep=0pt, parsep=0pt, topsep=0pt, partopsep=0pt]
    \item \textbf{Dataset Metadata \& Schema:} \{\{ dataset\_info \}\}
    \item \textbf{Data Features (Empirical Evidence):} \{\{ sample\_features \}\}
    \item \textbf{Synthesized User Persona:} \{\{ user\_info \}\} 
    \item \textbf{Synthesized Task Requirement:} \{\{ task\_info \}\}
\end{enumerate}

\vspace{1em}
\noindent \textbf{Evaluation Criteria (Score 1-5)}
\begin{enumerate}[itemsep=0pt, parsep=0pt, topsep=0pt, partopsep=0pt]
    \item \textbf{Role-Schema Alignment:} Does the persona's role derive its primary value from the dataset's specific domain?
    \item \textbf{Technical Feasibility:} Can the task be completed \textbf{strictly} using the available columns in Dataset Metadata \& Schema? Penalize if it assumes external data.
    \item \textbf{High-Stakes Significance:} Does the task solve a complex business problem, or is it a trivial exercise that doesn't justify a ``Principal'' level persona?
    \item \textbf{Persona Constraint Adherence:} Does the task respect the user's ``Zero-Coding'' limitation? It must require ``Cognitive Logic'' rather than ``Scripting Logic.''
    \item \textbf{Reasoning Depth:} Does the task require multi-dimensional or adversarial reasoning, or is it a simple ``find and summarize'' request?
    \item \textbf{Empirical Grounding:} Does the task explicitly reference or resolve the unique anomalies/values found in the ``Data Features'' context?
\end{enumerate}

\vspace{1em}
\hrule
\vspace{1em}

\noindent \textbf{Output Format}\\
Return ONLY a raw JSON object. Ensure the ``judgement'' is a critical, professional peer review.

\begin{verbatim}
{
  "judgement": "String",
  "scores": {
    "role_schema_alignment": 0,
    "task_feasibility": 0,
    "task_significance": 0,
    "persona_consistency": 0,
    "reasoning_depth": 0,
    "empirical_grounding": 0
  },
  "overall_score": 0.0,
  "improvement_suggestion": "A specific, actionable refinement to enhance technical grounding or professional fidelity."
}
\end{verbatim}

\vspace{1em}
\noindent \textbf{JSON Guardrail}\\
DO NOT include markdown code blocks (e.g., \texttt{```json}). \\
DO NOT include any text before or after the JSON object.

\end{minipage} \\
\bottomrule
\end{tabular}
\end{table*}

\subsubsection{AREAs Assistant Prompts}

The AREAs assistant prompt defines the agent's role, available context, and interaction policy during requirement elicitation. We present the complete prompts used in the Hybrid Interaction strategy, including all prompt components involved in the interaction pipeline. These prompts are shown in Tables~\ref{tab:prompt_examples_1}--\ref{tab:prompt_examples_4}. Since the prompts used in the other interaction strategies largely follow the same design principles and differ only in interaction-specific modifications, we omit them for brevity. Complete prompt implementations for all interaction strategies are provided in the accompanying code release.

\subsubsection{Simulated User Prompt}

The simulated user prompt is presented in Table~\ref{tab:prompt_examples_5}. The user's communication style is defined in Table~\ref{tab:prompt_user_commu_style}.

\subsubsection{Evaluation Prompts}
Our evaluation prompts are shown in Tables~\ref{tab:prompt_req_demp}--\ref{tab:prompt_examples_6}.

\subsubsection{User Interaction Fixed Prompt}
\label{sec:user_int_fixed_prompt}
For the fixed-question baseline, the assistant asks the following ten questions before generating the final requirement.    
\begin{enumerate}[label=\textbf{Q\arabic*.}, leftmargin=*, itemsep=0.3em, topsep=0.4em]
    \item \textbf{Task Objective:} What is the core objective of this task?
    \item \textbf{Input Structure:} What does the input data look like?
    \item \textbf{Input Scale:} Roughly how much input will there be at once?
    \item \textbf{Automation Scope:} Which parts should the AI handle automatically, and which, if any, will you continue to do manually?
    \item \textbf{Target Audience:} Who is the audience of the output?
    \item \textbf{Output Granularity:} Should the output be a single combined result or separate results for each input item?
    \item \textbf{Output Format:} Are there specific formatting requirements for the output?
    \item \textbf{Tone and Style:} What tone or style should the output use?
    \item \textbf{Source Attribution:} Should the output cite specific sources or synthesize the information without attribution?
    \item \textbf{External Knowledge:} Should the AI rely strictly on the provided information, or may it use outside knowledge to fill gaps?
\end{enumerate}
\subsection{Full Results}
\label{app:full_results}
Please refer to Tables~\ref{tab:main_results_app} and~\ref{tab:main_results_app_cont} for the full experimental results, including precision, recall, and F1 scores across all datasets.

\subsection{Artifact Licensing and Intended Use}

Our benchmark is constructed from multiple publicly available datasets, and we reviewed the license or usage terms associated with each upstream source. The datasets with explicit licenses or usage restrictions are: ESConv (\texttt{CC BY-NC 4.0}), Multi-News (\textit{non-commercial research and educational use}), FineWeb-Edu (\texttt{ODC-BY 1.0}), Reuters Financial News (\texttt{Apache-2.0}), FiscalNote BillSum (\texttt{CC0-1.0}), LessWrong (\texttt{MIT}), and Asclepius Synthetic Clinical Notes (\texttt{CC BY-NC-SA 4.0}). For the remaining datasets, including GovReport, PubMed Summarization, MediaSum, ArXiv Summarization, Patent Classification, ECTSum, UK Legislation, MATH-500, and ELI5, we could not identify an explicit redistribution license from the public dataset cards or associated repositories.

Because our artifact is derived from datasets with heterogeneous licensing conditions, we adopt a conservative release policy. When redistribution rights are unclear or when the underlying content may be copyrighted, we do not include any raw upstream data in the released artifact. This includes source documents, summaries, conversations, questions, and original examples. Our release instead contains dataset names, example identifiers or indices, and the annotations produced in this work. We also provide reconstruction scripts so that users can retrieve the original data from the upstream sources under their respective licenses and access conditions. The artifact is intended exclusively for non-commercial academic research use.

\subsection{Use of AI Assistants}
\label{sec:ai_assistants}

In accordance with the conference guidelines regarding the use of generative AI tools, we explicitly declare the involvement of AI assistants in the development of this work. 

For code development, we utilized \texttt{Claude Code} (Anthropic) as an interactive programming assistant to support software engineering, debugging, and the implementation of our evaluation pipeline. For manuscript polishing, we leveraged \texttt{Gemini} (Google) strictly for linguistic editing, grammatical refinement, and prose polishing to enhance the readability and stylistic fluency of the text.

Crucially, we emphasize that all core scientific contributions—including the conceptual framework design, the synthesis methodology of the benchmark, the execution of all empirical experiments, and the definitive interpretation of the results—were entirely conceived, directed, and authored by the human researchers. The authors maintain full intellectual ownership and accountability for the accuracy and integrity of all contents presented in this paper.

\begin{table*}[t]
\centering
\scriptsize

\begin{tabular}{
>{\raggedright\arraybackslash}p{0.92\textwidth}
}

\toprule

\textbf{No Interaction Prompt} \\

\midrule

\textbf{Role.} You are an expert requirement elicitation assistant responsible for transforming underspecified user instructions into high-fidelity task requirement specifications.

\textbf{Task Setting.}

- Zero-shot requirement inference  
- No user interaction  
- No data interaction  
- No external tools  
- No iterative refinement

\textbf{Objective.}

Generate structured and domain-aware specifications that are:

- Complete  
- Faithful to user intent  
- Actionable for downstream LLM agents  
- Written in a professional expert-level tone

\textbf{General Principles.}

- Preserve the original user intent  
- Expand vague concepts into operational definitions  
- Make only necessary and reasonable inferences  
- Avoid hallucinating unsupported constraints  
- Reflect uncertainty implicitly rather than explicitly

\textbf{Persona Usage.}
\begin{verbatim}
{% if persona is defined and persona %}
User persona/context:
{{ persona }}
{% endif %}
\end{verbatim}
Persona information may refine assumptions about expertise and priorities but should not override the original instruction.

\textbf{Required Output Structure.}
\textbf{1. Strategic Intent}

- Define the mission, role, and high-level objective  
- Frame the task as a professional responsibility  
- Clarify why the task matters and what failure would imply

\textbf{2. Data Interpretation \& Pre-processing}

- Define what the input represents  
- Identify primary signals versus noise  
- Describe how different parts of the input should be interpreted or prioritized

\textbf{3. Execution Steps (Reasoning Process)}

- Provide 3--5 structured reasoning or processing steps  
- Ensure each step reflects a concrete operational stage  
- Avoid generic reasoning steps

\textbf{4. Tone \& Style}

- Define expected vocabulary, tone, and expertise level  
- Specify the desired writing style

\textbf{5. Content Requirements}

- Specify required output components  
- Define key fields, sections, or deliverables that must appear

\textbf{6. Constraints \& Guardrails}

- Define explicit DO / DO NOT rules  
- Include restrictions on reasoning, style, or content when appropriate

\textbf{7. Evaluation Criteria}

- Define how output quality should be evaluated  
- Include dimensions such as correctness, completeness, faithfulness, and domain alignment

\textbf{Important Constraints.}

- Do not output JSON  
- Do not explicitly mention uncertainty  
- Maintain concise but information-dense writing  
- Target 300 - 350 words

User instruction:

\texttt{\{\{ user\_instruction \}\}}

\\

\midrule

\textbf{Router Prompt} \\

\midrule
\textbf{Role.}

You are a strategy router for an adaptive requirement elicitation agent.

Your task is to determine which interaction action should be executed at each iteration.

\textbf{Context.}

- Initial User Requirement: the original underspecified user request  
- Current Predicted Task Requirement: the agent’s current refined requirement specification  
- Elicitation Guideline: accumulated confirmed findings, pending confirmations, and unresolved gaps  
- Iteration Status: current iteration, maximum iterations, and remaining iterations

\textbf{Requirement Audit.}

Before routing, perform a mandatory audit comparing the current requirement specification against the original elevator pitch.

\textbf{Agent-Added Requirement Audit.}

For each requirement introduced beyond the original user request, classify it as:

- \texttt{[UNCONFIRMED-USER]}: subjective or scope assumptions requiring user confirmation  
- \texttt{[UNCONFIRMED-DATA]}: objective data-related claims requiring dataset verification  
- \texttt{[CONFIRMED]}: already validated through user interaction or data inspection

Any \texttt{[UNCONFIRMED-USER]} item must be confirmed or removed before final synthesis.

\textbf{Open Gap Identification.}

Identify missing information implied by the elevator pitch but not yet covered in the current requirement specification.

\textbf{Routing Policy.}

\textbf{Phase 1: Structured Bootstrap (Iterations 0--5).}

Follow a fixed alternating schedule:

- Iterations 0, 2, 4: \texttt{data\_interaction}  
- Iterations 1, 3, 5: \texttt{user\_interaction}

The bootstrap schedule overrides all other routing logic.

\textbf{Phase 2: Dynamic Routing (Iterations 6+).}

Apply routing decisions in the following priority order:

\textbf{Priority 1: Confirm Agent Inventions}

- If any \texttt{[UNCONFIRMED-USER]} item exists, choose \texttt{user\_interaction}  
- Present the most important unconfirmed assumption for user validation

\textbf{Priority 2: Fill Open Gaps}

- User intent or preference gaps $\rightarrow$ \texttt{user\_interaction}  
- Data behaviour or edge-case gaps $\rightarrow$ \texttt{data\_interaction}

\textbf{Priority 3: Verify Data Claims}

- If unresolved \texttt{[UNCONFIRMED-DATA]} items remain, choose \texttt{data\_interaction}

\textbf{Priority 4: Reflection \& Synthesis}

- If all audit items are confirmed and no gaps remain, choose \texttt{reflection\_synthesis}

\textbf{Global Constraints.}

- At least two \texttt{data\_interaction} steps must occur before synthesis  
- \texttt{reflection\_synthesis} is forbidden before iteration 15  
- If \texttt{current\_iteration >= max\_iterations - 1}, synthesis becomes mandatory  
- Avoid repeating the same action consecutively for the same unresolved item

\textbf{Required Output Format.}
Return valid JSON only:
\begin{verbatim}
{"action": "...", "reason": "...", "target_gap": "..."}
\end{verbatim}
\textbf{Output Constraints.}

- Do not output explanations outside JSON  
- The \texttt{target\_gap} field must reference the exact unresolved audit item or open gap  
- The \texttt{reason} field must explicitly justify the routing decision
\\
\bottomrule

\end{tabular}

\caption{
Representative prompts used in AREAs assistant.
We present representative prompts from the interaction framework.
Prompts used in other interaction strategies follow similar structures with interaction-specific modifications and are omitted for brevity.
The complete prompt implementations are provided in the accompanying code release.
}

\label{tab:prompt_examples_1}

\end{table*}

\newpage

\begin{table*}[t]
\centering
\scriptsize

\begin{tabular}{
>{\raggedright\arraybackslash}p{0.92\textwidth}
}

\toprule

\textbf{Data Analysis Prompt} \\

\midrule

\textbf{Role.}

You are a requirement elicitation assistant performing a data observation turn.

Your task is to inspect real dataset samples in light of the current task requirement and report grounded observations.

\textbf{Context.}

- Initial User Requirement: the original underspecified user request serving as the primary task anchor  
- Current Task Requirement: the agent’s current working hypothesis of the complete requirement specification  
- Data Samples: representative examples drawn from the actual dataset

\textbf{Objective.}

Analyze the dataset samples against both the initial user requirement and the current task requirement.

Your role is observational rather than generative:

- Do not rewrite the requirement specification  
- Do not refine or synthesize requirements  
- Do not generate structured task instructions

Instead, identify what the data:

- Confirms  
- Contradicts  
- Leaves ambiguous  
- Suggests may require future modification

\textbf{Observation Rules.}

Before reporting an observation, verify that:

- It directly relates to the user’s requested deliverable or failure mode  
- It reflects a recurring or salient data pattern rather than a one-off artifact  
- It concerns information not already fully captured in the current requirement  
- It can be stated concisely in a single sentence

\textbf{Observation Categories.}

\textbf{Confirmed Findings.}

- Report recurring data patterns that validate or ground an existing requirement  
- Emphasize concrete evidence or important nuances revealed by the data

\textbf{Gaps and Contradictions.}

- Identify mismatches between the current requirement and the observed data  
- Highlight underspecified areas not adequately addressed by the current requirement

\textbf{Potential Reasons to Modify the Guideline.}

- Flag data-grounded signals suggesting a requirement may need to be added, revised, or removed  
- Each observation must specify which requirement is affected and why the data supports reconsideration  
- These observations require future user confirmation

\textbf{Required Output Structure.}

\textbf{Confirmed Findings}

- One sentence per confirmed observation

\textbf{Gaps and Contradictions}

- One sentence per identified mismatch or underspecified area
\textbf{Potential Reasons to Modify the Guideline}
- One sentence describing the affected requirement and supporting evidence  
- Append \texttt{[PENDING USER CONFIRMATION]}

\textbf{Output Constraints.}

- Output plain text only  
- Do not output JSON  
- Do not generate markdown code fences  
- Do not rewrite the task requirement  
- Keep observations concise and evidence-grounded
\\
\midrule

\textbf{User interaction (Adaptive) Prompt} \\

\midrule
\textbf{Role.}

You are an AI assistant helping a user build an AI-powered system tailored to their needs.

Your task is to gather sufficient information to produce a precise and complete task requirement.

\textbf{Interaction Constraints.}

- Maximum iterations: \texttt{\{\{ max\_iterations \}\}}  
- Current iteration: \texttt{\{\{ current\_iteration \}\}}  
- Remaining iterations: \texttt{\{\{ max\_iterations - current\_iteration \}\}}

The interaction process must cover all critical task aspects within the remaining turns.

\textbf{Interaction Principles.}

- Prioritize breadth over depth  
- Avoid over-focusing on partially explored topics  
- Prefer unresolved high-impact ambiguities over minor refinements  
- Avoid repetitive or low-information questions

\textbf{Context.}

- Initial User Requirement: the original underspecified user request  
- Current Task Requirement: the agent's current refined understanding of the task  
- Elicitation Guideline: accumulated confirmed findings, pending data observations, and unresolved gaps  
- Conversation History: prior user-assistant interactions

\textbf{Objective.}

Ask exactly one focused question that maximizes elicitation value for the current interaction turn.

The prompt supports two interaction modes:

\textbf{Mode A: General Elicitation}

Use when major intent gaps remain unresolved.

Focus on:

- User goals and desired outcomes  
- Audience expertise and usage context  
- Success and failure criteria  
- Scope, format, or domain constraints

\textbf{Mode B: Data-Finding Confirmation}

Use when high-stakes pending data findings require explicit user confirmation before synthesis.

Focus on:

- Confirming or rejecting pending data-derived observations  
- Clarifying whether a data observation should become part of the requirement specification

\textbf{Mode Selection Policy.}

- Prefer Mode A when significant intent gaps remain  
- Use Mode B only for high-impact unresolved data findings  
- Avoid consecutive turns focused exclusively on data confirmation unless no intent gaps remain

\textbf{Question Selection Criteria.}

Before generating a question, evaluate:

- What critical intent information remains uncovered  
- Whether unresolved data findings materially affect the requirement specification  
- Which ambiguity has the highest downstream impact on synthesis quality  
- Which question provides the greatest information gain given the remaining iterations

\textbf{Required Output Structure.}

Return a JSON object containing:

\begin{verbatim}
{"analysis_reasoning": "...","identified_ambiguity": "...","question": "..."}
\end{verbatim}

\textbf{Field Requirements.}

- \texttt{analysis\_reasoning}: explain which interaction mode (A or B) was selected and why  
- \texttt{identified\_ambiguity}: describe the unresolved ambiguity addressed by the question  
- \texttt{question}: a single focused high-impact question

\textbf{Output Constraints.}

- Output valid JSON only  
- Do not generate explanations outside JSON  
- Ask exactly one question  
- Avoid broad or generic clarification questions  
- Do not repeat previously asked questions
\\
\bottomrule

\end{tabular}

\caption{
Representative prompts used in AREAs assistant.
We present representative prompts from the interaction framework.
Prompts used in other interaction strategies follow similar structures with interaction-specific modifications and are omitted for brevity.
The complete prompt implementations are provided in the accompanying code release.
}

\label{tab:prompt_examples_2}

\end{table*}

\newpage

\begin{table*}[t]
\centering
\scriptsize

\begin{tabular}{
>{\raggedright\arraybackslash}p{0.92\textwidth}
}

\toprule

\textbf{Guideline update Prompt} \\

\midrule

\textbf{Role.}

You are a requirement elicitation assistant responsible for maintaining an evolving Elicitation Guideline throughout the interaction process.

\textbf{Context.}

- Initial User Requirement: the original underspecified user request  
- Current Elicitation Guideline: the accumulated summary of confirmed findings, unresolved issues, and pending observations from previous turns  
- New Interaction: the latest interaction turn, which may be either a user interaction or a data interaction

\textbf{Objective.}

Update the Elicitation Guideline after each interaction turn.

The updated guideline should maintain a compact and structured summary of:

- Confirmed requirements  
- Pending data-derived findings requiring user validation  
- Explicitly rejected requirements  
- Remaining unresolved gaps

\textbf{Guideline Update Policy.}

\textbf{Confirmed Requirements.}

- Include requirements directly supported by explicit user feedback  
- Include pending data findings only after the user explicitly confirms them  
- Treat confirmed requirements as authoritative task constraints

\textbf{Pending Data Findings.}

- Include observations discovered through data interactions that have not yet been validated by the user  
- Mark every pending finding with \texttt{[PENDING USER CONFIRMATION]}  
- Pending findings remain provisional until explicitly confirmed or rejected

\textbf{Rejected Requirements.}

- Remove any pending finding explicitly rejected by the user  
- Add rejected findings to a dedicated rejected section  
- Mark rejected items with \texttt{[REJECTED — DELETE FROM TASK REQUIREMENT]}  
- Rejected requirements must be removed during future synthesis

\textbf{Open Gaps.}

- Track unresolved ambiguities, unanswered questions, and missing details.
- Include only unresolved gaps not covered by interactions or data evidence.

\textbf{Update Constraints.}

- Do not invent unsupported requirements  
- Keep each guideline item concise and limited to a single sentence  
- Preserve consistency with prior confirmed findings  
- Distinguish clearly between confirmed, pending, rejected, and unresolved information

\textbf{Required Output Structure.}

\textbf{Confirmed Requirements}

- One bullet per confirmed requirement

\textbf{Pending Data Findings (Awaiting User Confirmation)}

- One bullet per pending finding  
- Append \texttt{[PENDING USER CONFIRMATION]}

\textbf{Rejected Requirements}

- One bullet per rejected requirement  
- Append \texttt{[REJECTED — DELETE FROM TASK REQUIREMENT]}

\textbf{Open Gaps}

- One bullet per unresolved ambiguity or missing requirement

\textbf{Output Constraints.}

- Output plain text only  
- Do not output JSON  
- Do not generate markdown code fences  
- Preserve the exact required section structure  
- Keep the guideline concise and information-dense \\

\midrule

\textbf{Task requirement generation Prompt} \\

\midrule
\textbf{Role.}

You are an expert prompt engineer responsible for producing a final task requirement specification by synthesizing evidence collected through iterative user interactions and data interactions.

\textbf{Context.}

- Original User Requirement: the initial underspecified user request  
- Current Task Requirement Draft: the agent's current partially refined requirement specification  
- User Interactions: authoritative user feedback collected during elicitation  
- Data Interactions: observations, edge cases, and contradictions identified through real data inspection

\textbf{Objective.}

Generate a final task requirement specification that integrates all reliable evidence collected during the elicitation process.

The synthesis process must:

- Preserve requirements supported by strong evidence  
- Remove unsupported or contradicted requirements  
- Resolve inconsistencies across interaction sources  
- Produce a coherent, internally consistent, and complete specification

\textbf{Evidence-Gating Rules.}

Before retaining or introducing any requirement, verify that it is supported by at least one of the following evidence sources:
- Explicit user feedback from user interactions  
- Repeated observations appearing across multiple independent data interaction turns  
- Direct and unambiguous implications of the original user requirement

Requirements failing the evidence-gating rule must be removed entirely rather than weakened or softened.

\textbf{Evidence Integration Policy.}

- Treat user feedback as authoritative ground truth for intent and preference  
- Incorporate data observations only when corroborated across multiple turns or confirmed by user feedback  
- Preserve valid content from the current draft when supported by evidence  
- Remove any requirement explicitly marked for deletion during elicitation

\textbf{Conflict Resolution Priority.}

Resolve conflicts using the following hierarchy:

- User feedback  
- Multi-turn data reflections  
- Current task requirement draft  
- Original user requirement

\textbf{Synthesis Principles.}

- Prefer precision over excessive coverage  
- Remove speculative or weakly supported requirements  
- Maintain consistency across all sections  
- Integrate reasoning naturally into the final specification

\textbf{Required Output Structure.}

\textbf{1. Strategic Intent}

- Define the mission, role, and high-level objective  
- Clarify why the task matters and what failure would imply

\textbf{2. Data Interpretation \& Pre-processing}

- Define input structure and meaning  
- Identify important signals and noise patterns  
- Describe interpretation and prioritization strategies
\\
\bottomrule

\end{tabular}

\caption{
Prompts used in AREAs assistant.
We present representative prompts from the interaction framework.
Prompts used in other interaction strategies follow similar structures with interaction-specific modifications and are omitted for brevity.
The complete prompt implementations are provided in the accompanying code release.
}

\label{tab:prompt_examples_3}

\end{table*}

\newpage

\begin{table*}[t]
\centering
\scriptsize

\begin{tabular}{
>{\raggedright\arraybackslash}p{0.92\textwidth}
}

\toprule

\textbf{Task generation Prompt (Cont.)} \\

\midrule

\textbf{3. Execution Steps (Reasoning Process)}

- Provide 3--5 operational reasoning steps  
- Ensure steps are domain-specific and actionable

\textbf{4. Tone \& Style}

- Define expected tone, vocabulary, and expertise level  
- Specify the intended writing style

\textbf{5. Content Requirements}

- Define required output components and deliverables  
- Specify mandatory sections or fields

\textbf{6. Constraints \& Guardrails}

- Define explicit DO / DO NOT rules  
- Include restrictions on reasoning, style, or content

\textbf{7. Evaluation Criteria}

- Define evaluation dimensions such as correctness, completeness, faithfulness, and domain alignment

\textbf{Output Constraints.}

- Output plain text only  
- Do not output JSON  
- Do not generate markdown code fences  
- Do not explicitly mention uncertainty or inference status  
- Maintain concise but information-dense writing  
- The final specification must contain 300--350 words  
- Maintain a consistent expert-level tone throughout
\\
\bottomrule

\end{tabular}

\caption{
Prompts used in AREAs assistant.(Cont.)
}

\label{tab:prompt_examples_4}

\end{table*}

\begin{table*}[t]
\centering
\scriptsize

\begin{tabular}{
>{\raggedright\arraybackslash}p{0.92\textwidth}
}

\toprule

\textbf{Simulated user prompt} \\

\midrule
\textbf{Role.}

You are simulating a human user interacting with an AI assistant.
Your task is to provide persona-grounded feedback while intentionally withholding information unless the assistant asks directly relevant questions.

\textbf{Context.}

- User Profile: \texttt{\{\{ user\_profile \}\}}  
- Communication Style: \texttt{\{\{ communication\_habit \}\}}  
- Initial Instruction: \texttt{\{\{ init\_user\_instruction \}\}}  
- Hidden Actual Needs: \texttt{\{\{ gold\_task\_requirement \}\}}

The hidden actual needs represent information the user has not fully articulated yet and should only be revealed when directly prompted by relevant questions.

\textbf{Conversation Context.}

- Conversation History: \texttt{\{\{ conversation\_history \}\}}  
- Assistant Question: \texttt{\{\{ agent\_question \}\}}

\textbf{Grounding \& Disclosure Logic.}

Evaluate the assistant’s question against the hidden actual needs using the following grounding categories:

\textbf{Supported}

- The question directly aligns with the user’s actual needs  
- Reveal a small but meaningful detail

\textbf{Partially Supported}

- The question contains unsupported assumptions  
- Clarify only the grounded portion while hedging unsupported implications

\textbf{Unsupported}

- The question introduces ideas unrelated to the actual needs  
- Respond with uncertainty or lack of consideration

\textbf{Disclosure Principles.}

- Be vague when the question is vague  
- Be specific only when the question is specific  
- Never reveal the entire requirement specification at once

\textbf{Style \& Persona Constraints.}

- Maintain conversational and human-like responses  
- Avoid formal, academic, or highly structured language  
- Use natural imperfections and slightly vague phrasing  
- Do not use bullet points or numbered lists in responses  
- Adapt tone naturally to the assigned persona and expertise level

\textbf{Negative Constraints.}

- Do not ask clarification questions  
- Do not volunteer unprompted information  
- Follow communication-length constraints exactly  
- Avoid revealing hidden requirements unless directly triggered by the assistant's question

\textbf{Required Output Structure.}

Return a JSON object containing:

\begin{verbatim}
{"thought": "...","grounding": "...","feedback": "..."}
\end{verbatim}

\textbf{Field Requirements.}

- \texttt{thought}: brief reasoning about the support level and disclosure strategy  
- \texttt{grounding}: one of \texttt{supported}, \texttt{partially\_supported}, or \texttt{unsupported}  
- \texttt{feedback}: conversational user response following all persona and formatting constraints

\textbf{Output Constraints.}

- Output valid JSON only  
- Do not generate explanations outside JSON  
- Do not use question marks in the feedback field  
- Do not use bullet points or numbered lists in the feedback field
\\
\bottomrule

\end{tabular}

\caption{
Sumulated User Prompt
}

\label{tab:prompt_examples_5}

\end{table*}

\begin{table*}[t]
\centering
\scriptsize

\begin{tabular}{
>{\raggedright\arraybackslash}p{0.92\textwidth}
}

\toprule
\noindent {\small \textbf{Passive}}

\vspace{0.5em}
\noindent \hspace*{1em}\textbf{Metadata}
\begin{itemize}[itemsep=0pt, parsep=0pt, topsep=0pt, partopsep=0pt]
    \item \textbf{Tone:} Underspecified \& Vague
    \item \textbf{Description:} Correct grammar but limited expression. Not withholding information on purpose, but struggling to articulate specific needs.
\end{itemize}

\noindent \hspace*{1em}\textbf{Constraints}
\begin{itemize}[itemsep=0pt, parsep=0pt, topsep=0pt, partopsep=0pt]
    \item \textbf{Sentences:} one to two sentences
    \item \textbf{Words:} 15 to 35 words
    \item \textbf{Allow Questions:} false
    \item \textbf{Allow Volunteering:} false
\end{itemize}

\noindent \hspace*{1em}\textbf{Disclosure Logic}
\begin{itemize}[itemsep=0pt, parsep=0pt, topsep=0pt, partopsep=0pt]
    \item \textbf{Vague Ask:} Provide a vague answer.
    \item \textbf{Grounded Ask:} Reveal slightly more detail.
    \item \textbf{Assumption Ask:} Hedge and resist unsupported assumptions.
    \item \textbf{Global Rules:}
    \begin{itemize}[itemsep=0pt, parsep=0pt, topsep=0pt, partopsep=0pt]
        \item Reveal information incrementally only
        \item Never summarize full intent
        \item Keep latent requirements hidden
    \end{itemize}
\end{itemize}

\noindent \hspace*{1em}\textbf{Style Rules}
\begin{itemize}[itemsep=0pt, parsep=0pt, topsep=0pt, partopsep=0pt]
    \item \textbf{Do:} Use hedging (e.g., `I think', `maybe', `kind of'); Leave meaning loosely expressed; Maintain a slightly uncertain persona
    \item \textbf{Don't:} Give precise or complete answers; Elaborate or explain logic; Ask questions back
\end{itemize}

\vspace{1.5em}
\hrule
\vspace{1.5em}

\noindent {\small \textbf{Normall}}

\vspace{0.5em}
\noindent \hspace*{1em}\textbf{Metadata}
\begin{itemize}[itemsep=0pt, parsep=0pt, topsep=0pt, partopsep=0pt]
    \item \textbf{Tone:} Accurate \& Concise
    \item \textbf{Description:} Direct and clear. Addresses exactly what was asked with well-formed sentences without providing extra context.
\end{itemize}

\noindent \hspace*{1em}\textbf{Constraints}
\begin{itemize}[itemsep=0pt, parsep=0pt, topsep=0pt, partopsep=0pt]
    \item \textbf{Sentences:} two to three sentences
    \item \textbf{Words:} 30 to 60 words
    \item \textbf{Allow Questions:} false
    \item \textbf{Allow Volunteering:} false
\end{itemize}

\noindent \hspace*{1em}\textbf{Disclosure Logic}
\begin{itemize}[itemsep=0pt, parsep=0pt, topsep=0pt, partopsep=0pt]
    \item \textbf{Vague Ask:} Provide a vague answer.
    \item \textbf{Grounded Ask:} Reveal specific details relevant to the question.
    \item \textbf{Assumption Ask:} Answer cautiously; resist unsupported parts.
    \item \textbf{Global Rules:}
    \begin{itemize}[itemsep=0pt, parsep=0pt, topsep=0pt, partopsep=0pt]
        \item Reveal information incrementally only
        \item Never summarize full intent
        \item Focus strictly on the direct scope of the question
    \end{itemize}
\end{itemize}

\noindent \hspace*{1em}\textbf{Style Rules}
\begin{itemize}[itemsep=0pt, parsep=0pt, topsep=0pt, partopsep=0pt]
    \item \textbf{Do:} Use well-formed, complete sentences; Express points clearly; Address only the specific query
    \item \textbf{Don't:} Over-explain; Volunteer unrequested details; Ask questions back
\end{itemize}

\vspace{1.5em}
\hrule
\vspace{1.5em}

\noindent {\small \textbf{Active}}

\vspace{0.5em}
\noindent \hspace*{1em}\textbf{Metadata}
\begin{itemize}[itemsep=0pt, parsep=0pt, topsep=0pt, partopsep=0pt]
    \item \textbf{Tone:} Engaged \& Expressive
    \item \textbf{Description:} Conversational and helpful. Proactively discloses one relevant detail from the latent requirements when triggered.
\end{itemize}

\noindent \hspace*{1em}\textbf{Constraints}
\begin{itemize}[itemsep=0pt, parsep=0pt, topsep=0pt, partopsep=0pt]
    \item \textbf{Sentences:} three to five sentences
    \item \textbf{Words:} 50 to 100 words
    \item \textbf{Allow Questions:} false
    \item \textbf{Allow Volunteering:} true
\end{itemize}

\noindent \hspace*{1em}\textbf{Disclosure Logic}
\begin{itemize}[itemsep=0pt, parsep=0pt, topsep=0pt, partopsep=0pt]
    \item \textbf{Vague Ask:} Respond with a high-level overview.
    \item \textbf{Grounded Ask:} Reveal significant detail plus one related latent requirement.
    \item \textbf{Assumption Ask:} Clarify while resisting incorrect assumptions.
    \item \textbf{Global Rules:}
    \begin{itemize}[itemsep=0pt, parsep=0pt, topsep=0pt, partopsep=0pt]
        \item Disclose exactly one extra relevant detail to progress the conversation
        \item Never reveal the full requirement in one go
        \item Avoid formal specification tone
    \end{itemize}
\end{itemize}

\noindent \hspace*{1em}\textbf{Style Rules}
\begin{itemize}[itemsep=0pt, parsep=0pt, topsep=0pt, partopsep=0pt]
    \item \textbf{Do:} Sound engaged and conversational; Hint at core priorities without full disclosure; Add relevant context to the answer
    \item \textbf{Don't:} Use structured formats (bullet points, lists); Invent new constraints not in the gold requirement; Ask questions back
\end{itemize} \\
\bottomrule
\end{tabular}

\caption{
Three user communication styles: Passive, Normal, and Active
}

\label{tab:prompt_user_commu_style}

\end{table*}

\begin{table*}[t]
\centering
\scriptsize

\begin{tabular}{
>{\raggedright\arraybackslash}p{0.92\textwidth}
}

\toprule

\textbf{Task requirement decompose prompt} \\

\midrule
\textbf{Role.}

You are an expert evaluator for AI-Assisted User Needs Understanding (AUNU).

Your task is to decompose a task requirement specification into atomic requirement units.

\textbf{Context.}

- Task Requirement: \texttt{\{\{ task\_requirement \}\}}

\textbf{Objective.}

Transform the task requirement into a list of atomic requirement units representing the smallest independently meaningful requirements contained in the specification.

\textbf{Decomposition Principles.}

- Each atomic unit must express exactly one requirement, objective, constraint, instruction, or evaluation criterion  
- Split broad or compound statements into multiple units when they contain multiple independent requirements  
- Preserve the original semantic meaning of the specification  
- Maintain concise, normalized, and precise wording  
- Avoid redundant or overlapping units  
- Do not introduce unsupported assumptions or inferred requirements

\textbf{Atomic Unit Criteria.}

A valid atomic requirement unit should:

- Represent a single independently verifiable requirement  
- Remain understandable without additional decomposition  
- Avoid combining multiple constraints or objectives into one statement  
- Preserve the intent and specificity of the original text

\textbf{Required Output Structure.}

Return a JSON object containing:
\begin{verbatim}
{"atomic_units": ["..."]}
\end{verbatim}
\textbf{Output Constraints.}

- Output valid JSON only  
- Do not generate explanations outside JSON  
- Do not omit explicitly stated requirements  
- Do not generate inferred or hallucinated requirements  
- Keep each atomic unit concise and self-contained
\\
\bottomrule

\end{tabular}

\caption{
Evaluation prompt (Requirement decomposition)
}

\label{tab:prompt_req_demp}

\end{table*}

\newpage

\begin{table*}[t]
\centering
\scriptsize

\begin{tabular}{
>{\raggedright\arraybackslash}p{0.92\textwidth}
}

\toprule

\textbf{Task requirement classification prompt} \\

\midrule
\textbf{Role.}

You are an expert evaluator for AI-Assisted User Needs Understanding (AUNU).

Your task is to classify atomic requirement units into data-specified and user-specified categories.

\textbf{Context.}

- Atomic Requirement Units: a list of requirement units indexed by zero-based integer IDs

\textbf{Objective.}

Assign every atomic requirement unit to exactly one category:

- \texttt{data\_specified}  
- \texttt{user\_specified}

The classification should distinguish requirements derived from dataset-specific observations from requirements reflecting user intent, preferences, or quality expectations.

\textbf{Category Definitions.}

\textbf{Data-Specified Requirements.}

Requirements that can only be discovered through direct dataset inspection, including:

- Raw dataset structure  
- Field names or schema information  
- Delimiters or encoding artifacts  
- Dataset-specific formatting quirks  
- Observable structural properties unique to the dataset

Examples include delimiter rules, formatting artifacts, schema conventions, or parsing instructions tied to the raw data representation.

\textbf{User-Specified Requirements.}

Requirements reflecting the user’s intent, preferences, objectives, constraints, or evaluation standards, including:

- Output formatting preferences  
- Faithfulness or precision requirements  
- Analytical goals or evaluation criteria  
- Task-specific instructions or quality expectations

These requirements remain meaningful independently of the dataset’s raw structure.

\textbf{Classification Principles.}

- Assign every unit ID to exactly one category  
- Use \texttt{data\_specified} only when the requirement depends on dataset-specific observations unavailable without inspecting the raw data  
- Use \texttt{user\_specified} for instructions, preferences, constraints, or quality standards originating from the user  
- When uncertain, prefer \texttt{user\_specified}  
- If a unit mixes both categories, assign it to the dominant category

\textbf{Required Output Structure.}

Return a JSON object containing:

\begin{verbatim}

{"data_specified": [0, 3, 7],"user_specified": [1, 2, 4]}

\end{verbatim}

\textbf{Output Constraints.}
- Output valid JSON only  
- Do not generate explanations outside JSON  
- Every input ID must appear exactly once  
- Do not omit or duplicate IDs  
- Preserve the original integer indexing from the input
\\
\bottomrule

\end{tabular}

\caption{
Evaluation prompt (Requirement classification)
}

\label{tab:prompt_req_cls}

\end{table*}

\newpage

\begin{table*}[t]
\centering
\scriptsize

\begin{tabular}{
>{\raggedright\arraybackslash}p{0.92\textwidth}
}

\toprule

\textbf{Task requirement unit alignment prompt} \\

\midrule
\textbf{Role.}

You are an expert evaluator for AI-Assisted User Needs Understanding (AUNU).

Your task is to compare predicted atomic task requirements against ground-truth atomic task requirements.

\textbf{Context.}

- Ground-Truth Atomic Units: reference requirement units indexed by zero-based integer IDs  
- Predicted Atomic Units: generated requirement units indexed by zero-based integer IDs

\textbf{Objective.}

Evaluate how faithfully the predicted atomic units recover the original task specification represented by the ground-truth units.

The evaluation should identify:

- Correct semantic matches  
- Missing requirements  
- Hallucinated requirements  
- Misaligned or distorted interpretations  
- Critical missing requirements

\textbf{Alignment Principles.}

- Match units based on semantic equivalence rather than surface wording  
- A valid match preserves the same underlying intent and requirement semantics  
- Multiple predicted units may align with a single ground-truth unit when the original idea is decomposed across multiple predictions  
- Each ground-truth unit may appear at most once in the matched alignment list

\textbf{Missing Units.}

- Ground-truth units not recovered by any prediction must be classified as missing  
- Omitted details are not considered implicitly recovered

\textbf{Hallucinated Units.}

- Predicted units unsupported by the ground-truth specification must be classified as hallucinated

\textbf{Misaligned Units.}

- Predicted units that partially match but distort, weaken, or misinterpret a ground-truth requirement must be classified as misaligned  
- Misaligned predictions should not appear in the matched pairs list

\textbf{Critical Requirement Identification.}

- Identify ground-truth units representing essential objectives, constraints, or core requirements  
- Mark critical units absent from the prediction as critical missing requirements

\textbf{Evaluation Principles.}

- Apply strict evaluation standards  
- Do not reward vague or underspecified predictions when the ground truth is specific  
- Treat the ground-truth specification as the only source of truth  
- Do not infer omitted requirements implicitly  
- Prioritize faithful recovery of the original task specification

\textbf{Required Output Structure.}

Return a JSON object containing:

\begin{verbatim}

{"matched_pairs": [{"ground_truth": 0,"predicted": [2, 5]}],"missing_units": [1, 3],"hallucinated_units": [0, 4],

"misaligned_units": [{"predicted": 1,"ground_truth": 2}],"critical_units": [0, 1],"critical_missing"[1]}
\end{verbatim}
\textbf{Output Constraints.}
- Output valid JSON only  
- Do not generate explanations outside JSON  
- Use only integer IDs from the input lists  
- Do not duplicate IDs across categories inconsistently  
- Ensure every classification is semantically justified by the alignment rules \\

\bottomrule

\end{tabular}

\caption{
Evaluation prompt for requirement alignment
}

\label{tab:prompt_examples_6}

\end{table*}

\newpage

\begin{table*}[!t]
\centering
\scriptsize
\setlength{\tabcolsep}{1.6pt}
\resizebox{\textwidth}{!}{%
\begin{tabular}{llcccccccccccc}
\toprule
\multirow{3}{*}{Dataset}
& \multirow{3}{*}{Strategy}
& \multicolumn{12}{c}{Performance of each Backbone Model (\%)} \\
\cmidrule(lr){3-14}
& & \multicolumn{4}{c}{Claude-Sonnet-4.6}
& \multicolumn{4}{c}{Gemini-3.1-Pro}
& \multicolumn{4}{c}{GPT-5.4} \\
\cmidrule(lr){3-6}
\cmidrule(lr){7-10}
\cmidrule(lr){11-14}
& & Pr. & Re. & F1 avg. & F1 std.
  & Pr. & Re. & F1 avg. & F1 std.
  & Pr. & Re. & F1 avg. & F1 std. \\
\midrule
\multirow{5}{*}{MultiNews}
& No Interaction & $26.72$ & $47.36$ & $33.81$ & $6.85$ & $29.52$ & $46.07$ & $35.71$ & $12.32$ & $20.43$ & $54.07$ & $29.45$ & $7.75$ \\
& Data Interaction & $20.91$ & $39.72$ & $27.31$ & $12.76$ & $24.90$ & $35.98$ & $29.31$ & $11.40$ & $23.27$ & $51.66$ & $31.87$ & $8.61$ \\
& User Interaction (Fixed) & $27.86$ & $63.34$ & $38.42$ & $9.76$ & $36.75$ & $\mathbf{55.88}$ & $44.20$ & $12.20$ & $24.51$ & $65.66$ & $35.59$ & $9.77$ \\
& User Interaction (Adaptive) & $25.76$ & $\mathbf{64.53}$ & $36.50$ & $5.59$ & $36.86$ & $53.35$ & $43.37$ & $9.98$ & $29.16$ & $70.96$ & $40.93$ & $7.37$ \\
& Hybrid Interaction & $\mathbf{35.58}$ & $62.56$ & $\mathbf{44.60}$ & $12.55$ & $\mathbf{39.22}$ & $55.15$ & $\mathbf{45.43}$ & $10.70$ & $\mathbf{35.12}$ & $\mathbf{72.10}$ & $\mathbf{46.91}$ & $8.62$ \\
\midrule
\multirow{5}{*}{Clinical}
& No Interaction & $19.31$ & $31.52$ & $23.90$ & $7.95$ & $17.19$ & $26.18$ & $20.60$ & $6.67$ & $13.64$ & $37.63$ & $19.97$ & $4.14$ \\
& Data Interaction & $20.59$ & $36.83$ & $26.33$ & $10.06$ & $21.48$ & $31.33$ & $25.34$ & $6.51$ & $17.28$ & $38.97$ & $23.81$ & $7.53$ \\
& User Interaction (Fixed) & $25.82$ & $55.08$ & $34.94$ & $6.97$ & $32.53$ & $\mathbf{52.10}$ & $40.05$ & $7.57$ & $25.63$ & $\mathbf{67.14}$ & $36.83$ & $5.00$ \\
& User Interaction (Adaptive) & $25.01$ & $50.12$ & $33.18$ & $5.96$ & $28.64$ & $41.94$ & $33.83$ & $7.26$ & $27.18$ & $66.79$ & $38.55$ & $7.32$ \\
& Hybrid Interaction & $\mathbf{39.22}$ & $\mathbf{64.36}$ & $\mathbf{48.60}$ & $11.92$ & $\mathbf{40.01}$ & $51.96$ & $\mathbf{44.84}$ & $15.18$ & $\mathbf{32.46}$ & $61.23$ & $\mathbf{42.06}$ & $8.51$ \\
\midrule
\multirow{5}{*}{UK-Legal}
& No Interaction & $13.92$ & $24.95$ & $17.69$ & $4.91$ & $16.31$ & $27.70$ & $20.40$ & $5.45$ & $15.16$ & $43.46$ & $22.36$ & $6.62$ \\
& Data Interaction & $19.84$ & $35.25$ & $25.12$ & $10.53$ & $21.60$ & $31.72$ & $25.31$ & $5.74$ & $16.16$ & $40.23$ & $22.96$ & $5.58$ \\
& User Interaction (Fixed) & $32.31$ & $\mathbf{72.90}$ & $44.48$ & $7.58$ & $28.45$ & $\mathbf{57.08}$ & $37.97$ & $6.90$ & $28.27$ & $\mathbf{76.36}$ & $40.71$ & $9.34$ \\
& User Interaction (Adaptive)& $27.33$ & $61.64$ & $37.76$ & $6.18$ & $33.59$ & $49.67$ & $39.96$ & $5.95$ & $\mathbf{29.45}$ & $71.73$ & $\mathbf{41.32}$ & $8.61$ \\
& Hybrid Interaction & $\mathbf{39.14}$ & $70.51$ & $\mathbf{50.07}$ & $8.71$ & $\mathbf{34.12}$ & $49.10$ & $\mathbf{40.02}$ & $8.88$ & $27.32$ & $57.98$ & $36.87$ & $12.00$ \\
\midrule
\multirow{5}{*}{ESConv}
& No Interaction & $19.42$ & $33.76$ & $24.40$ & $9.42$ & $21.90$ & $33.67$ & $26.35$ & $8.65$ & $14.63$ & $40.28$ & $21.24$ & $4.17$ \\
& Data Interaction & $20.15$ & $37.07$ & $25.76$ & $6.17$ & $24.20$ & $36.15$ & $28.65$ & $6.69$ & $15.18$ & $37.10$ & $21.38$ & $5.33$ \\
& User Interaction (Fixed)  & $21.09$ & $57.75$ & $30.90$ & $13.57$ & $32.31$ & $\mathbf{58.45}$ & $41.62$ & $10.89$ & $25.86$ & $68.16$ & $37.08$ & $9.42$ \\
& User Interaction (Adaptive) & $\mathbf{26.35}$ & $\mathbf{65.36}$ & $\mathbf{37.42}$ & $7.29$ & $35.67$ & $56.27$ & $\mathbf{43.39}$ & $4.46$ & $28.64$ & $\mathbf{69.13}$ & $40.19$ & $6.97$ \\
& Hybrid Interaction & $25.16$ & $37.83$ & $29.87$ & $8.02$ & $\mathbf{36.67}$ & $51.52$ & $42.45$ & $6.23$ & $\mathbf{30.59}$ & $62.89$ & $\mathbf{40.92}$ & $7.01$ \\
\midrule
\multirow{5}{*}{ArXiv Sum}
& No Interaction & $18.74$ & $28.03$ & $22.36$ & $4.50$ & $22.73$ & $27.38$ & $24.67$ & $5.49$ & $16.26$ & $35.92$ & $22.33$ & $5.60$ \\
& Data Interaction & $23.90$ & $36.71$ & $28.83$ & $7.25$ & $20.72$ & $26.32$ & $22.92$ & $9.40$ & $20.40$ & $38.69$ & $26.67$ & $8.08$ \\
& User Interaction (Fixed) & $27.95$ & $\mathbf{63.72}$ & $38.86$ & $8.19$ & $31.48$ & $\mathbf{50.13}$ & $38.67$ & $13.11$ & $25.38$ & $\mathbf{69.64}$ & $37.20$ & $6.28$ \\
& User Interaction (Adaptive) & $29.10$ & $57.92$ & $38.61$ & $8.31$ & $34.04$ & $42.98$ & $37.90$ & $7.37$ & $\mathbf{29.55}$ & $60.75$ & $\mathbf{39.58}$ & $8.03$ \\
& Hybrid Interaction & $\mathbf{33.56}$ & $51.54$ & $\mathbf{40.34}$ & $9.72$ & $\mathbf{36.03}$ & $46.10$ & $\mathbf{40.31}$ & $12.93$ & $28.36$ & $53.75$ & $36.85$ & $9.09$ \\
\midrule
\multirow{5}{*}{GovReport}
& No Interaction & $26.21$ & $45.38$ & $32.95$ & $9.56$ & $26.25$ & $42.27$ & $32.07$ & $8.73$ & $19.44$ & $52.72$ & $28.24$ & $6.02$ \\
& Data Interaction & $22.48$ & $42.93$ & $29.32$ & $7.78$ & $28.37$ & $42.81$ & $33.79$ & $8.00$ & $18.05$ & $45.81$ & $25.81$ & $6.91$ \\
& User Interaction (Fixed) & $30.96$ & $\mathbf{75.83}$ & $43.97$ & $7.54$ & $28.36$ & $\mathbf{72.62}$ & $40.79$ & $5.27$ & $28.68$ & $\mathbf{72.32}$ & $41.07$ & $3.38$ \\
& User Interaction (Adaptive) & $33.92$ & $74.73$ & $\mathbf{46.31}$ & $5.21$ & $35.65$ & $55.56$ & $43.06$ & $8.06$ & $29.92$ & $68.71$ & $\mathbf{41.68}$ & $8.37$ \\
& Hybrid Interaction & $\mathbf{33.95}$ & $56.17$ & $42.21$ & $7.96$ & $\mathbf{38.67}$ & $54.28$ & $\mathbf{44.67}$ & $8.82$ & $\mathbf{31.15}$ & $63.76$ & $41.47$ & $6.90$ \\
\midrule
\multirow{5}{*}{MediaSum}
& No Interaction & $22.33$ & $31.89$ & $26.13$ & $8.70$ & $24.87$ & $31.69$ & $27.63$ & $5.63$ & $19.68$ & $42.79$ & $26.86$ & $8.45$ \\
& Data Interaction & $22.06$ & $34.86$ & $26.91$ & $8.45$ & $23.35$ & $29.10$ & $25.71$ & $5.69$ & $16.43$ & $32.74$ & $21.76$ & $5.83$ \\
& User Interaction (Fixed) & $25.63$ & $\mathbf{65.60}$ & $36.86$ & $7.94$ & $32.63$ & $\mathbf{55.67}$ & $41.14$ & $13.58$ & $28.72$ & $\mathbf{64.56}$ & $39.66$ & $4.96$ \\
& User Interaction (Adaptive) & $27.70$ & $50.91$ & $35.61$ & $8.13$ & $\mathbf{39.16}$ & $47.56$ & $\mathbf{42.57}$ & $13.94$ & $\mathbf{36.90}$ & $45.57$ & $40.50$ & $9.42$ \\
& Hybrid Interaction & $\mathbf{33.71}$ & $47.15$ & $\mathbf{38.85}$ & $8.06$ & $34.70$ & $41.34$ & $37.59$ & $11.17$ & $36.72$ & $53.28$ & $\mathbf{43.49}$ & $7.67$ \\
\midrule
\multirow{5}{*}{Patent}
& No Interaction & $13.13$ & $25.85$ & $17.34$ & $9.28$ & $17.19$ & $25.82$ & $20.48$ & $8.53$ & $13.72$ & $36.62$ & $19.90$ & $6.17$ \\
& Data Interaction & $19.94$ & $36.57$ & $25.63$ & $9.41$ & $23.47$ & $33.52$ & $27.43$ & $8.48$ & $19.22$ & $43.11$ & $26.48$ & $10.04$ \\
& User Interaction (Fixed) & $24.05$ & $\mathbf{67.57}$ & $35.47$ & $13.79$ & $31.15$ & $\mathbf{64.28}$ & $41.96$ & $13.75$ & $22.18$ & $\mathbf{68.29}$ & $33.48$ & $6.96$ \\
& User Interaction (Adaptive) & $30.16$ & $63.53$ & $\mathbf{40.59}$ & $11.35$ & $\mathbf{37.98}$ & $52.23$ & $\mathbf{43.70}$ & $12.21$ & $23.83$ & $61.72$ & $34.37$ & $8.82$ \\
& Hybrid Interaction & $\mathbf{30.71}$ & $51.24$ & $37.86$ & $12.92$ & $33.51$ & $50.22$ & $39.75$ & $5.23$ & $\mathbf{27.52}$ & $61.69$ & $\mathbf{37.77}$ & $7.49$ \\
\midrule
\multirow{5}{*}{PubMed}
& No Interaction & $20.67$ & $35.41$ & $25.95$ & $7.93$ & $22.23$ & $32.59$ & $26.13$ & $8.89$ & $16.84$ & $44.07$ & $24.30$ & $7.05$ \\
& Data Interaction & $19.62$ & $36.03$ & $25.22$ & $5.62$ & $26.59$ & $37.55$ & $30.98$ & $7.92$ & $18.25$ & $41.21$ & $25.18$ & $5.33$ \\
& User Interaction (Fixed) & $26.01$ & $\mathbf{70.14}$ & $37.95$ & $8.59$ & $22.04$ & $\mathbf{65.03}$ & $32.92$ & $9.55$ & $25.70$ & $\mathbf{77.77}$ & $38.63$ & $8.30$ \\
& User Interaction (Adaptive) & $27.74$ & $61.64$ & $38.16$ & $11.15$ & $27.06$ & $62.06$ & $37.51$ & $8.05$ & $26.03$ & $65.24$ & $37.09$ & $9.28$ \\
& Hybrid Interaction & $\mathbf{32.05}$ & $54.96$ & $\mathbf{40.31}$ & $10.64$ & $\mathbf{37.35}$ & $54.01$ & $\mathbf{43.98}$ & $9.53$ & $\mathbf{29.23}$ & $58.54$ & $\mathbf{38.82}$ & $6.46$ \\
\midrule
\multirow{5}{*}{ECTSum}
& No Interaction & $18.26$ & $30.16$ & $22.60$ & $7.26$ & $24.14$ & $36.28$ & $28.72$ & $7.33$ & $18.24$ & $48.19$ & $26.27$ & $4.30$ \\
& Data Interaction & $19.61$ & $35.29$ & $25.05$ & $7.28$ & $22.88$ & $34.84$ & $27.56$ & $9.51$ & $18.67$ & $45.79$ & $26.31$ & $6.97$ \\
& User Interaction (Fixed)& $26.31$ & $\mathbf{64.33}$ & $37.35$ & $5.81$ & $30.88$ & $\mathbf{52.58}$ & $38.91$ & $11.76$ & $26.89$ & $\mathbf{67.89}$ & $38.43$ & $8.07$ \\
& User Interaction (Adaptive) & $24.05$ & $56.39$ & $33.58$ & $5.48$ & $35.00$ & $50.06$ & $41.09$ & $8.49$ & $27.69$ & $65.92$ & $38.85$ & $4.13$ \\
& Hybrid Interaction & $\mathbf{30.77}$ & $53.27$ & $\mathbf{38.76}$ & $6.78$ & $\mathbf{36.33}$ & $48.61$ & $\mathbf{41.34}$ & $10.93$ & $\mathbf{31.91}$ & $63.31$ & $\mathbf{42.24}$ & $5.68$ \\
\midrule
\multirow{5}{*}{FineWeb-Edu}
& No Interaction & $17.27$ & $26.09$ & $20.69$ & $9.26$ & $21.56$ & $30.82$ & $25.31$ & $11.15$ & $15.60$ & $36.60$ & $21.72$ & $6.89$ \\
& Data Interaction & $17.04$ & $28.86$ & $21.36$ & $9.26$ & $26.78$ & $36.51$ & $30.73$ & $11.21$ & $16.66$ & $37.35$ & $22.99$ & $8.13$ \\
& User Interaction (Fixed) & $25.70$ & $\mathbf{59.48}$ & $35.89$ & $9.50$ & $26.86$ & $\mathbf{57.05}$ & $36.52$ & $9.53$ & $22.31$ & $\mathbf{62.45}$ & $32.88$ & $9.66$ \\
& User Interaction (Adaptive) & $27.27$ & $54.87$ & $36.18$ & $10.46$ & $31.67$ & $43.40$ & $36.43$ & $7.71$ & $24.62$ & $57.93$ & $34.37$ & $6.10$ \\
& Hybrid Interaction & $\mathbf{30.06}$ & $52.44$ & $\mathbf{38.14}$ & $8.71$ & $\mathbf{36.93}$ & $49.12$ & $\mathbf{42.04}$ & $9.58$ & $\mathbf{25.69}$ & $53.02$ & $\mathbf{34.52}$ & $6.91$ \\

\bottomrule
\end{tabular}%
}
\caption{Full evaluation results on the first 11 datasets across three backbone models. Precision (Pr.), recall (Re.), and F1 average (F1 avg.) columns report mean scores; F1 std reports the corresponding standard deviation. Bold indicates the best mean score for each dataset, model, and metric.}
\label{tab:main_results_app}
\end{table*}

\begin{table*}[t]
\centering
\scriptsize
\setlength{\tabcolsep}{1.6pt}
\resizebox{\textwidth}{!}{%
\begin{tabular}{llcccccccccccc}
\toprule
\multirow{3}{*}{Dataset}
& \multirow{3}{*}{Strategy}
& \multicolumn{12}{c}{Performance of each Backbone Model (\%)} \\
\cmidrule(lr){3-14}
& & \multicolumn{4}{c}{Claude-Sonnet-4.6}
& \multicolumn{4}{c}{Gemini-3.1-Pro}
& \multicolumn{4}{c}{GPT-5.4} \\
\cmidrule(lr){3-6}
\cmidrule(lr){7-10}
\cmidrule(lr){11-14}
& & Pr. & Re. & F1 avg. & F1 std
  & Pr. & Re. & F1 avg. & F1 std
  & Pr. & Re. & F1 avg. & F1 std \\
\midrule
\multirow{5}{*}{MATH-500}
& No Interaction & $19.76$ & $30.96$ & $23.77$ & $7.99$ & $21.02$ & $31.09$ & $24.76$ & $7.94$ & $15.79$ & $37.78$ & $22.15$ & $9.85$ \\
& Data Interaction & $18.56$ & $34.97$ & $24.05$ & $7.78$ & $28.95$ & $39.65$ & $33.06$ & $8.30$ & $16.51$ & $38.18$ & $22.91$ & $5.86$ \\
& User Interaction (Fixed) & $36.50$ & $\mathbf{69.47}$ & $47.22$ & $9.25$ & $38.74$ & $42.87$ & $40.70$ & $7.82$ & $28.77$ & $\mathbf{70.26}$ & $40.59$ & $5.68$ \\
& User Interaction (Adaptive) & $27.19$ & $55.87$ & $36.19$ & $9.24$ & $35.48$ & $50.63$ & $41.20$ & $13.63$ & $26.86$ & $63.97$ & $37.53$ & $11.13$ \\
& Hybrid Interaction & $\mathbf{39.91}$ & $64.96$ & $\mathbf{49.08}$ & $9.38$ & $\mathbf{38.98}$ & $\mathbf{55.25}$ & $\mathbf{45.43}$ & $6.05$ & $\mathbf{35.50}$ & $68.43$ & $\mathbf{46.37}$ & $7.55$ \\
\midrule
\multirow{5}{*}{Reuters}
& No Interaction & $18.47$ & $34.05$ & $23.71$ & $7.22$ & $20.36$ & $32.17$ & $24.88$ & $7.50$ & $15.90$ & $45.64$ & $23.49$ & $3.32$ \\
& Data Interaction & $19.34$ & $39.47$ & $25.86$ & $4.33$ & $23.37$ & $35.33$ & $27.82$ & $7.82$ & $15.44$ & $41.03$ & $22.33$ & $6.83$ \\
& User Interaction (Fixed) & $25.78$ & $\mathbf{68.51}$ & $37.46$ & $5.94$ & $32.47$ & $\mathbf{55.76}$ & $41.04$ & $9.72$ & $29.47$ & $\mathbf{77.10}$ & $42.48$ & $3.87$ \\
& User Interaction (Adaptive) & $29.81$ & $65.85$ & $\mathbf{40.82}$ & $10.35$ & $35.16$ & $50.78$ & $41.23$ & $11.48$ & $\mathbf{31.99}$ & $73.23$ & $\mathbf{44.31}$ & $8.33$ \\
& Hybrid Interaction & $\mathbf{32.01}$ & $54.78$ & $40.16$ & $9.62$ & $\mathbf{36.66}$ & $51.52$ & $\mathbf{42.51}$ & $9.31$ & $30.82$ & $60.55$ & $40.51$ & $10.26$ \\
\midrule
\multirow{5}{*}{ELI5}
& No Interaction & $19.90$ & $36.24$ & $25.38$ & $4.22$ & $18.71$ & $29.35$ & $22.68$ & $7.15$ & $15.33$ & $42.78$ & $22.45$ & $4.72$ \\
& Data Interaction & $21.32$ & $43.88$ & $28.52$ & $6.02$ & $25.32$ & $38.49$ & $29.89$ & $6.87$ & $19.57$ & $48.65$ & $27.78$ & $7.30$ \\
& User Interaction (Fixed) & $\mathbf{33.37}$ & $\mathbf{66.82}$ & $\mathbf{44.25}$ & $6.61$ & $\mathbf{38.46}$ & $\mathbf{55.18}$ & $\mathbf{45.01}$ & $11.12$ & $24.34$ & $63.97$ & $35.22$ & $8.25$ \\
& User Interaction (Adaptive) & $24.94$ & $57.96$ & $34.56$ & $7.67$ & $31.89$ & $46.79$ & $37.43$ & $9.24$ & $\mathbf{27.72}$ & $\mathbf{70.24}$ & $\mathbf{39.51}$ & $7.36$ \\
& Hybrid Interaction & $26.81$ & $49.73$ & $34.49$ & $10.47$ & $29.29$ & $46.43$ & $35.66$ & $8.65$ & $23.75$ & $55.17$ & $32.99$ & $7.38$ \\
\midrule
\multirow{5}{*}{BillSum}
& No Interaction & $17.37$ & $31.37$ & $21.92$ & $6.96$ & $20.21$ & $30.32$ & $23.93$ & $7.64$ & $12.39$ & $33.53$ & $17.95$ & $3.65$ \\
& Data Interaction & $20.75$ & $38.14$ & $26.51$ & $5.93$ & $16.76$ & $25.82$ & $20.10$ & $7.30$ & $15.63$ & $34.86$ & $21.43$ & $6.62$ \\
& User Interaction (Fixed) & $24.43$ & $\mathbf{53.71}$ & $33.58$ & $10.24$ & $23.06$ & $57.18$ & $32.87$ & $13.36$ & $25.65$ & $\mathbf{67.28}$ & $36.84$ & $8.30$ \\
& User Interaction (Adaptive) & $25.85$ & $50.50$ & $33.95$ & $7.44$ & $23.64$ & $\mathbf{61.08}$ & $34.08$ & $13.43$ & $28.82$ & $62.46$ & $\mathbf{39.04}$ & $9.01$ \\
& Hybrid Interaction & $\mathbf{29.79}$ & $46.59$ & $\mathbf{36.13}$ & $8.65$ & $\mathbf{31.26}$ & $42.84$ & $\mathbf{35.66}$ & $7.64$ & $\mathbf{29.27}$ & $55.03$ & $37.88$ & $7.83$ \\
\midrule
\multirow{5}{*}{LessWrong}
& No Interaction & $15.92$ & $26.32$ & $19.67$ & $5.98$ & $16.09$ & $21.83$ & $18.35$ & $5.54$ & $11.89$ & $29.69$ & $16.89$ & $4.39$ \\
& Data Interaction & $17.12$ & $27.56$ & $21.03$ & $6.43$ & $20.14$ & $30.18$ & $23.94$ & $8.32$ & $11.93$ & $26.74$ & $16.43$ & $4.09$ \\
& User Interaction (Fixed) & $29.02$ & $\mathbf{61.39}$ & $39.18$ & $6.70$ & $36.24$ & $51.49$ & $42.52$ & $11.95$ & $27.48$ & $\mathbf{70.13}$ & $39.27$ & $4.76$ \\
& User Interaction (Adaptive) & $28.43$ & $57.04$ & $37.66$ & $8.74$ & $39.20$ & $\mathbf{52.46}$ & $\mathbf{44.69}$ & $9.40$ & $28.90$ & $65.38$ & $39.89$ & $7.19$ \\
& Hybrid Interaction & $\mathbf{34.79}$ & $55.74$ & $\mathbf{42.62}$ & $9.19$ & $\mathbf{40.06}$ & $50.44$ & $44.50$ & $7.36$ & $\mathbf{31.77}$ & $60.91$ & $\mathbf{41.51}$ & $8.44$ \\
\bottomrule
\end{tabular}%
}
\caption{Continuation of full evaluation results on the remaining four datasets across three backbone models. Precision (Pr.), Recall (Re.), and F1 average (F1 avg.) columns report mean scores; F1 std reports the corresponding standard deviation. Bold indicates the best mean score for each dataset, model, and metric.}
\label{tab:main_results_app_cont}
\end{table*}

\end{document}